\documentclass[aps,preprint,floats,epsf,epsfig,nofootinbib,letter]{revtex4}
\usepackage{epsfig}
\usepackage{graphicx}
\usepackage{dcolumn}
\usepackage{bm}
\usepackage{subfigure}
\usepackage{hyperref}                 
\usepackage{amsmath}

\begin{document}
\def\be{\begin{eqnarray}}
\def\en{\end{eqnarray}}
\def\non{\nonumber}
\def\la{\langle}
\def\ra{\rangle}
\def\Br{{\mathcal B}}
\def\A{{\mathcal A}}
\def\B{{\cal B}}
\def\D{{\cal D}}
\def\Bbar{\overline{\cal B}}
\def\bfB{{\rm\bf B}}
\def\bfBB{{\rm\bf B}\overline{\rm\bf B}}
\def\BB{{{\cal B} \overline {\cal B}}}
\def\BD{{{\cal B} \overline {\cal D}}}
\def\DB{{{\cal D} \overline {\cal B}}}
\def\DD{{{\cal D} \overline {\cal D}}}
\def\sq{\sqrt}


\title{Two-body Baryonic $B_{u,d,s}$ and $B_c$ to Charmless Final State Decays}

\author{Chun-Khiang Chua}
\affiliation{Department of Physics and Center for High Energy Physics,
Chung Yuan Christian University,
Chung-Li, Taiwan 320, Republic of China}

\date{\today}

\begin{abstract}
We study the rates and direct CP violations of two-body baryonic $\bar B_{u,d,s}\to {{\rm\bf B}\overline{\rm\bf B}}'$ and $B^-_c\to{{\rm\bf B}\overline{\rm\bf B}}'$ decays, where the final state baryons include low-lying octet and decuplet baryons. We incorporate topological amplitude formalism and the factorization approach. Asymptotic relations at large $m_b$ are used to simplify decay amplitudes. Using the most up-to-date data on $\bar B^0\to p\bar p$ and $B^-\to\Lambda\bar p$ decay rates as inputs, rates and direct CP violations of $\bar B_{u,d,s}\to {{\rm\bf B}\overline{\rm\bf B}}'$ decays are revised and predicted.
It is interesting that the results on rates satisfy all existing experimental bounds and some are close to the bounds. 
Factorization diagrams contribute to penguin-exchange, exchange, annihilation and penguin-annihilation amplitudes. 
Although the resulting penguin-exchange factorization amplitudes are sizable, the rest suffer from severe chiral suppression and are sensitive to non-factorizable contributions.
As the $\bar B_s\to p\bar p$ decay is governed by exchange and penguin annihilation amplitudes, the rate predicted in factorization calculation is very rare, but 
it can be enhanced by including non-factorizable contributions. 
The case where the rate is enhanced to saturate the present experimental bound through the enhancement on exchange or penguin-annihilation amplitudes is discussed.
As annihilation modes $B^-_c\to{{\rm\bf B}\overline{\rm\bf B}}'$ decays from factorization calculation are found to be 
very rare, 
but they can be enhanced by including non-factorizable contributions as well. 
Small direct CP violations of pure penguin modes in $\Delta S=-1$ $\bar B_{u,d,s}\to{{\rm\bf B}\overline{\rm\bf B}}'$ decays are robust predictions of the SM, while vanishing direct CP violations of exchange modes in $\bar B_{u,d,s}\to{{\rm\bf B}\overline{\rm\bf B}}'$ decays and in all $B^-_c\to{{\rm\bf B}\overline{\rm\bf B}}'$ decay modes are null tests of the SM.
\end{abstract}


\maketitle


\vfill\eject

\section{Introduction}

Two body baryonic $B$ decays with octet and decuplet baryons have been searched experimentally for some time. 
The present situation is summarized in Table~\ref{tab: expt}~\cite{LHCb:2016nbc, Belle:2007gob,Belle:2007lbz,Belle:2007oni,LHCb:2017swz,LHCb:2022lff,Belle:2019abe,CLEO:1989xsn, PDG}. 
So far only the $\bar B^0\to p\bar p$ and $B^-\to\Lambda \bar p$ modes have been observed \cite{LHCb:2016nbc, LHCb:2017swz}. 
As most of the bounds have not been updated over a decade, experimental progress in this sector from LHCb and Belle~II in near future is anticipated.

Theoretically two body baryonic $B$ decays have beed studied in various approaches, including 
pole model~\cite{Deshpande:1987nc,Jarfi:1990ej,Cheng:2001tr,Cheng:2001ub},
sum rule~\cite{Chernyak:ag}, diquark
model~\cite{Ball:1990fw,Chang:2001jt}, flavor symmetry ~\cite{Gronau:1987xq,He:re,Sheikholeslami:fa,Luo:2003pv,Chua:2003it,Chua:2013zga,Chua:2016aqy},  factorization ~\cite{He:2006vz,Hsiao:2014zza,Hsiao:2019wyd,Jin:2021onb} and some other calculations~\cite{Cheng:2014qxa}.
For some recent reviews, see \cite{review,Huang:2021qld}.

In this work we will employ the approach of refs. \cite{Chua:2003it,Chua:2013zga,Chua:2016aqy}, 
which made use of the well established topological amplitude formalism~\cite{Zeppenfeld:1980ex,Chau:tk,Chau:1990ay,Gronau:1994rj,Gronau:1995hn,Chiang:2004nm,Cheng:2014rfa,Savage:ub} and asymptotic relations~\cite{Brodsky:1980sx} in the large $m_b$ limit. 
Note that the approach successfully predicted the $B^-\to \Lambda \bar p$ rate \cite{Chua:2013zga} using the data of $\bar B^0\to p\bar p$ decay~\cite{Aaij:2013fta}. 

We shall extend the previous study in several aspects. 
First, additional topological amplitudes will be introduced in $\bar B_{q=u,d,s}\to\bfBB'$ decays with $\bfB$ denoting low lying octet and decuplet baryons. 
Second, some of the topological amplitudes have factorization contributions, which can be calculated using factorization approach. 
For some important progress of diagrammatic approach with factorization assisted, one is referred to refs. \cite{Li:2012cfa, Qin:2013tje}.
Note that $\bar B_{u,d,s}$ decaying to low lying octet baryon pairs have been studied in refs.~\cite{Hsiao:2014zza,Jin:2021onb} using factorization approach, 
but our formalism is different and, consequently, we will be able to extend the study to include all low lying octet and decuplet baryon pairs.
Third, we will study $B^-_c\to \bfBB'$ decays and will give predictions on rates and direct CP violations.

There are accumulating speculations of new physics effects in rare $B$ decays, see, for example, \cite{Altmannshofer:2021qrr,Cornella:2021sby} from some recent discussions. 
Any test of the Standard Model (SM) should be welcomed. 
In this work we try to identify some robust predictions from SM and null tests of the SM in rare $B$ decays in the baryonic sector.

The layout of this paper is as following. We give the formalism in Sec. II, which is followed by numerical results on rates and direct CP violations of $\bar B_{u,d,s}\to\bfBB'$ and $B^-_c\to\bfBB'$ decays in Sec III. 
Sec. IV is devoted to discussions and conclusions. We end this paper by two appendices.

\begin{table}[t!]
\caption{\label{tab: expt}
Experimental results of $\bar B_{u,d,s}\to\bfBB'$ branching ratios. The upper limits are at 90\% confidence level.
}
\footnotesize{
\begin{ruledtabular}
\begin{tabular}{lcccr}
Mode
          & LHCb
          & Belle
          & CLEO
          & PDG~\cite{PDG}
          \\
\hline $B^-\to\Lambda\bar p$
          & $(2.4^{+1.0}_{-0.8}\pm 0.3)\times 10^{-7}$ \cite{LHCb:2016nbc}
          & $<3.2\times 10^{-7}$ \cite{Belle:2007gob}
          & 
          & $(2.4^{+1.0}_{-0.9})\times 10^{-7}$
          \\
$B^-\to\Sigma^{*0}\bar p$
          & 
          & $<4.7\times 10^{-7}$ \cite{Belle:2007lbz}
          &
          & $<4.7\times 10^{-7}$
          \\
$B^-\to\Lambda\overline{\Delta^+}$
          & 
          & $<8.2\times 10^{-7}$ \cite{Belle:2007lbz}
          & 
          & $<8.2\times 10^{-7}$
          \\
$B^-\to\Delta^0\bar p$
          & 
          & $<1.38\times 10^{-6}$ \cite{Belle:2007oni}
          & 
          & $< 1.38\times 10^{-6}$
          \\
$B^-\to p \overline{\Delta^{++}}$
          & 
          & $<1.4\times 10^{-7}$ \cite{Belle:2007oni}
          & 
          & $<1.4\times 10^{-7}$
          \\          
\hline $\bar B^0\to p\bar p$
          & $(1.27\pm 0.13\pm 0.05\times 0.03)\times 10^{-8}$ \cite{LHCb:2022lff} 
          &  $<1.1\times 10^{-7}$ \cite{Belle:2007gob}
          &
          & $(1.25\pm 0.32)\times 10^{-8}$
          \\
$\bar B^0\to\Sigma^{*+}\bar p$
          & 
          & $<2.6\times 10^{-7}$ \cite{Belle:2007lbz}
          &
          & $<2.6\times 10^{-7}$
          \\
$\bar B^0\to p\overline{\Delta^+}, \Delta^-\bar p$
          & 
          & $<1.6\times 10^{-6}$ \cite{Belle:2019abe}
          &
          & $<1.6\times 10^{-6}$
          \\          
$\bar B^0\to \Lambda\overline{\Delta^0}$
          & 
          & $<9.3\times 10^{-7}$ \cite{Belle:2007lbz}
          &
          & $<9.3\times 10^{-7}$
          \\
$\bar B^0\to \Lambda\overline{\Lambda}$
          & 
          & $<3.2\times 10^{-7}$ \cite{Belle:2007gob}
          &
          & $<3.2\times 10^{-7}$
          \\
$\bar B^0\to \Delta^0\overline{\Delta^0}$
          & 
          &
          & $<1.5\times 10^{-3}$ \cite{CLEO:1989xsn}
          & $<1.5\times 10^{-3}$
          \\ 
$\bar B^0\to \Delta^{++}\overline{\Delta^{++}}$
          & 
          &
          & $<1.1\times 10^{-4}$ \cite{CLEO:1989xsn}
          & $<1.1\times 10^{-4}$
          \\ 
\hline $\bar B^0_s\to p\bar p$
          & $<4.4\times 10^{-9}$  \cite{LHCb:2022lff}
          & 
          &
          & $<1.5\times 10^{-8}$
          \\                                                         
\end{tabular}
\end{ruledtabular}
}
\end{table}

\section{Formalism}

\subsection{Topological amplitudes}

\begin{figure}[t]
\centering
 \subfigure[]{
  \includegraphics[width=0.35\textwidth]{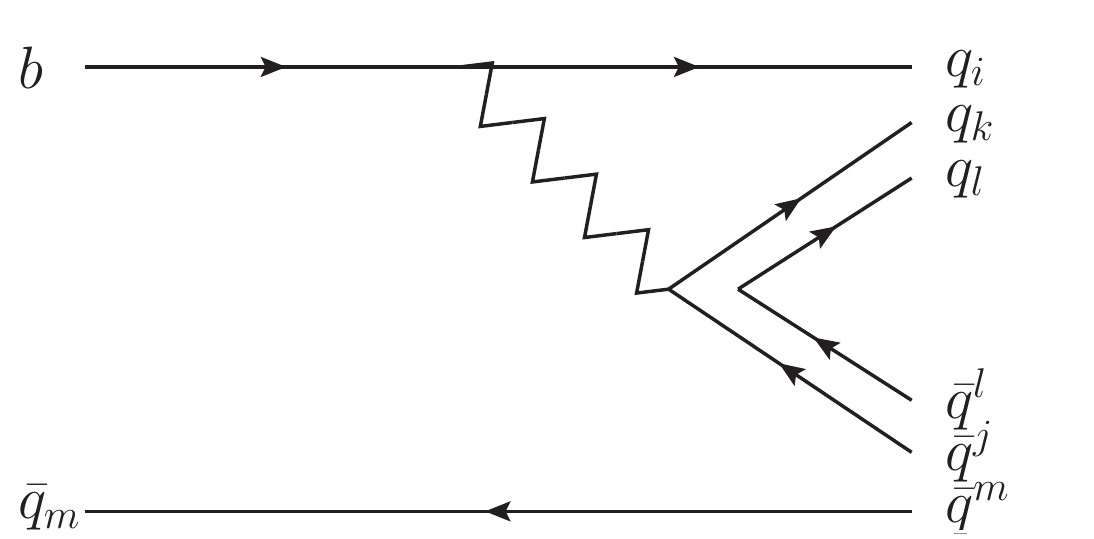}
}
\hspace{12pt}
\subfigure[]{
  \includegraphics[width=0.35\textwidth]{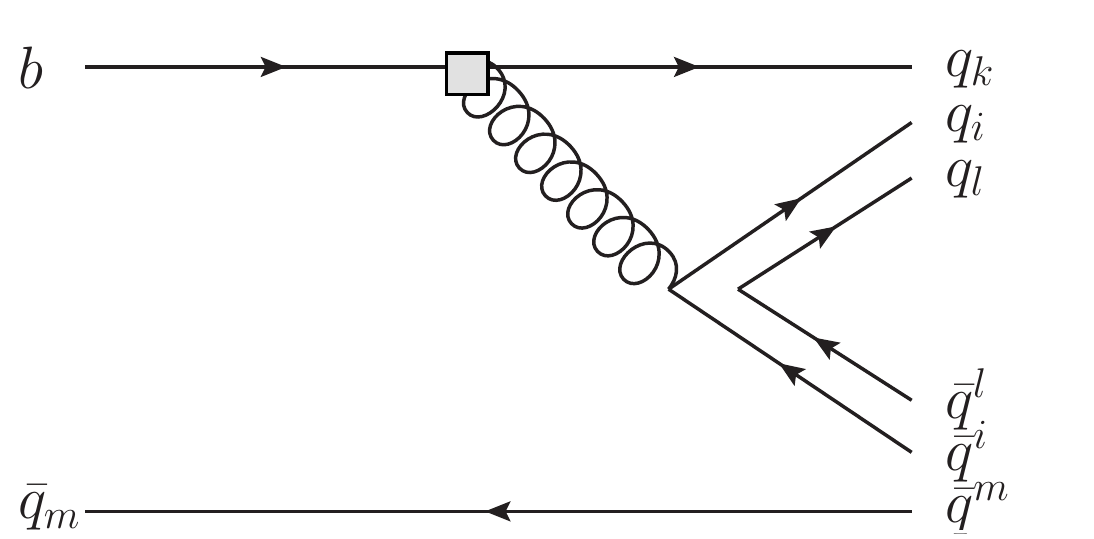}
}\\\subfigure[]{
  \includegraphics[width=0.35\textwidth]{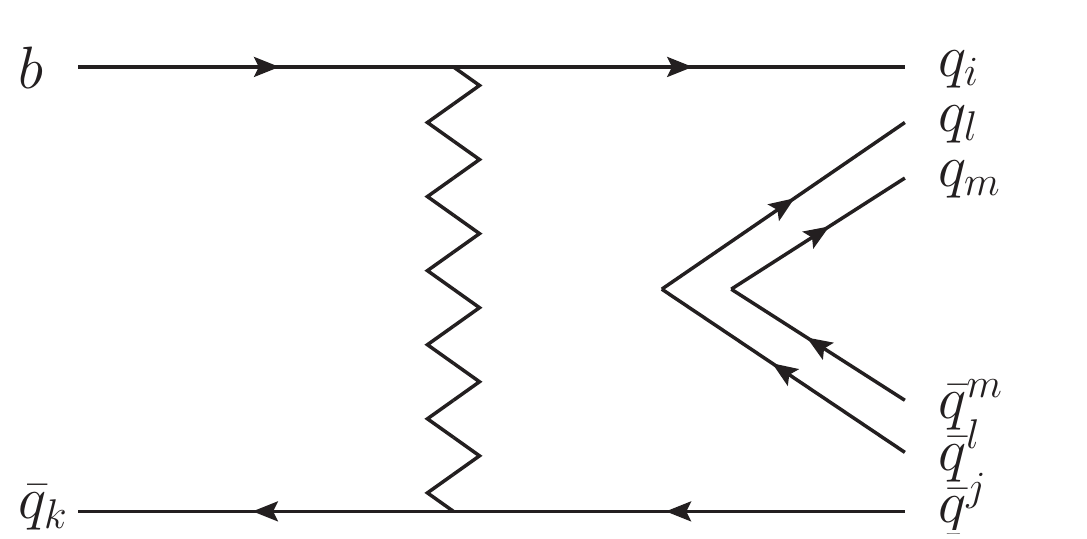}
}
\hspace{12pt}
\subfigure[]{
  \includegraphics[width=0.35\textwidth]{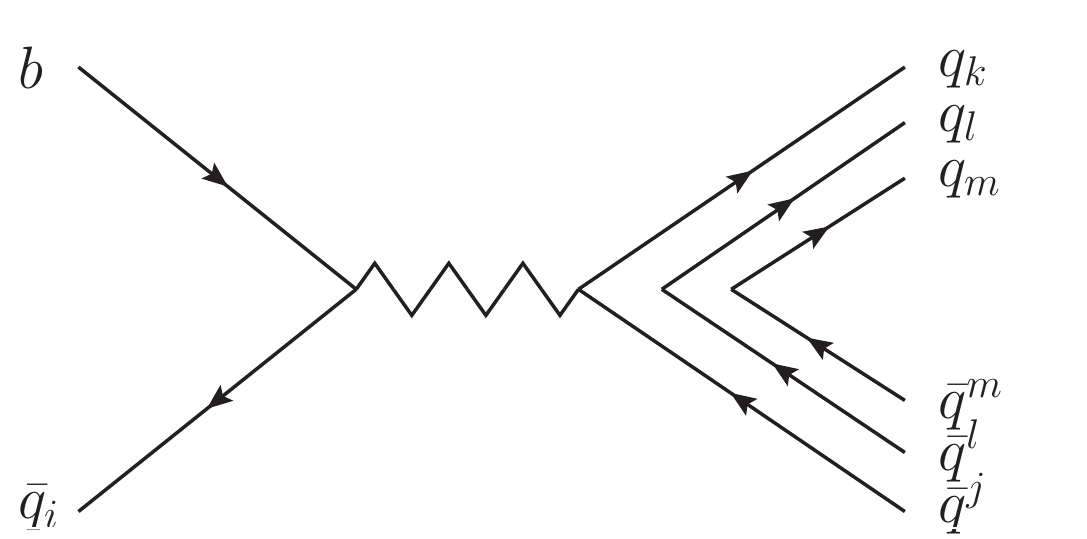}
}
\\\subfigure[]{
  \includegraphics[width=0.35\textwidth]{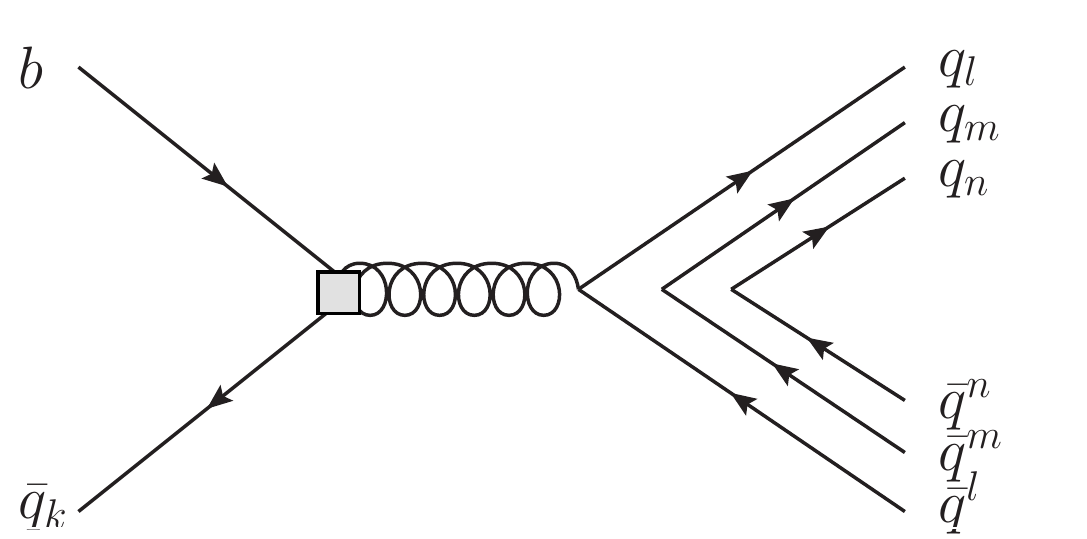}
}
\hspace{12pt}
\subfigure[]{
  \includegraphics[width=0.35\textwidth]{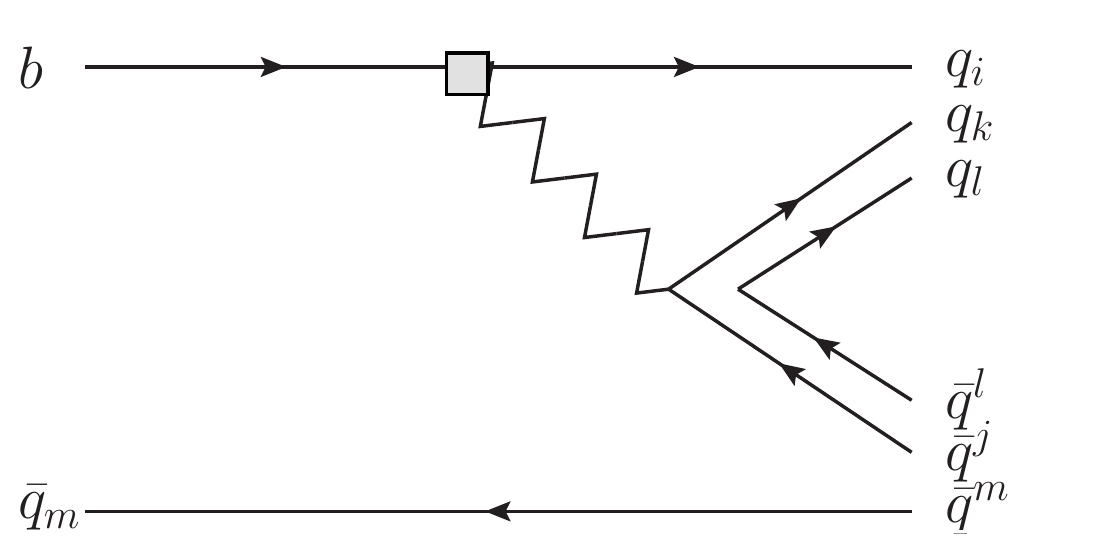}
}
\\\subfigure[]{
  \includegraphics[width=0.35\textwidth]{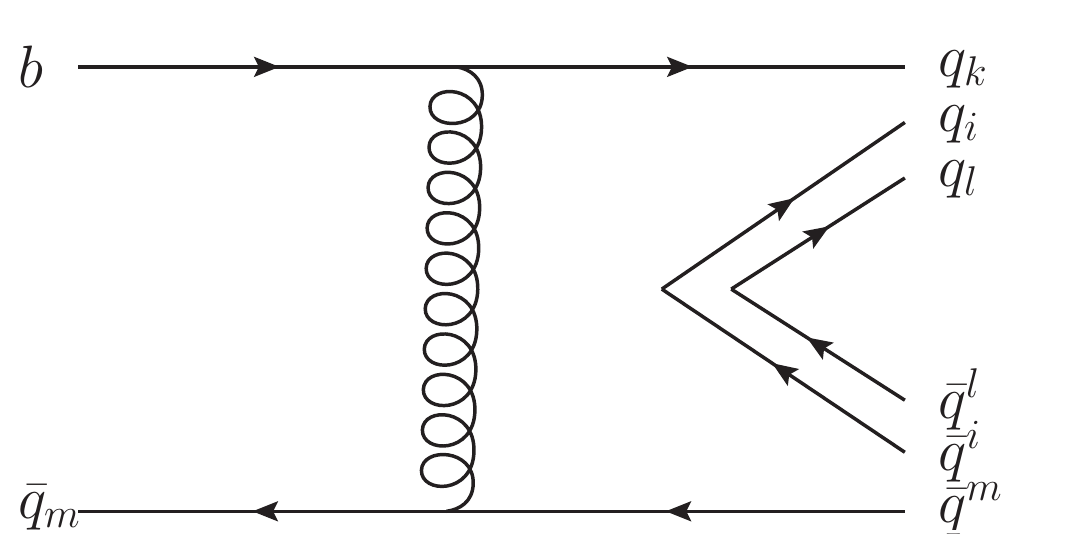}
}
\caption{Topological diagrams of 
  (a) $T$ (tree), (b) $P$ (penguin), (c) $E$ ($W$-exchange),
  (d) $A$ (annihilation), (e) $PA$ (penguin annihilation), (f) $P_{EW}$ (electroweak penguin)
  and (g) $PE$ (penguin-exchange)
  amplitudes in $\overline B$ to baryon pair decays. 
  These are flavor flow diagrams. 
} \label{fig:TA}
\end{figure}

The effective weak Hamiltonian for charmless $\bar B_{u,d,s}$ decays is given by~\cite{Buras}
\be
H_{\rm eff}=\frac{G_f}{\sq2}
                   \Big\{\sum_{r=u,c} V_{qb}V^*_{uq}[c_1 O^r_1+c_2 O^r_2]
                         -V_{tb} V^*_{tq}\sum_{i=3}^{10} c_i O_i\Big\}+{\rm H.c.},
\label{eq: H_eff1}                         
\en
where we have $q=d,s$, and
\be
O^r_1=(\bar r b)_{V-A}(\bar q r)_{V-A},
\quad
O^r_2=(\bar r_\alpha b_\beta)_{V-A}(\bar q_\beta r_\alpha)_{V-A}, 
\non\\
O_{3(5)}=(\bar q b)_{V-A}\sum_{q'}(\bar q' q')_{V\mp A},
\quad
O_{4(6)}=(\bar q_\alpha b_\beta)_{V-A}\sum_{q'}(\bar q_\beta' q_\alpha')_{V\mp A},
\non\\
O_{7(9)}=\frac{3}{2}(\bar q b)_{V-A}\sum_{q'}e_{q'}(\bar q' q')_{V\pm A},
\quad
O_{8(10)}=\frac{3}{2}(\bar q_\alpha b_\beta)_{V-A}\sum_{q'}e_{q'}(\bar q_\beta' q_\alpha')_{V\pm A},
\label{eq: H_eff2}   
\en
with $O_{3-6}$ the QCD penguin operators, $O_{7-10}$ the electroweak penguin operators, and $(\bar q' q)_{V\pm A}\equiv \bar q'\gamma_\mu(1+\pm\gamma_5)q$. The next-to-leading order Wilson coefficients,  
\be
c_1 = 1.081,\, 
c_2 = -0.190,\,
c_3 = 0.014,\,
c_4 = -0.036,\,
c_5 = 0.009,\, 
c_6 = -0.042,
\non\\ 
c_7 = -0.011\alpha_{EM},\, 
c_8 = 0.060\alpha_{EM},\, 
c_9 = -1.254\alpha_{EM},\,
c_{10} = 0.223\alpha_{EM},
\en
are evaluated in the naive dimensional regularization scheme at scale $\mu=4.2$GeV~\cite{Beneke:2001ev}.

We follow the approach of \cite{Chua:2003it,Chua:2013zga,Chua:2016aqy} to decompose $\bar B_q\to \BB$, $\BD$, $\DB$ and $\DD$ decay amplitudes, with $q=u,d,s$, $\B$ and $\D$ denoting low-lying octet and decuplet baryons, into topological amplitudes. 
We have tree ($T$), penguin ($P$), electroweak penguin($P_{EW}$), $W$-exchange ($E$) annihilation, penguin-annihilation ($PA$) and penguin-exchange ($PE$) amplitudes, see Fig.~\ref{fig:TA} for the corresponding diagrams.
For $\Delta S=0$ transition, the
tree (${\cal O}_T=O_{1,2}$), penguin (${\cal O}_P=O_{3-6}$) and electroweak penguin (${\cal O}_{EWP}=O_{7-10}$) operators in Hamiltonian has the following flavor structure,
\begin{eqnarray}
&&{\cal O}_T\sim (\bar u b )(\bar d u )
 =H^{ik}_j (\bar q_i b) (\bar q_k q^j),
\quad
{\cal O}_P\sim(\bar d b) (\bar q_i  q^i )
 =H^k (\bar q_k b) (\bar q_i q^i) ,
\non\\
&&{\cal O}_{EWP}\sim Q_j(\bar d b) ( \bar q_j q^j)
 =H_{EW}{}^{ik}_j (\bar q_i b) (\bar q_k q^j), 
\en
with
\be
H^{12}_1=1=H^2,H_{EW}{}^{2k}_j=Q_j\delta^k_j\,\,\,
{\rm otherwise}\,\,\,H^{ik}_j=H_{EW}{}^{ik}_j=H^k=0.
\end{eqnarray}

Following ref. \cite{Chua:2003it} by suitably matching the $q_k q_i q_l$ flavor to decuplet and octet baryon fields, 
we obtain the following effective Hamiltonian for $\bar B_q\to \DD, \BD, \DB$ and $\BB$ decays,
\begin{eqnarray}
H^\DD_{\rm eff}&=& 6\,T_{{\cal D}\overline{\cal D}}\,\overline B_m
H^{ik}_j \overline {\cal D}_{ikl} {\cal D}^{ljm}
              +6P_{{\cal D}\overline{\cal D}}\,\overline B_m H^k
             \overline {\cal D}_{kil} {\cal D}^{lim}
             +6E_\DD\,\overline B_k H^{ik}_j
              \overline {\cal D}_{ilm} {\cal D}^{mlj}
\nonumber\\
            && +6A_\DD\,\overline B_i H^{ik}_j
               \overline {\cal D}_{klm} {\cal D}^{mlj}
              +2PA_\DD\,\overline B_k H^k
               \overline {\cal D}_{lmn} {\cal D}^{nml}
               +6\,P_{EW{\cal D}\overline{\cal D}}\,\overline B_m
                  H^{ik}_{EW\,j} \overline {\cal D}_{ikl} {\cal D}^{ljm}
\non\\
&&+6PE_{{\cal D}\overline{\cal D}}\,\overline B_m H^k
             \overline {\cal D}_{kil} {\cal D}^{lim},
\label{eq: DD}
\end{eqnarray}
\begin{eqnarray}
H^\BD_{\rm eff}&=& -\sqrt{6}\,T_{1\BD}\,
                        \overline B_m H^{ik}_j
                        \epsilon_{ika}\overline {\cal B}^a_l
                        {\cal D}^{ljm}
               -2\sqrt{6}\,T_{2\BD}\,
                        \overline B_m H^{ik}_j
                        \epsilon_{akl}\overline {\cal B}^a_i
                       {\cal D}^{ljm}
\nonumber\\
           && -\sqrt{6} P_\BD\,
                        \overline B_m H^k
                        \epsilon_{kia} \overline {\cal B}^a_{l}
                       {\cal D}^{lim}
                  -\sq6 E_\BD\,
                        \overline B_k H^{ik}_j
                        \epsilon_{ila}\overline {\cal B}^a_m
                        {\cal D}^{mlj}
                  -\sq6 A_\BD\,
                        \overline B_i H^{ik}_j 
                        \epsilon_{kla}\overline{\cal B}^a_m
                        {\cal D}^{mlj}
                        \non\\
&&-\sqrt{6}\,P_{1EW\BD}\,
                        \overline B_m H_{EW}{}^{ik}_j
                        \epsilon_{ika}\overline {\cal B}^a_l
                        {\cal D}^{ljm}
               -2\sqrt{6}\,P_{2EW\BD}\,
                        \overline B_m H_{EW}{}^{ik}_j
                        \epsilon_{akl}\overline {\cal B}^a_i
                       {\cal D}^{ljm}
\nonumber\\
           && 
               -\sqrt{6} PE_\BD\,
                        \overline B_m H^k
                        \epsilon_{kia} \overline {\cal B}^a_{l}
                       {\cal D}^{lim} ,                      
\label{eq: BD}
\end{eqnarray}
\begin{eqnarray}
H^\DB_{\rm eff}&=& -\sqrt{6}\,T_{1\DB}\,
                        \overline B_m H^{ik}_j \overline{\cal D}_{ikl}
                        \epsilon^{ljb} {\cal B}^m_b
               +\sqrt{6}\,T_{2\DB}\,
                        \overline B_m H^{ik}_j \overline{\cal D}_{ikl}
                        \epsilon^{bjm} {\cal B}^l_b
\nonumber\\
           && +\sqrt{6} P_\DB\,
                        \overline B_m H^k \overline{\cal D}_{kil} 
                        \epsilon^{bim}{\cal B}^l_{b}
                   +\sq6 E_\DB\,
                        \overline B_k H^{ik}_j \overline{\cal D}_{ilm}
                        \epsilon^{bli} {\cal B}^m_b
                     +\sq6 A_\DB\,
                        \overline B_i H^{ik}_j \overline{\cal D}_{klm}
                        \epsilon^{bli} {\cal B}^m_b   
\non\\
           &&-\sqrt{6}\,P_{1EW\DB}\,
                        \overline B_m H_{EW}{}^{ik}_j \overline{\cal D}_{ikl}
                        \epsilon^{ljb} {\cal B}^m_b
               +\sqrt{6}\,P_{2EW\DB}\,
                        \overline B_m H_{EW}{}^{ik}_j \overline{\cal D}_{ikl}
                        \epsilon^{bjm} {\cal B}^l_b
\nonumber\\
           &&                +\sqrt{6} PE_\DB\,
                        \overline B_m H^k \overline{\cal D}_{kil} 
                        \epsilon^{bim}{\cal B}^l_{b},                       
\label{eq: DB}
\end{eqnarray}
and
\begin{eqnarray}
H^\BB_{\rm eff}&=& T_{1\BB}\,
                        \overline B_m H^{ik}_j
                        \epsilon_{ika}\overline {\cal B}^a_l
                        \epsilon^{ljb} {\cal B}^m_b
                   -T_{2\BB}\,
                        \overline B_m H^{ik}_j
                        \epsilon_{ika}\overline {\cal B}^a_l
                        \epsilon^{bjm} {\cal B}^l_b
\non\\
            && +2T_{3\BB}\,
                        \overline B_m H^{ik}_j
                        \epsilon_{akl}\overline {\cal B}^a_i
                        \epsilon^{ljb} {\cal B}^m_b
                 -2T_{4\BB}\,
                        \overline B_m H^{ik}_j
                        \epsilon_{akl}\overline {\cal B}^a_i
                        \epsilon^{bjm} {\cal B}^l_b
\non\\
           && -5 P_{1\BB}\,
                        \overline B_m H^k
                        \epsilon_{kia} \overline {\cal B}^a_{l}
                        \epsilon^{lib} {\cal B}^m_{b}
                 -P_{2\BB}\,
                        \overline B_m H^k
                        \epsilon_{kia} \overline {\cal B}^a_{l}
                        \epsilon^{bim} {\cal B}^l_{b}
\non\\
         &&-5 E_{1\BB}\,
                        \overline B_k H^{ik}_j
                        \epsilon_{ila}\overline {\cal B}^a_m
                        \epsilon^{mlb} {\cal B}^j_b
              -E_{2\BB}\,
                        \overline B_k H^{ik}_j
                        \epsilon_{ila}\overline {\cal B}^a_m
                        \epsilon^{blj} {\cal B}^l_m        
\non\\
         &&-5 A_{1\BB}\,
                        \overline B_i H^{ik}_j
                        \epsilon_{kla}\overline {\cal B}^a_m
                        \epsilon^{mlb} {\cal B}^j_b
                -A_{2\BB}\,
                        \overline B_i H^{ik}_j
                        \epsilon_{kla}\overline {\cal B}^a_m
                        \epsilon^{blj} {\cal B}^l_m 
\non\\
            && +P_{1EW\BB}\,
                        \overline B_m H_{EW}{}^{ik}_j
                        \epsilon_{ika}\overline {\cal B}^a_l
                        \epsilon^{ljb} {\cal B}^m_b
                 -P_{2EW\BB}\,
                        \overline B_m H_{EW}{}^{ik}_j
                        \epsilon_{ika}\overline {\cal B}^a_l
                        \epsilon^{bjm} {\cal B}^l_b
\non\\
            &&+2P_{3EW\BB}\,
                        \overline B_m H_{EW}{}^{ik}_j
                        \epsilon_{akl}\overline {\cal B}^a_i
                        \epsilon^{ljb} {\cal B}^m_b
                 -2P_{4EW\BB}\,
                        \overline B_m H_{EW}{}^{ik}_j
                        \epsilon_{akl}\overline {\cal B}^a_i
                        \epsilon^{bjm} {\cal B}^l_b 
\non\\
              &&-PA_\BB\,
                        \overline B_k H^k
                        \epsilon_{lma}\overline {\cal B}^a_n
                        \epsilon^{nmb} {\cal B}^l_b
                  -5 PE_{1\BB}\,
                        \overline B_m H^k
                        \epsilon_{kia} \overline {\cal B}^a_{l}
                        \epsilon^{lib} {\cal B}^m_{b}
\non\\
           &&
                 -PE_{2\BB}\,
                        \overline B_m H^k
                        \epsilon_{kia} \overline {\cal B}^a_{l}
                        \epsilon^{bim} {\cal B}^l_{b}.                                    
\label{eq: BB}
\end{eqnarray}
with ${\overline B}_m=\left(
B^- ,\overline B {}^0, \overline B {}^0_s\right)$,
${\cal D}^{111}=\Delta^{++}$,
${\cal D}^{112}=\Delta^{+}/\sqrt3$, 
${\cal D}^{122}=\Delta^0/\sq3$,
${\cal D}^{222}=\Delta^-$, 
${\cal D}^{113}=\Sigma^{*-}/\sq3$,
${\cal D}^{123}=\Sigma^{*0}/\sq6$, 
${\cal D}^{223}=\Sigma^{*-}/\sq3$,
${\cal D}^{133}=\Xi^{*0}/\sq3$, 
${\cal D}^{233}=\Xi^{*-}/\sq3$
${\cal D}^{333}=\Omega^-$,
and
\begin{eqnarray}
{\mathcal B}= \left(
\begin{array}{ccc}
{{\Sigma^0}\over\sqrt2}+{{\Lambda}\over\sqrt6}
       &{\Sigma^+}
       &{p}
       \\
{\Sigma^-}
       &-{{\Sigma^0}\over\sqrt2}+{{\Lambda}\over\sqrt6}
       &{n}
       \\
{\Xi^-}
       &{\Xi^0}
       &-\sqrt{2\over3}{\Lambda}
\end{array}
\right), \label{eq: octet}
\end{eqnarray}
(see, for example~\cite{text}). Note that the penguin exchange amplitudes, $PE$ are new and the coefficients of $PA$ are adjusted (by a factor of $1/3$) for later purpose. 

The above formalism can be extended to study $B^-_c\to\bfBB'$ decays. The Hamiltonian governing the decays has the following flavor structure, 
\begin{eqnarray}
&&{\cal O}_T\sim (\bar c b )(\bar d u )
 =H^{ck}_j (\bar c b) (\bar q_k q^j),
\quad
H^{c2}_1=1,\,\,\,
{\rm otherwise}\,\,\,H^{ck}_j=0.
\end{eqnarray}
Hence the effective Hamiltonian for $\bar B_c\to \DD, \BD, \DB$ and $\BB$ decays can be constructed similarly giving
\begin{eqnarray}
H^{c\DD}_{\rm eff}&=& 
             6A^c_\DD\,\overline B_c H^{ck}_j
               \overline {\cal D}_{klm} {\cal D}^{mlj},
\label{eq: Bc DD}
\end{eqnarray}
\begin{eqnarray}
H^{c\DB}_{\rm eff}&=&\sq6 A^c_\DB\,
                        \overline B_c H^{ck}_j \overline{\cal D}_{klm}
                        \epsilon^{bli} {\cal B}^m_b,                      
\label{eq: Bc DB}
\end{eqnarray}
\begin{eqnarray}
H^{c\BD}_{\rm eff}&=& 
                  -\sq6 A^c_\BD\,
                        \overline B_c H^{ck}_j 
                        \epsilon_{kla}\overline{\cal B}^a_m
                        {\cal D}^{mlj},                      
\label{eq: Bc BD}
\end{eqnarray}
and
\begin{eqnarray}
H^{c\BB}_{\rm eff}&=& -5 A^c_{1\BB}\,
                        \overline B_c H^{ck}_j
                        \epsilon_{kla}\overline {\cal B}^a_m
                        \epsilon^{mlb} {\cal B}^j_b
                -A^c_{2\BB}\,
                        \overline B_c H^{ck}_j
                        \epsilon_{kla}\overline {\cal B}^a_m
                        \epsilon^{blj} {\cal B}^l_m.                                    
\label{eq: Bc BB}
\end{eqnarray}

The above results for $\bar B_{u,d,s}$ and $B^-_c$ decays are for $\Delta S=0$ transitions. 
In the case of $\Delta S=-1$ transition,
we put a prime in topological amplitudes and use 
$H^{13}_1=1=H^3$, $H_{EW}{}^{3k}_j=Q_j\delta^k_j$ and $H^{c3}_1=1$ for non-vanishing elements, instead.

The $\bar B_q, \bar B_c\to \BB', \BD', \DB'$ and $\DD'$ decay amplitudes obtained using these effective Hamiltonian are collected in App.~\ref{App: Topological Amplitudes}. Since the flavor flow structures of penguin exchange diagrams and penguin diagrams are identical, see Fig. \ref{fig:TA} (b) and (g), these two topological amplitudes always occur in the combination of $P^{(\prime)}_i+PE^{(\prime)}_i$ in the decay amplitudes.
It should be noted that although the above constructions make use of SU(3) symmetry, they are use as tools, as bookkeeping devices, to obtain flavor flow structure of the decay amplitudes. 
Once the flavor flow structure is obtained, SU(3) breaking effects, through masses, decay constants and so on, in these topological amplitudes can be imposed.
Note that annihilation diagrams only exist in $B^-$ and $B^-_c$ decays, while exchange diagrams only exist in $B^0_d$ and $B^0_s$ decays and penguin-annihilation diagrams only exist in $\bar B^0_{d(s)}\to \BB$ and $\DD$ decays, where the final state anti-baryon is the anti-particle of the associated final state baryon.


\subsection{Factorization contributions to $A^{(\prime)}$, $E^{(\prime)}$, $PE^{(\prime)}$ and $PA^{(\prime)}$}

A typical factorizable $\bar B\to \bfBB'$ decay amplitude has the following expression, which is similar to the mesonic case \cite{Beneke:2001ev} :
\begin{eqnarray}
 A_{\rm fac}\left(\bar B\to \bfBB'\right) &=&\frac{G_F}{\sqrt2}\bigg\{ V_{ub}
V_{uq}^*[
   a_1(\bar q u)_{V-A}\otimes(\bar u b)_{V-A}
  +a_2(\bar u u)_{V-A}\otimes(\bar q b)_{V-A}]
\nonumber\\
&&- V_{tb} V_{tq}^*\Big[
   a_3\sum_{q^\prime}(\bar q^\prime q^\prime)_{V-A}\otimes(\bar q b)_{V-A}
  +a_4\sum_{q^\prime} (\bar q q^\prime)_{V-A}\otimes(\bar q^\prime
b)_{V-A}
\nonumber \\
&&\quad
  +a_5\sum_{q^\prime}(\bar q^\prime q^\prime)_{V+A}\otimes(\bar q b)_{V-A}
 -2a_6\sum_{q^\prime} (\bar q q^\prime)_{S+P}\otimes(\bar q^\prime
b)_{S-P}
\Big]\bigg\}\,,
\label{eq:  factorization}
\end{eqnarray}
where ${\cal O}_1\otimes{\cal O}_2$ is the shorthand of 
\be
{\cal O}_1\otimes{\cal O}_2
\equiv
\langle \bfBB'|{\cal O}_1|0\rangle\langle 0|{\cal O}_2|\bar B\rangle
\en 
and we neglect the contributions from electroweak penguin operators in the factorization calculation.
Note that in factorization calculation the electroweak penguin operators contribute to electroweak-exchange and electroweak-annihilation diagrams, which are negligible comparing to the topological amplitudes generated from tree and strong penguin operators, and these topological amplitudes are not considered in this work. 
In the leading order, the coefficients
$a_i$ are defined in terms of the effective Wilson coefficients $c_i$ as
\be
a^{\rm LO}_{i={\rm odd}}&\equiv& c_i+c_{i+1}/N_c,
\non\\
 a^{\rm LO}_{i={\rm even}}&\equiv& c_i+c_{i-1}/N_c.
\label{eq:  ai LO} 
\en
Contributions beyond the leading order will neglected in this work.
It will be useful to express the above factorization amplitudes according to the decaying mesons, giving
\begin{eqnarray}
A_{\rm fac}\left(B^-\to \bfBB'\right) &=&\frac{G_F}{\sqrt2}\bigg\{ V_{ub}
V_{uq}^*
   a_1(\bar q u)_{V-A}\otimes(\bar u b)_{V-A}- V_{tb} V_{tq}^*\Big[
  a_4(\bar q u)_{V-A}\otimes(\bar u b)_{V-A}
\nonumber \\
&&\quad
 -2a_6 (\bar q u)_{S+P}\otimes(\bar u b)_{S-P}
\Big]\bigg\}\,,
\label{eq:  factorization Bm}
\end{eqnarray}
\begin{eqnarray}
 A_{\rm fac}\left(\bar B_d\to \bfBB'\right) &=&\frac{G_F}{\sqrt2}\bigg\{ V_{ub}
V_{uq}^* a_2(\bar u u)_{V-A}\otimes(\bar q b)_{V-A}\delta_{q d}
\nonumber\\
&&- V_{tb} V_{tq}^*\Big[
   a_3\sum_{q^\prime}(\bar q^\prime q^\prime)_{V-A}\otimes(\bar q b)_{V-A} \delta_{q d}
  +a_4 (\bar q d)_{V-A}\otimes(\bar d b)_{V-A}
\nonumber \\
&&\quad
  +a_5\sum_{q^\prime}(\bar q^\prime q^\prime)_{V+A}\otimes(\bar q b)_{V-A}\delta_{qd}
 -2a_6(\bar q d)_{S+P}\otimes(\bar d b)_{S-P}
\Big]\bigg\}\,,
\label{eq:  factorization Bd}
\end{eqnarray}
\begin{eqnarray}
 A_{\rm fac}\left(\bar B_s\to \bfBB'\right) &=&\frac{G_F}{\sqrt2}\bigg\{ V_{ub}
V_{uq}^*a_2(\bar u u)_{V-A}\otimes(\bar q b)_{V-A}\delta_{qs}
\nonumber\\
&&- V_{tb} V_{tq}^*\Big[
   a_3\sum_{q^\prime}(\bar q^\prime q^\prime)_{V-A}\otimes(\bar q b)_{V-A}\delta_{qs}
  +a_4 (\bar q s)_{V-A}\otimes(\bar s b)_{V-A}
\nonumber \\
&&\quad
  +a_5\sum_{q^\prime}(\bar q^\prime q^\prime)_{V+A}\otimes(\bar q b)_{V-A}\delta_{qs}
 -2a_6 (\bar q s)_{S+P}\otimes(\bar s b)_{S-P}
\Big]\bigg\}\,,
\label{eq:  factorization Bs}
\end{eqnarray}
and
\begin{eqnarray}
A_{\rm fac}\left(B^-_c\to \bfBB'\right) &=&\frac{G_F}{\sqrt2}
V_{cb} V_{uq}^*
   a_1(\bar q u)_{V-A}\otimes(\bar c b)_{V-A}.
\label{eq: factorization Bc}
\end{eqnarray}

By comparing these amplitudes with the topological amplitudes given in App.~\ref{App: Topological Amplitudes}, we have the following correspondence between topological amplitudes and the factorization amplitudes:
\be
c^A_{i,B^-_{u}\to \bfBB'} A_{i,B^-_{u}\to \bfBB'}&=& \frac{G_F}{\sqrt2}V_{ub} V_{ud}^* \, a_1(\bar d u)_{V-A}\otimes(\bar u b)_{V-A},
\non\\
c^{A^c}_{i,B^-_{c}\to \bfBB'} A^c_{i,B^-_{c}\to \bfBB'}&=& \frac{G_F}{\sqrt2}V_{cb} V_{ud}^* a_1(\bar d u)_{V-A}\otimes(\bar c b)_{V-A},
\non\\
c^E_{i,\bar B_d\to\bfBB'}
E_{i,\bar B_d\to\bfBB'}&=& \frac{G_F}{\sqrt2} V_{ub} V_{ud}^* a_2(\bar u u)_{V-A}\otimes(\bar d b)_{V-A},
\non\\
c^{PE}_{i,\bar B_{q''}\to \bfBB'}
PE_{i,\bar B_{q''}\to \bfBB'}&=& -\frac{G_F}{\sqrt2}V_{tb} V_{td}^* \, [a_4 (\bar d q'')_{V-A}\otimes(\bar q'' b)_{V-A}
-2a_6 (\bar d q'')_{S+P}\otimes(\bar q'' b)_{S-P}],
\non\\
c^{PA}_{B^-_{d}\to \bfBB'}
PA_{B^-_{d}\to \bfBB'}&=& -\frac{G_F}{\sqrt2}V_{tb} V_{td}^* \, [ a_3\sum_{q^\prime}(\bar q^\prime q^\prime)_{V-A}\otimes(\bar d b)_{V-A}
\non\\
&&\quad
+a_5\sum_{q^\prime}(\bar q^\prime q^\prime)_{V+A}\otimes(\bar d b)_{V-A}],
\label{eq: TA=fac Delts S=0}
\en
for $\Delta S=0$ transition, and
\be
c^{A'}_{i,B^-_{u}\to \bfBB'}
A'_{i,B^-_{u}\to \bfBB'}&=& \frac{G_F}{\sqrt2}V_{ub} V_{us}^* \, a_1(\bar s u)_{V-A}\otimes(\bar u b)_{V-A},
\non\\
c^{A^{\prime c}}_{i,B^-_{c}\to \bfBB'}
A^{\prime c}_{i,B^-_{c}\to \bfBB'}&=& \frac{G_F}{\sqrt2}V_{cb} V_{us}^* a_1(\bar s u)_{V-A}\otimes(\bar c b)_{V-A},
\non\\
c^{E'}_{i,\bar B_s\to\bfBB'}
E'_{i,\bar B_s\to\bfBB'}&=& \frac{G_F}{\sqrt2} V_{ub} V_{us}^* a_2(\bar u u)_{V-A}\otimes(\bar s b)_{V-A},
\non\\
c^{PE'}_{i,\bar B_{q''}\to \bfBB'}
PE'_{i,\bar B_{q''}\to \bfBB'}&=& -\frac{G_F}{\sqrt2}V_{tb} V_{ts}^* \, [a_4 (\bar s q'')_{V-A}\otimes(\bar q'' b)_{V-A}
-2a_6 (\bar s q'')_{S+P}\otimes(\bar q'' b)_{S-P}],
\non\\
c^{PA'}_{B^-_{s}\to \bfBB'}
PA'_{B^-_{s}\to \bfBB'}&=& -\frac{G_F}{\sqrt2}V_{tb} V_{ts}^* \, [ a_3\sum_{q^\prime}(\bar q^\prime q^\prime)_{V-A}\otimes(\bar s b)_{V-A}
\non\\
&&\quad
+a_5\sum_{q^\prime}(\bar q^\prime q^\prime)_{V+A}\otimes(\bar s b)_{V-A}],
\label{eq: TA=fac Delts S=-1}
\en
for $\Delta S=-1$ transition, 
where the constants $c$ are the Clebsch-Gordan coefficients accompanying with the topological amplitudes 
in the corresponding $\bar B_{q}\to \bfBB'$ decay amplitudes as shown in App.~\ref{App: Topological Amplitudes} and summations over $i$, if necessary, are understood. 

\begin{table}[t!]
\caption{\label{tab: e1e2e3}
The coefficients $(e_1,e_2,e_3)$ and $r_{\bfBB}$ for $\DD$ and $\BB$ final states. 
}
\begin{ruledtabular}
\begin{tabular}{lcclcc}
$\bfBB$ 
          & $(e_1,e_2,e_3)$
          & $r_{\bfBB}$
          & $\bfBB$ 
          & $(e_1,e_2,e_3)$
          & $r_{\bfBB}$
          \\
\hline $\Delta^{++}\overline{\Delta^{++}}$
          & $(1,0,0)$
          & 0.022
          & $\Delta^+\overline{\Delta^+}$
          & $\frac{1}{3}(2,1,0)$
          & 0.030
          \\
$\Delta^0\overline{\Delta^0}$
          & $\frac{1}{3}(1,2,0)$
          & 0.038
          & $\Delta^-\overline{\Delta^-}$
          & $(0,1,0)$
          & 0.047
          \\
$\Sigma^{*+}\overline{\Sigma^{*+}}$
          & $\frac{1}{3}(2,0,1)$
          & 0.325
          & $\Sigma^{*0}\overline{\Sigma^{*0}}$
          & $\frac{1}{3}(1,1,1)$
          & 0.333
          \\
$\Sigma^{*-}\overline{\Sigma^{*-}}$
          & $\frac{1}{3}(0,2,1)$
          & 0.342
          & $\Xi^{*0}\overline{\Xi^{*0}}$
          & $\frac{1}{3}(1,0,2)$
          & 0.628
          \\
$\Xi^{*-}\overline{\Xi^{*-}}$
          & $\frac{1}{3}(0,1,2)$
          & 0.637
          & $\Omega^{-}\overline{\Omega^{-}}$
          & $(0,0,1)$
          & 0.932
          \\          
\hline $p\,\overline p$
          & $\frac{1}{3}(4,-1,0)$
          & 0.013
&$n\,\overline n$
          & $\frac{1}{3}(-1,4,0)$
          & 0.055
          \\
$\Sigma^+\,\overline{\Sigma^+}$
          & $\frac{1}{3}(4,0,-1)$
          & $-0.282$
          &$\Sigma^0\,\overline{\Sigma^0}$
          & $\frac{1}{3}(2,2,-1)$
          & $-0.265$
          \\
$\Sigma^-\,\overline{\Sigma^-}$
          & $\frac{1}{3}(0,4,-1)$
          & $-0.248$
          &$\Lambda\,\overline{\Lambda}$
          & $(0,0,1)$
          & $0.932$
          \\          
$\Xi^0\,\overline {\Xi^0}$
          & $\frac{1}{3}(-1,0,4)$
          & 1.235
&$\Xi^-\,\overline {\Xi^-}$
          & $\frac{1}{3}(0,-1,4)$
          & 1.226
          \\
\end{tabular}
\end{ruledtabular}
\end{table}

Using 
\be
\la 0| (\bar q b)_{V-A}|\bar B_q\ra=-if_{B_q} p_\mu
\en
and equations of motions,
 the matrix elements in the above equations can be evaluated as\be
\la\bfBB'| (\bar q q')_{V\mp A}|0\ra
\la 0| (\bar q'' b)_{V-A}|\bar B_{q''}\ra
&=&-i f_{B_{q''}}[(m_q-m_{\bar q'}) \la \bfBB'| (\bar q q')_{S}|0\ra
\non\\
&&\quad\hspace{18pt}
\mp (m_q+m_{\bar q'}) \la \bfBB'| (\bar q q')_{P}|0\ra],
\non\\
\la\bfBB'| (\bar q q')_{S+P}|0\ra
\la 0| (\bar q'' b)_{S-P}|\bar B_{q''}\ra
&=& i f_{B_{q''}}\frac{m^2_{B_{q''}}}{m_b+m_{q''}} \la \bfBB'| (\bar q q')_{S+P}|0\ra.
\label{eq: MM}
\en
Hence the above factorization amplitudes can all be expressed 
in terms of the $\la \bfBB'| (\bar q q')_{S,P}|0\ra$ matrix elements.
For later purpose, we define
\be
r_{\bfBB}\equiv\frac{\la \bfBB|\sum_{q'} m_{q'} (\bar q' q')_{S-P}|0\ra }{(\sum_{q'} m_{q'})\la \bfBB|\sum_{q'} (\bar q' q')_{S-P}|0\ra}.
\label{eq: r0}
\en

In the large $m_B$ limit, there are asymptotic relations between the matrix elements of $ \la \bfBB'| (\bar q q')_{S,P}|0\ra$, see App.~\ref{App: asym}.
Consequently, the $r_{\bfBB}$ defined in Eq. (\ref{eq: r0}) reduces to
\be
r_{\bfBB}
=\frac{m_u e_1+m_d e_2+m_s e_3}{m_u+m_d+m_s},
\label{eq: r}
\en
with $e_i$ some constants.
The $(e_1, e_2, e_3)$ and $r_\BB$ for $\bar B\to \DD$ and $\BB$ decays
are given in Table~\ref{tab: e1e2e3}.
Furthermore, in the large $m_B$ limit, the matrix elemets $ \la \bfBB'| (\bar q q')_{S,P}|0\ra$ are related,  
giving the following asymptotic relations,
\be
A^{(\prime)}_{1,2\,\BB'}
&=&A^{(\prime)}_{\BD'}
=A^{(\prime)}_{\DB'}
=A^{(\prime)}_{\DD'}=A^{(\prime)}_{B^-_{u}},
\non\\
A^{(\prime)c}_{1,2\,\BB'}
&=&A^{(\prime)c}_{\BD'}
=A^{(\prime)c}_{\DB'}
=A^{(\prime)c}_{\DD'}=A^{(\prime)c}_{B^-_{c}},
\non\\
E^{(\prime)}_{1,2\,\BB'}
&=&E^{(\prime)}_{\BD'}
=E^{(\prime)}_{\DB'}
=E^{(\prime)}_{\DD'}=E^{(\prime)}_{\bar B^0(\bar B^0_s)},
\non\\
PE^{(\prime)}_{1,2\,\BB'}
&=&PE^{(\prime)}_{\BD'}
=PE^{(\prime)}_{\DB'}
=PE^{(\prime)}_{\DD'}=PE^{(\prime)}_{\bar B_{q''}},
\non\\
PA^{(\prime)}_{\BB}
&=&PA^{(\prime)}_{\DD}=PA^{(\prime)}_{\bar B^0_{d(s)}},
\label{eq:asymptoticrelationsnew}
\en
with
\be
A_{B^-_{u}}
&=& -i f_{B_u}\frac{G_F}{\sqrt2}V_{ub} V_{ud}^* \, a_1 f(m^2_{B_u}) \bar{u}(p_{\rm\bf B})[(m_d-m_u) -(m_d+m_u) \gamma_5] v(p_{\overline{\rm\bf B}^\prime})\,,
\non\\
A^c_{B^-_{c}}
&=&  -i f_{B_c}\frac{G_F}{\sqrt2}V_{cb} V_{ud}^* a_1 f(m^2_{B_c}) \bar{u}(p_{\rm\bf B})[(m_d-m_u) -(m_d+m_u) \gamma_5] v(p_{\overline{\rm\bf B}^\prime}),
\non\\
E_{\bar B_d}
&=& i f_{B_d} \frac{G_F}{\sqrt2} V_{ub} V_{ud}^* a_2 2 m_u f(m^2_{B_d}) \bar{u}(p_{\rm\bf B}) \gamma_5 v(p_{\overline{\rm\bf B}^\prime}),
\non\\
PE_{\bar B_{q''}}
&=&  i f_{B_{q''}}\frac{G_F}{\sqrt2}V_{tb} V_{td}^* \, f(m^2_{B_{q''}})
\bar u(p_{\rm\bf B}) 
\bigg[\bigg(a_4 (m_d-m_{q''})+ 2a_6 \frac{m^2_{B_{q''}}}{m_b+m_{q''}} \bigg)
\non\\
&&\qquad\qquad\qquad\qquad\qquad
+\bigg(-a_4(m_d+m_{q''})+2a_6 \frac{m^2_{B_{q''}}}{m_b+m_{q''}} \bigg) \gamma_5 \bigg] v(p_{\overline{\rm\bf B}^\prime}),
\non\\
PA_{\bar B_{d}}
&=&- i f_{B_d}\frac{G_F}{\sqrt2}V_{tb} V_{td}^* \, f(m^2_{B_d}) (a_3-a_5)  2 (m_u+m_d+m_s) r_{\bfBB} \, \bar u(p_{\rm\bf B})\gamma_5 v(p_{\overline{\rm\bf B}}),
\label{eq: TA=fac asym Delta S=0}
\en
and
\be
A'_{B^-_{u}}
&=& -i f_{B_u}\frac{G_F}{\sqrt2}V_{ub} V_{us}^* \, a_1 f(m^2_{B_u}) \bar{u}(p_{\rm\bf B})[(m_s-m_u) -(m_s+m_u) \gamma_5] v(p_{\overline{\rm\bf B}^\prime})\,,
\non\\
A^{\prime c}_{B^-_{c}}
&=& -i f_{B_c}\frac{G_F}{\sqrt2}V_{cb} V_{us}^* \, a_1 f(m^2_{B_c}) \bar{u}(p_{\rm\bf B})[(m_s-m_u) -(m_s+m_u) \gamma_5] v(p_{\overline{\rm\bf B}^\prime})\,,
\non\\
E'_{\bar B_s}
&=& i f_{B_s} \frac{G_F}{\sqrt2} V_{ub} V_{us}^* a_2 2 m_u f(m^2_{B_s}) \bar{u}(p_{\rm\bf B}) \gamma_5 v(p_{\overline{\rm\bf B}^\prime}),
\non\\
PE'_{\bar B_{q''}}
&=&i f_{B_{q''}}\frac{G_F}{\sqrt2}V_{tb} V_{ts}^* f(m^2_{B_{q''}})\, 
\bar{u_{\rm\bf B}}(p_{\rm\bf B})
\bigg[\bigg(a_4(m_s-m_{q''})+2a_6 \frac{m^2_{B_{q''}}}{m_b+m_{q''}}\bigg)
\non\\
&&\qquad\qquad\qquad\qquad\qquad
+\bigg( -a_4(m_s+m_{q''})+2a_6 \frac{m^2_{B_{q''}}}{m_b+m_{q''}}\bigg) \gamma_5\bigg] 
v(p_{\overline{\rm\bf B}^\prime}),
\non\\
PA'_{\bar B_{s}}
&=&-if_{B_s}\frac{G_F}{\sqrt2}V_{tb} V_{ts}^* \, ( a_3-a_5) f(m^2_{B_s})\, 2(m_u+m_d+m_s) r_{\bfBB}\bar{u}(p_{\rm\bf B}) \gamma_5 v(p_{\overline{\rm\bf B}^\prime}).
\label{eq: TA=fac asym Delta S=-1}
\en
Note that the Clebsch-Gordan coefficients in Eqs.~(\ref{eq: TA=fac Delts S=0}) and (\ref{eq: TA=fac Delts S=-1}) canceled out in the above equations. 
Furthermore, $A^{(\prime)}$, $E^{(\prime)}$ and $PA^{(\prime)}$ are proportional to light quark masses and are vanishing in the chiral limit, while $PE^{(\prime)}$ are not vanishing. 
The chiral limits of these topological amplitudes are consistent with the findings in \cite{Chua:2013zga, Chua:2016aqy}, except $PE^{(\prime)}$, which were not considered in \cite{Chua:2013zga, Chua:2016aqy}.

\begin{table}[t!]
\caption{\label{tab: fq2}
Central values of the form factors $f(q^2)$ at various $B$ meson masses.
}
\begin{ruledtabular}
\begin{tabular}{lclc}
$B_q$ 
          & $f(m^2_{B_q})$
          & $B_q$ 
          & $f(m^2_{B_q})$
           \\
\hline 
$B^-$
          & $-0.0012$
          & $\bar B^0$
          & $-0.0012$
          \\
$\bar B^0_s$
          & $-0.0011$
         & $B^-_c$
          & $-0.0005$
          \\
\end{tabular}
\end{ruledtabular}
\end{table}

Through SU(3) symmetry the matrix element $\la \Lambda \overline p|\bar s \gamma_\mu u|0\ra$ can be related to proton electromagnetic (EM) form factors \cite{Chua:2002yd}.
The time-like proton EM form factors were fitted in ref. \cite{Chua:2001vh} using data from ref. \cite{EM}.
By employing the fitted proton electromagnetic form factors from ref. \cite{Chua:2001vh} and by matching with the following matrix element,
\be
\la \Lambda \overline p| \bar s u |0\ra=3\sqrt{\frac{3}{2}} f(q^2) \bar u(p_\Lambda) v(p_{\overline p}),
\en
in the asymptotic limit, we obtain
\be
f(q^2)= -\frac{1}{3}\frac{m_\Lambda-m_p}{m_s-m_u} \,G^p_M(q^2),
\en
where $G^p_M$ is the time-like magnetic form factor of proton.
The central values of the form factors $f(q^2)$ at various $B$ meson masses, by using the $G^p_M(q^2)$ in ref. \cite{Chua:2001vh} and the quark masses at $\mu=4.2$ GeV (see the next section for the quark masses used) are shown in Table~\ref{tab: fq2}.
Note that the values of the form factors (except the one for $B^-_c$) in the table are of the same order to those obtained in a recent work using MIT-bag model calculation \cite{Jin:2021onb}.

\subsection{Specifying topological amplitudes}

In the large $m_B$ limit, the chirality nature of weak and strong interactions provides asymptotic relations~\cite{Brodsky:1980sx} giving~\cite{Chua:2003it,Chua:2013zga,Chua:2016aqy}:
 \begin{eqnarray}
 T^{(\prime)}
 &=&T^{(\prime)}_{1\BB,2\BB,3\BB,4\BB}
 =T^{(\prime)}_{1\BD,2\BD}
 =T^{(\prime)}_{1\DB,2\DB}
 =T^{(\prime)}_{\DD},
 \non\\
 P^{(\prime)}
 &=&P^{(\prime)}_{1\BB,2\BB}
 =P^{(\prime)}_{{\cal B} \overline{\cal D}}
 =P^{(\prime)}_{{\cal D} \overline{\cal B}}
 =P^{(\prime)}_{{\cal D} \overline{\cal D}},
 \non\\
 P^{(\prime)}_{EW}
 &=&P^{(\prime)}_{1EW\BB,2EW\BB,3EW\BB,4EW\BB}
 =P^{(\prime)}_{1EW\BD,2EW\BD}
 =P^{(\prime)}_{1EW\DB,2EW\DB}
 =P^{(\prime)}_{EW\DD},
 \label{eq:asymptoticrelations}
 \en
and those shown in Eq. (\ref{eq:asymptoticrelationsnew}).
Note that the relations on $PE^{(\prime)}$, $E^{(\prime)}$ $A^{(\prime)}$ and $PA^{(\prime)}$ in 
Eq.~(\ref{eq:asymptoticrelationsnew}) are new.

The tree, penguin and electroweak penguin amplitudes are estimated to be~\cite{Chua:2016aqy}
\be
T^{(\prime)}&=&V_{ub} V^*_{ud(s)}\frac{G_f}{\sq2}(c_1+c_2) \chi \bar u'(1-\gamma_5)v,
\non\\
P^{(\prime)}&=&-V_{tb} V^*_{td(s)}\frac{G_f}{\sq2}[c_3+c_4+\kappa_1 c_5+\kappa_2 c_6]\chi \bar u'(1-\gamma_5)v,
\non\\
P^{(\prime)}_{EW}&=&-\frac{3}{2}V_{tb} V^*_{td(s)}\frac{G_f}{\sq2}[c_9+c_{10}+\kappa_1c_7+\kappa_2 c_8]\chi \bar u'(1-\gamma_5)v,
\label{eq:asymptotic1}
\en
where $c_i$ are the next-to-leading order Wilson coefficients.
The parameters $\kappa_i$ are expected to be of ${\cal O}(1)$ and, for simplicity, we assume $\kappa_1=\kappa_2=\kappa$.
We will extract $\chi$ and $\kappa$ from the latest data on $\overline B{}^0\to p\bar p$ and $B^-\to\Lambda\bar p$  decay rates.
From the above equation we see that $\chi$ and $\kappa$ are correlated.
In our numerical study we assume $\chi$ and $\kappa$ to be real and positive for simplicity. 
This assumption will be relaxed in the uncertainty estimation by introducing a relative phase between $T^{(\prime)}$ and $P^{(\prime)}+PE^{(\prime)}$ (recall that $P$ and $PE$ always come together in this combination) and we will see that the phase does not sizably affect the $\bar B^0\to p\bar p$ and $B^-\to\Lambda\bar p$ rates.  

Following ref.~\cite{Chua:2016aqy} we apply the following corrections to $T^{(\prime)}_i$, $P^{(\prime)}_i$ and $P^{(\prime)}_{EWi}$ to the asymptotic relations, Eq.~(\ref{eq:asymptoticrelations}), to account for the finite $m_B$ effects, which are estimated to be ${\cal O}(m_{\bf B}/m_B)$ with $m_{\bf B}$ the baryon mass, giving
\be
T^{(\prime)}_i=(1+r_{t,i}^{(\prime)}) T^{(\prime)},
\,
P^{(\prime)}_i=(1+r_{p,i}^{(\prime)}) P^{(\prime)}_,
\,
P^{(\prime)}_{EW i}=(1+r_{ewp,i}^{(\prime)}) P^{(\prime)}_{EW},
\label{eq:correction0}
\en
and
\be
 |r_{t,i}^{(\prime)}|, |r_{p,i}^{(\prime)}|, |r_{ewp,i}^{(\prime)}|\leq m_p/m_B.
\label{eq:correction1}
\en
The above parameters $r^{(\prime)}$ can have phases
and the $r^{(\prime)}$ for $\bar u' u$ and $\bar u' \gamma_5 u$ terms are independent. 
Likewise,
for penguin exchange amplitudes, we use
\be
PE^{(\prime)}_i\equiv(1+r_{pe,i}^{(\prime)}) PE^{(\prime)}_{\bar B_{q''}}
\label{eq:correction2}
\en
to estimate the corrections to the asymptotic relations in Eq.~(\ref{eq:asymptoticrelationsnew}).
Note that the amplitudes are proportional to the form factor $f(m^2_B)$ in Table\ref{tab: fq2} and the above corrections should include the uncertainty in the form factors.  
We assign $|r^{(\prime)}_{pe,i}|\leq 0.5$.

\begin{figure}[t]
\centering
  \includegraphics[width=0.7\textwidth]{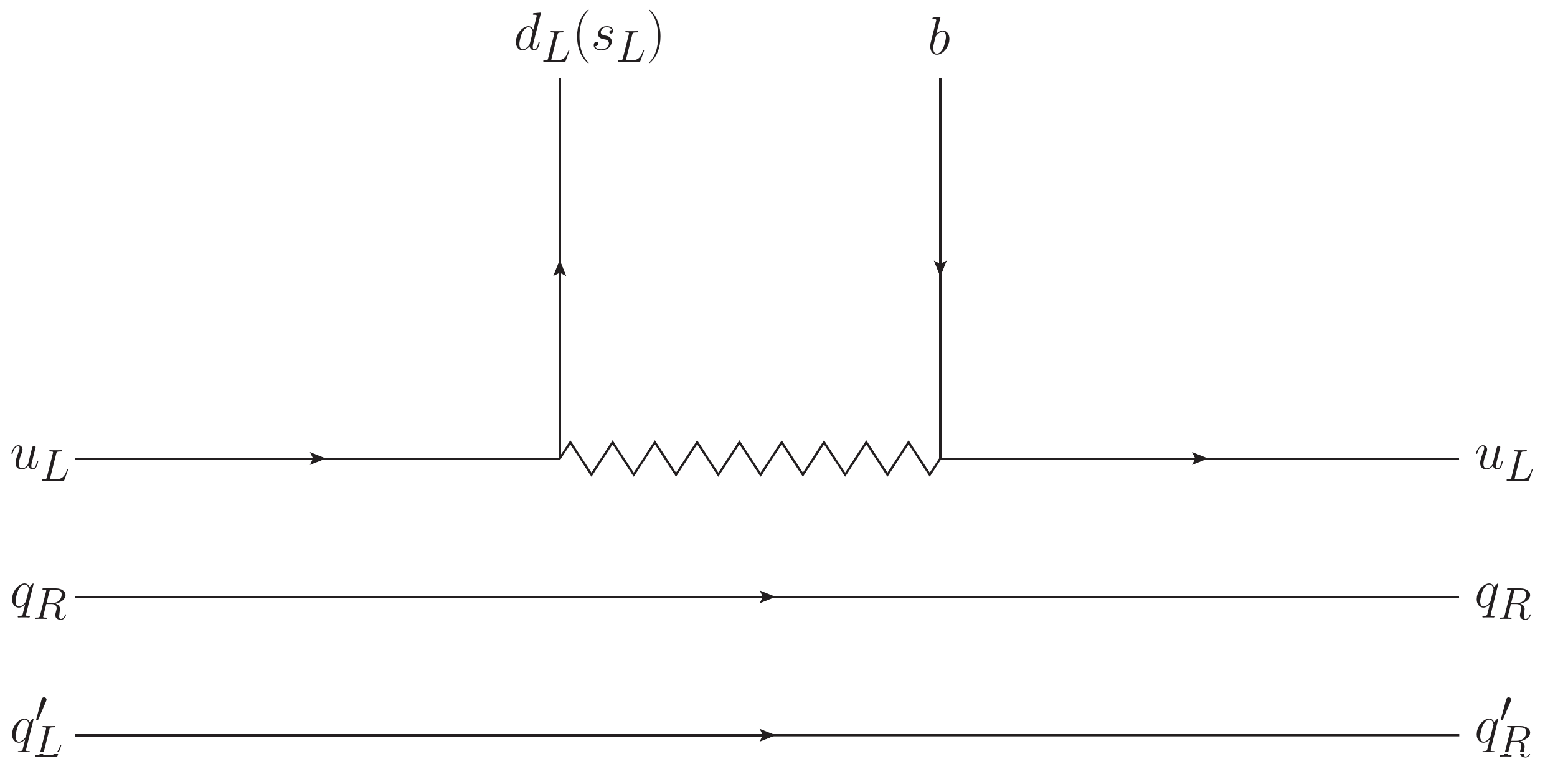}
\caption{The chiral structure of an exchange diagram without chiral flip (for simplicity, we plot a spacelike diagram). It needs a chiral flip to contribute to $\bar B$ decays.
} \label{fig: Exchange}
\end{figure}

For annihilation, exchange and penguin annihilation-amplitudes, the situation is more complicate.
For example, as shown in Fig.~\ref{fig: Exchange}, an exchange diagram without chiral flip cannot produce $\bar u v$ or $\bar u\gamma_5 v$ structure and hence cannot contribute to $\bar B_{d,s}$ decays. 
To overcome that we need to introduce chiral flip. There are two ways to generate a chiral flip, either by quark mass or by baryon mass. 
As one can see from Eqs. (\ref{eq: TA=fac asym Delta S=0}) and (\ref{eq: TA=fac asym Delta S=-1}), the factorization amplitudes in annihilation, 
exchange and penguin-annihilation amplitudes use quark masses to generate the chiral flips, 
while it is possible to have non-factorization contributions generating the chiral flip through $\Lambda_{\rm QCD}/m_b$. 
Indeed, it is well known that the majority of the mass of a baryon does not come from quark mass, but from strong interaction. 
Therefore, $\Lambda_{\rm QCD}$ can play the role of baryon mass in chiral flip.  
To estimate the corrections to the asymptotic relations in Eq.~(\ref{eq:asymptoticrelationsnew}) for annihilation, exchange and penguin-annihilation amplitudes,
we use the following equations,
\be
A^{(\prime)}_i&\equiv&(1+r_{a,i}^{(\prime)}) A^{(\prime)}_{B^-_{u}}+\eta^{(\prime)}_{a,i} \frac{\Lambda_{\rm QCD}}{m_b} T^{(\prime)},
\non\\
A^{c(\prime)}_i&\equiv&(1+r_{a,i}^{c(\prime)}) A^{c(\prime)}_{B^-_{c}}+\eta^{c(\prime)}_{a,i} \frac{\Lambda_{\rm QCD}}{m_b} T^{(\prime)}_c,
\non\\
E^{(\prime)}_i&\equiv&(1+r_{e,i}^{(\prime)}) E^{(\prime)}_{\bar B^0(\bar B^0_s)}+\eta^{(\prime)}_{e,i} \frac{\Lambda_{\rm QCD}}{m_b} T^{(\prime)},
\non\\
PA^{(\prime)}_{\BB,\DD}&\equiv& (1+r_{pa,i}^{(\prime)}) PA^{(\prime)}_{\bar B^0_{d(s)}}+\eta^{(\prime)}_{pa} \frac{\Lambda_{\rm QCD}}{m_b} P^{(\prime)},
\label{eq:correction3}
\en
where terms with $\Lambda_{\rm QCD}/m_b$ are estimations of non-factorization contributions.
Since $T^{(\prime)}$ and $P^{(\prime)}$ are non-factorizable, it is natural to use them in the above estimation.
Note that $T^{(\prime)}_c$ is $T^{(\prime)}$ but with the Cabibbo–Kobayashi–Maskawa (CKM) factor $V_{ub}$ replaced by $V_{cb}$.  
We assign $|r^{(\prime)}_{a,i}|$, $|r^{(\prime)}_{e,i}|$, $|r^{(\prime)}_{pa,i}|$, $|r^{c(\prime)}_{a,i}|\leq 0.5$ and $|\eta^{(\prime)}_{a,i}|$, $|\eta^{c(\prime)}_{a,i}|$, $|\eta^{(\prime)}_{e,i}|$ and $|\eta^{(\prime)}_{pa}|\leq 1$. 
Numerically we take $\Lambda_{\rm QCD}=292$ MeV \cite{ParticleDataGroup:2018ovx}.
Note that we will separate penguin, penguin-exchange and penguin-annihilation amplitudes into $u$-penguin and $c$-penguin contributions and their $r$s and $\eta$s will be varying separately.
Furthermore, all $r$s and $\eta$s for $\bar u' u$ and $\bar u' \gamma_5 u$ terms are independent.

We are now ready to perform numerical study using the above equations and the amplitudes given in App.~\ref{App: Topological Amplitudes}.

\section{Numerical Results on Rates and Direct $CP$ Asymmetries}

Numerical results on rates and direct $CP$ asymmetries will be presented in this section.
In our numerical study, masses of mesons and baryons are taken from ref. \cite{PDG}.
In addition, quark masses and decay constants are taken from the central values given in ref. \cite{PDG}, explicitly, we use
$m_u=1.86$ MeV, $m_d=4.02$ MeV, $m_s=79.98$ MeV, $m_b=4.2$ GeV, at $\mu=4.2$ GeV, 
$f_{B_u}=190$ MeV, $f_{B_d}=190$ MeV and $f_{B_s}=230$ MeV.
For the $B_c$ decay constant, we follow ref. \cite{Narison:2014ska} and use $f_{B_c}=436$ MeV. 
CKM matrix elements are from the latest fit in ref. \cite{CKMfitter}.

\subsection{Sizes of topological amplitudes}

Using the recent data on the $\overline B{}^0\to p\bar p$ rate and the $B^-\to \Lambda\bar p$ rate, 
the unknown parameters $\chi$ and $\kappa$ in asymptotic amplitudes , Eq. (\ref{eq:asymptotic1}), are fitted to be~\footnote{As noted previously $\chi$ and $\kappa$ are correlated. We find that with $\chi=(4.495+0.354 R \cos\lambda-0.003 R\sin\lambda)\times 10^{-3}$ GeV and $\kappa=1.441-0.000 R\cos\lambda+0.573 R\sin\lambda$, the experimental rates $\B(\overline B{}^0\to p\bar p)=(1.27\pm0.14)\times 10^{-8}$ and $\B(B^-\to \Lambda\bar p)=(2.4^{+1.0}_{-0.9})\times 10^{-7}$ can be reproduced with $0\leq R\leq 1$ and $0\leq \lambda\leq 2\pi$. Note that $R=0$ does not correspond to the experimental central values of the decay rates.}
\be
\chi=(4.50^{+0.25}_{-0.26})\times 10^{-3}~{\rm GeV}^2,
\qquad
\kappa=1.47^{+0.55}_{-0.60}.
\label{eq:chikappa}
\en
These values are similar to those given in ref. \cite{Chua:2016aqy}, 
where the values were $\chi=(5.08^{+1.12}_{-1.02})\times 10^{-3}~{\rm GeV}^2$
and
$\kappa=1.92^{+0.39}_{-0.46}$. 
The value of $\chi$ is reduced as the experimental $\overline B{}^0\to p\bar p$ rate is reduced. While $\kappa$ is reduced as $PE'$ also contributes to the $B^-\to \Lambda\bar p$ rate and hence reduces the contribution from $P'$. Note that the $\kappa$ in the above equation is closer to 1 and hence agrees better with our expectation.

As noted previously $P^{(\prime)}$ and $PE^{(\prime)}$ always come in the combination of $P^{(\prime)}+PE^{(\prime)}$. 
Therefore it is $T^{(\prime)}$ and $P^{(\prime)}+PE^{(\prime)}$ that are determined from the data.
It will be useful to see their ratios.
The penguin-tree (tree-penguin) and penguin-exchange-penguin ratios 
for $\Delta S=0$ ($-1$) transitions are found to be
\be
\left|\frac{P_{\bar B_d}+PE_{\bar B_d}}{T_{\bar B_d}}\right|=0.26\pm 0.05,
\quad
\left|\frac{T'_{B^-}}{P'_{B^-}+PE'_{B^-}}\right|=0.19^{+0.05}_{-0.03},
\label{eq:P/T}
\en
and
\be
\left|\frac{PE^{(\prime)}_{B^-}}{P^{(\prime)}_{B^-}+PE^{(\prime)}_{B^-}}\right|=0.27^{+0.07}_{-0.04},
\quad
\left|\frac{PE^{(\prime)}_{\bar B_d}}{P^{(\prime)}_{\bar B_d}+PE^{(\prime)}_{\bar B_d}}\right|=0.27^{+0.07}_{-0.04},
\quad
\left|\frac{PE^{(\prime)}_{\bar B_s}}{P^{(\prime)}_{\bar B_s}+PE^{(\prime)}_{\bar B_s}}\right|=0.30^{+0.07}_{-0.04},
\label{eq:P/T}
\en
where the errors reflect the uncertainties in $\chi$ and $\kappa$ and we keep only the dominant contribution in $PE^{(\prime)}$.
Note that the $|PE^{(\prime)}|/|P^{(\prime)}+PE^{(\prime)}|$ ratio is about 30\%.
It is interesting that the $PE^{(\prime)}$ from factorization contribution is non-negligible comparing to 
$P^{(\prime)}$. This agrees with some early studies, although they did not identify the contribution as $PE^{(\prime)}$ \cite{Hsiao:2014zza,Jin:2021onb}. 

We now discuss the sizes of annihilation, exchange and penguin-annihilation amplitudes with respect to the sizes of penguin-exchange amplitudes in factorization amplitudes.
Using Eqs. (\ref{eq: TA=fac asym Delta S=0}) and (\ref{eq: TA=fac asym Delta S=-1}), we see that the ratio of annihilation and penguin-exchange factorization amplitudes for $\Delta S=0$ transition is given by
\be
\frac{A_{B^-}}{PE_{B^-}}\simeq -\frac{V_{ub} V^*_{ud}}{V_{tb} V^*_{td}} \frac{a_1}{a_6} \frac{m_b+m_u}{2 m^2_{B^-}}
\frac{\bar{u}(p_{\rm\bf B})[(m_d-m_u) -(m_d+m_u) \gamma_5] v(p_{\overline{\rm\bf B}^\prime})}
{\bar{u}(p_{\rm\bf B})(1+\gamma_5) v(p_{\overline{\rm\bf B}^\prime})},
\en
where we keep only the dominant term in $PE$.
Note that the form factor $f(q^2)$ are cancelled out in the ratio. Furthermore, in the decay rate, the $\bar u v$ and $\bar u\gamma_5 v$ terms in the decay amplitude do not interfere. Hence it is legitimate to consider their ratios separately, namely
\be
\frac{(A_{B^-})_S}{(PE_{B^-})_S}&\simeq& -\frac{V_{ub} V^*_{ud}}{V_{tb} V^*_{td}} \frac{a_1}{a_6} \frac{m_b+m_u}{2 m^2_{B^-}}
\frac{\bar{u}(p_{\rm\bf B})[(m_d-m_u)] v(p_{\overline{\rm\bf B}^\prime})}
{\bar{u}(p_{\rm\bf B})v(p_{\overline{\rm\bf B}^\prime})}
\non\\
&=& -\frac{V_{ub} V^*_{ud}}{V_{tb} V^*_{td}} \frac{a_1}{a_6} \frac{(m_b+m_u)(m_d-m_u)}{2 m^2_{B^-}},
\en
and
\be
\frac{(A_{B^-})_P}{(PE_{B^-})_P}&\simeq& -\frac{V_{ub} V^*_{ud}}{V_{tb} V^*_{td}} \frac{a_1}{a_6} \frac{m_b+m_u}{2 m^2_{B^-}}
\frac{\bar{u}(p_{\rm\bf B})[-(m_d+m_u) \gamma_5] v(p_{\overline{\rm\bf B}^\prime})}
{\bar{u}(p_{\rm\bf B})\gamma_5 v(p_{\overline{\rm\bf B}^\prime})}
\non\\
&=&\frac{V_{ub} V^*_{ud}}{V_{tb} V^*_{td}} \frac{a_1}{a_6} \frac{(m_b+m_u)(m_d+m_u)}{2 m^2_{B^-}}.
\en
It should be noted that the ratio of $|(PE_{\bar B_q})_S|$ and $|(PE_{\bar B_q})_P|$ are of order 1, as
\be
\frac{|(PE_{\bar B_q})_S|}{|(PE_{\bar B_q})_P|}=\sqrt{\frac{m^2_{B_q}-(m_{\bf B}+m_{\bf {\overline B}'})^2}{m^2_{B_q}-(m_{\bf B}-m_{\bf {\overline B}'})^2}}
=0.8\sim 0.9,
\en
for the modes we are considering in this work.

Similarly the ratios of exchange and penguin-exchange amplitudes and penguin-annihilation and penguin-exchange factorization amplitudes are
\be
\frac{(E_{\bar B_d})_S}{(PE_{\bar B_d})_S}=0,
\quad
\frac{(E_{\bar B_d})_P}{(PE_{\bar B_d})_P}\simeq
\frac{V_{ub} V^*_{ud}}{V_{tb} V^*_{td}} \frac{a_2}{a_6} \frac{(m_b+m_d)m_u}{m^2_{\bar B_d}},
\en
and
\be
\frac{(PA_{\bar B_d})_S}{(PE_{\bar B_d})_S}=0,
\quad
\frac{(PA_{\bar B_d})_P}{(PE_{\bar B_d})_P}\simeq
 \frac{a_5-a_3}{a_6} \frac{(m_b+m_d)(m_u+m_d+m_s)}{m^2_{\bar B_d}} r_{\bfBB},
\en
with $r_{\bfBB}$ defined in Eq. (\ref{eq: r}) and its value given in Table~\ref{tab: e1e2e3}.
For $\Delta S=-1$ transition, we have the following expression for ratios of topological factorization amplitudes,
\be
\frac{(A'_{B^-})_S}{(PE'_{B^-})_S}&\simeq&-\frac{V_{ub} V^*_{us}}{V_{tb} V^*_{ts}} \frac{a_1}{a_6} \frac{(m_b+m_u)(m_s-m_u)}{2 m^2_{B^-}},
\non\\
\frac{(A'_{B^-})_P}{(PE'_{B^-})_P}&\simeq&\frac{V_{ub} V^*_{us}}{V_{tb} V^*_{ts}} \frac{a_1}{a_6} \frac{(m_b+m_u)(m_s+m_u)}{2 m^2_{B^-}},
\en
\be
\frac{(E'_{\bar B_s})_S}{(PE'_{\bar B_s})_S}=0,
\quad
\frac{(E'_{\bar B_s})_P}{(PE'_{\bar B_s})_P}\simeq
\frac{V_{ub} V^*_{us}}{V_{tb} V^*_{ts}} \frac{a_2}{a_6} \frac{(m_b+m_s)m_u}{m^2_{\bar B_s}},
\en
and
\be
\frac{(PA'_{\bar B_s})_S}{(PE'_{\bar B_s})_S}=0,
\quad
\frac{(PA'_{\bar B_s})_P}{(PE'_{\bar B_s})_P}\simeq
 \frac{a_5-a_3}{a_6} \frac{(m_b+m_s)(m_u+m_d+m_s)}{m^2_{\bar B_s}} r_{\bfBB}.
\en

Numerically we obtain the following ratios of sizes for topological factorization amplitudes,
\be
\bigg|\frac{(A_{B^-})_S}{(PE_{B^-})_S}\bigg| &\simeq& 0.0018,
\quad
\bigg|\frac{(A_{B^-})_P}{(PE_{B^-})_P}\bigg| \simeq 0.0048,
\non\\
\bigg|\frac{(E_{\bar B_d})_S}{(PE_{\bar B_d})_S}\bigg|&=& 0,
\qquad\quad
\bigg|\frac{(E_{\bar B_d})_P}{(PE_{\bar B_d})_P}\bigg|\simeq 0.0005,
\non\\
\bigg|\frac{(PA_{\bar B_d})_S}{(PE_{\bar B_d})_S}\bigg| &=& 0,
\qquad\quad
\bigg|\frac{(PA_{\bar B_d})_P}{(PE_{\bar B_d})_P}\bigg|\simeq 0.0012\, |r_{\bfBB}|,
\label{eq: A E PA per PE} 
\en
for $\Delta S=0$ transition, and
\be
\bigg|\frac{(A'_{B^-})_S}{(PE'_{B^-})_S}\bigg| &\simeq& 0.0031,
\quad
\bigg|\frac{(A'_{B^-})_P}{(PE'_{B^-})_P}\bigg|\simeq 0.0032,
\non\\
\bigg|\frac{(E'_{\bar B_s})_S}{(PE'_{\bar B_s})_S}\bigg|&=& 0,
\qquad\quad
\bigg|\frac{(E'_{\bar B_s})_P}{(PE'_{\bar B_s})_P}\bigg|\simeq 2.4\times 10^{-5},
\non\\
\bigg|\frac{(PA'_{\bar B_s})_S}{(PE'_{\bar B_s})_S}\bigg|&=& 0,
\qquad\quad
\bigg|\frac{(PA'_{\bar B_s})_P}{(PE'_{\bar B_s})_P}\bigg|\simeq 0.0011\, |r_{\bfBB}|,
\label{eq: A' E' PA' per PE'}
\en
for $\Delta S=-1$ transition. 
These ratios are very small.

As the $B^-_c\to\bfBB'$ decays are governed by annihilation diagrams, it is useful to have some estimations on the sizes of the annihilation factorization amplitudes. 
Using Eqs. (\ref{eq: TA=fac asym Delta S=0}) and (\ref{eq: TA=fac asym Delta S=-1}), we have
\be
\bigg|\frac{(A^{(\prime)c}_{B^-_c})_{S,P}}{(A^{(\prime)}_{B^-_u})_{S,P}}\bigg|
&=&
\frac{p_c(B_c)}{p_c(B_u)}
\sqrt{
\frac{m^2_{B_c}-(m_{\bf B}\pm m_{\bf {\overline B}'})^2}
{m^2_{B_u}-(m_{\bf B}\pm m_{\bf {\overline B}'})^2}}
\frac{f_{B_c} f(m^2_{B_c})}{f_{B_u}f(m^2_{B_u})}
\bigg|\frac{V_{cb}}{V_{ub}}\bigg|
\non\\
&\simeq &
\frac{p_c(B_c)}{p_c(B_u)}
\sqrt{
\frac{m^2_{B_c}-(m_{\bf B}\pm m_{\bf {\overline B}'})^2}
{m^2_{B_u}-(m_{\bf B}\pm m_{\bf {\overline B}'})^2}}
\times 10.9
\non\\
&\simeq&
{\cal O}(10).
\label{eq: Ac per A}
\en
Hence, annihilation factorization amplitudes in $B^-_c$ decays are greater than those in $B^-_u$ decays by roughly one order of magnitude.
It should also be noted that although the lifetimes of $B_{u,d,s}$ are more or less similar, the lifetime of $B_c$ is only about one third of their typical lifetime 
providing a factor of 3 suppression in $B^-_c$ branching ratios. 

As these factorization contributions suffer from severe chiral suppression, the non-factorizable contributions become important and non-negligible. 

\subsection{Numerical Results on Rates}

Predictions on $\bar B_q\to\bfBB'$ and $B^-_c\to \bfBB'$ decay rates will given in this section. 
Before we present our result, it will be useful to remind us the detection sensitivities of various baryonic final states. 
In Table \ref{tab: sensitivity}, we show some final states of the baryons with unsuppressed branching ratios.
These final states affect the detectability of the baryons. The detectability of the baryons in decreasing order are 
final states with all charged states, final states involving $\pi^0$ or $\gamma$ and final states involving $n$~\cite{Chua:2013zga,PDG}. 
Baryons are grouped accordingly in the table. 

\begin{table}[t!]
\caption{\label{tab: sensitivity}
Baryons are grouped according to their detectability. Some final states where baryons can decay with unsuppressed branching ratios are shown.
}
\begin{ruledtabular}
\begin{tabular}{lr}
Baryons
          & Final states
           \\
\hline 
$p$, $\Delta^{++,0}$, $\Lambda$, $\Xi^-$, $\Sigma^{*\pm}$, $\Xi^{*0}$, $\Omega^-$
          & all charged particles 
          \\
          &($\Delta^{++,0},\Lambda\to p\pi^\pm$;  $\Xi^-$, $\Sigma^{*\pm}$, $\Xi^{*0}\to p(\pi^+)^n(\pi^-)^{n'}$; $\Omega^-\to p\pi^-K^-$) 
          \\
$\Delta^+$, $\Sigma^{+}$, $\Xi^0$, $\Sigma^{*0}$, $\Xi^{*-}$
          & involving $\pi^0$ 
          \\
          &($\Delta^+,\Sigma^{+}\to p\pi^0$; $\Xi^0, \Sigma^{*0}\to\Lambda\pi^0$, $\Xi^{*-}\to\Lambda\pi^0\pi^-$)
          \\
$\Sigma^{0}$
          & involving $\gamma$ 
          \\
          & ($\Sigma^{0}\to\Lambda\gamma$) 
          \\   
$n$, $\Delta^-$, $\Sigma^-$
          & involving $n$ 
          \\
         & ($\Delta^-,\Sigma^-\to n\pi^-$)
         \\      
\end{tabular}
\end{ruledtabular}
\end{table}

\begin{table}[t!]
\caption{\label{tab: BBDS=0, -1} Decay rates $\overline B_q\to\BB$ decays for $\Delta S=0$ and $-1$ transitions. 
See text for the sources of the uncertainties.
Occasionally the last uncertainties are shown to larger decimal place. 
The experimental $\overline B{}^0\to p\bar p$ and $B^-\to \Lambda \bar p$ rates are inputs.
The rank indicates the detectablilty of the mode, where more asterisks are more favorable.}
\begin{ruledtabular}
\centering
{\begin{tabular}{llll}
Mode (rank)
          & ${\mathcal B}(10^{-8})$
          & Mode (rank)
          & ${\mathcal B}(10^{-8})$
          \\
\hline $B^-\to n\overline{p}$
          & $3.39^{+0.86}_{-0.78}{}^{+0}_{-0.23}{}^{+2.81}_{-1.93}{}^{+1.06}_{-0.76}$ %
          & $\overline B{}^0_s\to p\overline{\Sigma^{+}}$ (**)
          & $1.26^{+0.14}_{-0.14}{}^{+0}_{-0.05}{}^{+1.84}_{-1.01}\pm0$%
           \\
$B^-\to \Sigma^{0}\overline{\Sigma^{+}}$ (*)
          & $3.01^{+0.42}_{-0.39}{}^{+0}_{-0.18}{}^{+2.57}_{-1.78}{}^{+1.07}_{-0.87}$%
          & $\overline B{}^0_s\to n\overline{\Sigma^{0}}$
          & $0.59^{+0.07}_{-0.07}{}^{+0.02}_{-0}{}^{+0.29}_{-0.22}\pm0$%
           \\ 
$B^-\to \Sigma^{-}\overline{\Sigma^{0}}$
          & $0.57^{+0.24}_{-0.22}{}^{+0.00}_{-0}{}^{+0.68}_{-0.38}{}^{+0.12}_{-0.00}$%
          &$\overline B{}^0_s\to n\overline{\Lambda}$
          & $2.95^{+0.54}_{-0.50}{}^{+0}_{-0.20}{}^{+2.65}_{-1.80}\pm0$%
           \\            
$B^-\to \Sigma^{-}\overline{\Lambda}$
          & $0.43^{+0.18}_{-0.16}{}^{+0.00}_{-0}{}^{+0.31}_{-0.21}{}^{+0.04}_{-0.00}$%
          & $\overline B{}^0_s\to \Sigma^{0}\overline{\Xi^{0}}$ (*)
          & $9.77^{+1.06}_{-1.07}{}^{+0}_{-0.44}{}^{+5.63}_{-4.36}\pm0$%
           \\ 
$B^-\to \Xi^{-}\overline{\Xi^{0}}$
          & $0.07^{+0.03}_{-0.03}{}^{+0.00}_{-0}{}^{+0.05}_{-0.03}{}^{+0.01}_{-0.00}$%
          & $\overline B{}^0_s\to \Sigma^{-}\overline{\Xi^{-}}$ 
          & $1.76^{+0.72}_{-0.65}\pm 0{}^{+1.32}_{-0.88}\pm0$%
           \\ 
$B^-\to \Lambda\overline{\Sigma^+}$
          & $0.43^{+0.18}_{-0.16}{}^{+0.00}_{-0}{}^{+0.53}_{-0.21}{}^{+0.04}_{-0.00}$%
          & $\overline B{}^0_s\to \Lambda\overline{\Xi^0}$
          & $0.11^{+0.04}_{-0.04}\pm 0{}^{+0.85}_{-0.10}\pm0$%
           \\ 
$\overline B{}^0\to p\overline{p}$ (***)
          & $1.27^{+0.14}_{-0.14}{}^{+0}_{-0.05}{}^{+1.85}_{-1.02}{}^{+1.32}_{-0.84}$%
          & $\overline B{}^0\to \Sigma^{+}\overline{\Sigma^{+}}$
          & $0.00\pm0\pm0\pm0^{+0.22}_{-0.00}$%
           \\ 
$\overline B{}^0\to n\overline{n}$
          & $6.09^{+0.85}_{-0.80}{}^{+0}_{-0.36}{}^{+5.23}_{-3.62}{}^{+0.60}_{-0.57}$%
          & $\overline B{}^0\to \Sigma^{0}\overline{\Sigma^{0}}$ (*)
          & $1.39^{+0.19}_{-0.18}{}^{+0}_{-0.08}{}^{+1.19}_{-0.82}{}^{+0.63}_{-0.50}$%
           \\
$\overline B{}^0\to \Xi^{0}\overline{\Xi^{0}}$
          & $0.00\pm0\pm0\pm0^{+0.01}_{-0.00}$%
          & $\overline B{}^0\to \Sigma^{-}\overline{\Sigma^{-}}$
          & $1.05^{+0.45}_{-0.40}{}^{+0.00}_{-0}{}^{+1.25}_{-0.71}{}^{+0.09}_{-0.08}$%
           \\      
$\overline B{}^0\to \Xi^{-}\overline{\Xi^{-}}$
          & $0.06^{+0.03}_{-0.02}{}^{+0.00}_{-0}{}^{+0.04}_{-0.03}\pm 0.02$%
          & $\overline B{}^0\to \Sigma^{0}\overline{\Lambda}$ (*)
           & $3.54^{+0.39}_{-0.39}{}^{+0}_{-0.13}{}^{+1.88}_{-1.48}{}^{+0.47}_{-0.44}$%
           \\
$\overline B{}^0\to \Lambda\overline{\Lambda}$
          & $0.00\pm0\pm0{}^{+0.24}_{-0}{}^{+0.03}_{-0.00}$%
          & $\overline B{}^0\to \Lambda\overline{\Sigma^{0}}$
           & $0.20^{+0.09}_{-0.08}{}^{+0}_{-0.00}{}^{+0.17}_{-0.10}{}^{+0.01}_{-0.00}$%
           \\ 
\hline $B^-\to \Sigma^{0}\overline{p}$
          & $0.82^{+0.36}_{-0.32}{}^{+0}_{-0.25}{}^{+0.71}_{-0.49}\pm0.01$%
          & $\overline B{}^0\to \Sigma^{+}\overline{p}$ (**)
          & $1.75^{+0.68}_{-0.62}{}^{+0.59}_{-0}{}^{+1.24}_{-0.85}\pm0$%
           \\
$B^-\to \Sigma^{-}\overline{n}$
          & $1.70^{+0.73}_{-0.65}{}^{+0}_{-0.00}{}^{+1.10}_{-0.83}\pm 0.02$%
          & $\overline B{}^0\to \Sigma^{0}\overline{n}$
          & $1.01^{+0.35}_{-0.32}{}^{+0.65}_{-0}{}^{+0.73}_{-0.45}\pm0$%
           \\ 
$B^-\to \Xi^{0}\overline{\Sigma^{+}}$ (*)
          & $40.32^{+16.97}_{-15.25}{}^{+2.70}_{-0}{}^{+26.22}_{-19.67}{}^{+0.45}_{-0.44}$%
          &$\overline B{}^0\to \Xi^{0}\overline{\Sigma^{0}}$ (*)
          & $18.68^{+7.83}_{-7.06}{}^{+1.26}_{-0}{}^{+12.15}_{-9.11}\pm0$%
           \\            
$B^-\to\Xi^{-}\overline{\Sigma^{0}}$ (**)
          & $19.80^{+8.44}_{-7.57}{}^{+0}_{-0.00}{}^{+12.93}_{-9.70}{}^{+0.24}_{-0.23}$%
          & $\overline B{}^0\to \Xi^{0}\overline{\Lambda}$ (**)
          & $2.48^{+0.97}_{-0.88}{}^{+0.84}_{-0}{}^{+4.33}_{-2.13}\pm0$%
           \\ 
$B^-\to \Xi^{-}\overline{\Lambda}$ (***)
          & $2.41^{+1.03}_{-0.92}{}^{+0}_{-0.00}{}^{+4.27}_{-2.14}{}^{+0.07}_{-0.06}$%
          & $\overline B{}^0\to  \Xi^{-}\overline{\Sigma^{-}}$
          & $36.68^{+15.64}_{-14.02}\pm 0{}^{+23.95}_{-17.96}\pm0$%
           \\ 
$B^-\to \Lambda\overline{p}$ (***)
          & $24.00^{+10.00}_{-9.00}{}^{+2.69}_{-0}{}^{+19.37}_{-13.65}{}^{+0.31}_{-0.30}$%
          & $\overline B{}^0\to \Lambda\overline{n}$
          & $23.00^{+9.31}_{-8.41}{}^{+5.19}_{-0}{}^{+18.59}_{-13.01}\pm0$%
           \\              
$\overline B{}^0_s\to p\overline{p}$
          & $0.00\pm0\pm0{}^{+0.07}_{-0.00}$%
          & $\overline B{}^0_s\to \Sigma^{+}\overline{\Sigma^{+}}$ (*)
          & $1.83^{+0.69}_{-0.63}{}^{+0.60}_{-0}{}^{+1.33}_{-0.92}{}^{+0.67}_{-0.54}$%
           \\
$\overline B{}^0_s\to n\overline{n}$
          & $0.00\pm0\pm0{}^{+0.05}_{-0.00}$%
          & $\overline B{}^0_s\to \Sigma^{0}\overline{\Sigma^{0}}$ (*)
          & $1.74^{+0.69}_{-0.62}{}^{+0.28}_{-0}{}^{+1.19}_{-0.87}{}^{+0.62}_{-0.52}$%
           \\
$\overline B{}^0_s\to \Xi^{0}\overline{\Xi^{0}}$ (*)
          & $26.38^{+10.47}_{-9.50}{}^{+4.27}_{-0}{}^{+28.15}_{-17.95}{}^{+2.08}_{-2.00}$%
          & $\overline B{}^0_s\to \Sigma^{-}\overline{\Sigma^{-}}$
          & $1.66^{+0.68}_{-0.62}{}^{+0}_{-0.00}{}^{+1.10}_{-0.82}{}^{+0.55}_{-0.47}$%
           \\ 
$\overline B{}^0_s\to \Xi^{-}\overline{\Xi^{-}}$ (***)
          & $25.22^{+10.37}_{-9.36}{}^{+0}_{-0.04}{}^{+26.95}_{-17.26}{}^{+1.97}_{-1.90}$%
          & $\overline B{}^0_s\to \Sigma^{0}\overline{\Lambda}$
          & $0.04^{+0.00}_{-0.00}{}^{+0.05}_{-0}{}^{+0.05}_{-0.03}\pm 0.01$%
           \\    
$\overline B{}^0_s\to \Lambda\overline{\Lambda}$ (***)
          & $16.08^{+6.38}_{-5.79}{}^{+2.60}_{-0}{}^{+13.21}_{-9.23}{}^{+1.73}_{-1.64}$%
          & $\overline B{}^0_s\to \Lambda\overline{\Sigma^{0}}$
          & $0.04^{+0.00}_{-0.00}{}^{+0.05}_{-0}{}^{+0.02}_{-0.01}\pm 0.01$%
           \\                                                                                                                                                                                                             
\end{tabular}
}
\\
\end{ruledtabular}
\end{table}

\begin{table}[t!]
\caption{\label{tab: BDDS=0, -1} Same as Table~\ref{tab: BBDS=0, -1}, but for $\overline B_q\to\BD$ modes.
}
\begin{ruledtabular}
\centering
\begin{tabular}{llll}
Mode (rank)
          & ${\mathcal B}(10^{-8})$
          & Mode (rank)
          & ${\mathcal B}(10^{-8})$
          \\
\hline $B^-\to p\overline{\Delta^{++}}$(***)
          & $5.88^{+0.61}_{-0.61}{}^{+0}_{-0.30}{}^{+9.26}_{-4.88}{}^{+0.71}_{-0.67}$%
          & $\overline B{}^0_s\to p\overline{\Sigma^{*+}}$(***)
          & $1.82^{+0.18}_{-0.19}{}^{+0}_{-0.10}{}^{+2.89}_{-1.51}\pm0$%
           \\
$B^-\to n\overline{\Delta^+}$
          & $1.79^{+0.19}_{-0.19}{}^{+0.09}_{-0}{}^{+0.98}_{-0.67}{}^{+0.23}_{-0.22}$%
          & $\overline B{}^0_s\to n\overline{\Sigma^{*0}}$
          & $0.82^{+0.09}_{-0.09}{}^{+0.04}_{-0}{}^{+0.46}_{-0.31}\pm0$%
           \\ 
$B^-\to\Sigma^0\overline{\Sigma^{*+}}$(**)
          & $2.70^{+0.30}_{-0.29}{}^{+0}_{-0.08}{}^{+1.46}_{-1.15}{}^{+0.18}_{-0.17}$%
          &$\overline B{}^0_s\to \Sigma^{0}\overline{\Xi^{*0}}$(**)
          & $2.45^{+0.27}_{-0.27}{}^{+0}_{-0.08}{}^{+1.36}_{-1.06}\pm0$%
           \\            
$B^-\to\Sigma^-\overline{\Sigma^{*0}}$
          & $0.10^{+0.03}_{-0.02}{}^{+0.00}_{-0}{}^{+0.09}_{-0.06}{}^{+0.00}_{-0.00}$%
          & $\overline B{}^0_s\to \Sigma^{-}\overline{\Xi^{*-}}$
          & $0.20\pm0.05\pm 0{}^{+0.20}_{-0.12}\pm0$%
           \\ 
$B^-\to\Xi^{-}\overline{\Xi^{*0}}$
          & $0.15^{+0.04}_{-0.04}{}^{+0.00}_{-0}{}^{+0.15}_{-0.09}{}^{+0.01}_{-0.00}$%
          & $\overline B{}^0_s\to \Xi^{-}\overline{\Omega^-}$
          & $0.47^{+0.12}_{-0.11}\pm 0{}^{+0.47}_{-0.29}\pm0$%
           \\ 
$B^-\to\Lambda\overline{\Sigma^{*+}}$
          & $0.30^{+0.08}_{-0.08}{}^{+0.00}_{-0}{}^{+0.54}_{-0.18}{}^{+0.01}_{-0.00}$%
          & $\overline B{}^0_s\to \Lambda\overline{\Xi^{*0}}$
          & $0.31^{+0.08}_{-0.07}\pm 0{}^{+0.54}_{-0.19}\pm0$%
           \\ 
$\overline B{}^0\to p\overline{\Delta^+}$(**)
          & $1.82^{+0.19}_{-0.19}{}^{+0}_{-0.09}{}^{+2.86}_{-1.51}{}^{+0.22}_{-0.21}$%
          & $\overline B{}^0\to \Sigma^{+}\overline{\Sigma^{*+}}$
          & $0.00\pm0\pm0{}^{+0.01}_{-0.00}$%
           \\
$\overline B{}^0\to n\overline{\Delta^0}$
          & $1.66^{+0.18}_{-0.18}{}^{+0.08}_{-0}{}^{+0.91}_{-0.62}{}^{+0.21}_{-0.20}$%
          & $\overline B{}^0\to \Sigma^{0}\overline{\Sigma^{*0}}$(*)
          & $1.25^{+0.14}_{-0.14}{}^{+0}_{-0.04}{}^{+0.68}_{-0.53}\pm0.08$%
           \\
$\overline B{}^0\to \Xi^{0}\overline{\Xi^{*0}}$
          & $0.00\pm0\pm0{}^{+0.01}_{-0.00}$%
          & $\overline B{}^0\to \Sigma^{-}\overline{\Sigma^{*-}}$
          & $0.18^{+0.05}_{-0.05}\pm 0{}^{+0.17}_{-0.10}\pm0$%
           \\      
$\overline B{}^0\to \Xi^{-}\overline{\Xi^{*-}}$
          & $0.14^{+0.04}_{-0.04}\pm 0{}^{+0.14}_{-0.08}\pm0$%
          & $\overline B{}^0\to \Lambda\overline{\Sigma^{*0}}$
          & $0.14^{+0.04}_{-0.04}{}^{+0}_{-0.00}{}^{+0.25}_{-0.08}{}^{+0.01}_{-0.00}$%
           \\  
\hline $B^-\to \Sigma^+\overline{\Delta^{++}}$(**)
          & $15.68^{+3.92}_{-3.81}{}^{+3.39}_{-0}{}^{+14.09}_{-9.39}\pm0.09$%
          & $\overline B{}^0\to \Sigma^{+}\overline{\Delta^+}$(*)
          & $4.85^{+1.21}_{-1.18}{}^{+1.05}_{-0}{}^{+4.35}_{-2.90}\pm0$%
           \\
$B^-\to \Sigma^0\overline{\Delta^+}$(*)
          & $10.11^{+2.60}_{-2.52}{}^{+1.09}_{-0}{}^{+8.78}_{-6.02}\pm0.07$%
          & $\overline B{}^0\to \Sigma^{0}\overline{\Delta^0}$(**)
          & $9.37^{+2.41}_{-2.34}{}^{+1.01}_{-0}{}^{+8.14}_{-5.58}\pm0$%
           \\ 
$B^-\to \Sigma^-\overline{\Delta^0}$
          & $4.91^{+1.29}_{-1.25}{}^{+0}_{-0.00}{}^{+4.18}_{-2.90}\pm0.04$%
          & $\overline B{}^0\to \Sigma^{-}\overline{\Delta^-}$
          & $13.64^{+3.60}_{-3.48}\pm0{}^{+11.63}_{-8.08}\pm0$%
           \\            
$B^-\to \Xi^{0}\overline{\Sigma^{*+}}$ (**)
          & $3.92^{+1.07}_{-1.02}{}^{+0}_{-0.80}{}^{+3.86}_{-2.55}\pm0.04$%
          & $\overline B{}^0\to \Xi^{0}\overline{\Sigma^{*0}}$ (*) 
          & $1.82^{+0.49}_{-0.48}{}^{+0}_{-0.37}{}^{+1.79}_{-1.18}\pm0$%
           \\ 
$B^-\to \Xi^{-}\overline{\Sigma^{*0}}$
          & $2.02^{+0.53}_{-0.51}{}^{+0}_{-0.00}{}^{+1.72}_{-1.19}\pm0.02$%
          & $\overline B{}^0\to \Xi^{-}\overline{\Sigma^{*-}}$(***)
          & $3.73^{+0.98}_{-0.95}\pm0{}^{+3.18}_{-2.21}\pm0$%
           \\ 
$B^-\to \Lambda\overline{\Delta^{+}}$
          & $0.10^{+0.01}_{-0.01}{}^{+0.13}_{-0}{}^{+0.04}_{-0.03}\pm0$%
          & $\overline B{}^0\to \Lambda\overline{\Delta^0}$
          & $0.09^{+0.01}_{-0.01}{}^{+0.12}_{-0}{}^{+0.04}_{-0.03}\pm0$%
           \\
$\overline B{}^0_s\to p\overline{\Delta^+}$
          & $0.00\pm0\pm0{}^{+0.0004}_{-0.00}$%
          & $\overline B{}^0_s\to \Sigma^{+}\overline{\Sigma^{*+}}$(**)
          & $5.36^{+1.26}_{-1.23}{}^{+1.09}_{-0}{}^{+4.93}_{-3.27}\pm0.03$%
           \\
$\overline B{}^0_s\to n\overline{\Delta^0}$
          & $0.00\pm0\pm0{}^{+0.0004}_{-0.00}$%
          & $\overline B{}^0_s\to \Sigma^{0}\overline{\Sigma^{*0}}$(*)
          & $5.19^{+1.25}_{-1.22}{}^{+0.52}_{-0}{}^{+4.62}_{-3.15}\pm0.02$%
           \\
$\overline B{}^0_s\to \Xi^{0}\overline{\Xi^{*0}}$(**)
          & $4.00^{+1.02}_{-0.99}{}^{+0}_{-0.77}{}^{+4.01}_{-2.63}\pm0.03$%
          & $\overline B{}^0_s\to \Sigma^{-}\overline{\Sigma^{*-}}$ 
          & $5.03^{+1.24}_{-1.21}\pm0{}^{+4.41}_{-3.03}\pm0$%
           \\ 
$\overline B{}^0_s\to \Xi^{-}\overline{\Xi^{*-}}$(**)
          & $4.11^{+1.01}_{-0.99}\pm0{}^{+3.60}_{-2.47}\pm0$%
          & $\overline B{}^0_s\to \Lambda\overline{\Sigma^{*0}}$
          & $0.05^{+0.01}_{-0.01}{}^{+0.06}_{-0}{}^{+0.02}_{-0.02}\pm0.01$%
           \\                                                                                                                                                                              
\end{tabular}
\end{ruledtabular}
\end{table}

\begin{table}[t!]
\caption{\label{tab: DBDS=0, -1} Same as Table~\ref{tab: BBDS=0, -1}, but for $\overline B_q\to\DB$ modes.
}
\begin{ruledtabular}
\centering
\begin{tabular}{llll}
Mode
          & ${\mathcal B}(10^{-8})$
          & Mode
          & ${\mathcal B}(10^{-8})$
          \\
\hline $B^-\to \Delta^0\overline{p}$(***)
          & $1.68^{+0.19}_{-0.19}{}^{+0.05}_{-0}{}^{+0.82}_{-0.62}{}^{+0.23}_{-0.22}$%
          & $\overline B{}^0_s\to \Delta^+\overline{\Sigma^{+}}$(*)
          & $1.50^{+0.16}_{-0.17}{}^{+0}_{-0.06}{}^{+0.81}_{-0.64}\pm0$%
           \\
$B^-\to \Delta^-\overline{n}$
          & $0.30^{+0.13}_{-0.12}{}^{+0.00}_{-0}{}^{+0.22}_{-0.15}{}^{+0.03}_{-0.00}$%
          & $\overline B{}^0_s\to \Delta^0\overline{\Sigma^{0}}$(**)
          & $0.91^{+0.13}_{-0.12}{}^{+0}_{-0.05}{}^{+0.61}_{-0.45}\pm0$%
           \\ 
$B^-\to \Sigma^{*0}\overline{\Sigma^{+}}$ (*)
          & $0.65^{+0.07}_{-0.07}{}^{+0.02}_{-0}{}^{+0.32}_{-0.24}{}^{+0.09}_{-0.08}$%
          &$\overline B{}^0_s\to \Delta^-\overline{\Sigma^{-}}$
          & $0.27^{+0.11}_{-0.10}\pm0{}^{+0.20}_{-0.13}\pm0$%
           \\            
$B^-\to \Sigma^{*-}\overline{\Sigma^{0}}$
          & $0.04^{+0.02}_{-0.01}{}^{+0.00}_{-0}{}^{+0.03}_{-0.02}\pm 0.00$%
          & $\overline B{}^0_s\to \Sigma^{*0}\overline{\Xi^{0}}$(*)
          & $2.33^{+0.26}_{-0.26}{}^{+0}_{-0.05}{}^{+1.08}_{-0.88}\pm0$%
           \\ 
$B^-\to \Xi^{*-}\overline{\Xi^{0}}$
          & $0.06^{+0.03}_{-0.02}{}^{+0.00}_{-0}{}^{+0.05}_{-0.03}{}^{+0.01}_{-0.00}$%
          & $\overline B{}^0_s\to \Sigma^{*-}\overline{\Xi^{-}}$
          & $0.07\pm0.03\pm0{}^{+0.05}_{-0.04}\pm0$%
           \\ 
$B^-\to \Sigma^{*-}\overline{\Lambda}$
          & $0.12^{+0.05}_{-0.05}{}^{+0.00}_{-0}{}^{+0.09}_{-0.06}{}^{+0.01}_{-0.00}$%
          & $\overline B{}^0_s\to \Delta^0\overline{\Lambda}$(***)
          & $2.14^{+0.24}_{-0.24}{}^{+0}_{-0.01}{}^{+0.83}_{-0.69}\pm0$%
           \\ 
$\overline B{}^0\to \Delta^+\overline{p}$(**)
          & $1.66^{+0.18}_{-0.18}{}^{+0}_{-0.06}{}^{+0.88}_{-0.69}{}^{+0.22}_{-0.20}$%
          & $\overline B{}^0\to \Sigma^{*+}\overline{\Sigma^{+}}$
          & $0.00\pm0\pm0\pm0{}^{+0.01}_{-0.00}$%
           \\
$\overline B{}^0\to \Delta^0\overline{n}$
          & $6.24^{+0.70}_{-0.69}{}^{+0}_{-0.14}{}^{+2.86}_{-2.32}{}^{+0.43}_{-0.41}$%
          & $\overline B{}^0\to \Sigma^{*0}\overline{\Sigma^{0}}$
          & $0.30^{+0.03}_{-0.03}{}^{+0.01}_{-0}{}^{+0.15}_{-0.11}\pm0.04$%
           \\
$\overline B{}^0\to \Xi^{*0}\overline{\Xi^{0}}$
          & $0.00\pm0\pm0\pm0{}^{+0.01}_{-0.00}$%
          & $\overline B{}^0\to \Sigma^{*-}\overline{\Sigma^{-}}$
          & $0.07^{+0.03}_{-0.03}\pm0{}^{+0.05}_{-0.04}\pm0$%
           \\      
$\overline B{}^0\to \Xi^{*-}\overline{\Xi^{-}}$
          & $0.06^{+0.02}_{-0.02}\pm0{}^{+0.04}_{-0.03}\pm0$%
          & $\overline B{}^0\to \Sigma^{*0}\overline{\Lambda}$(**)
           & $1.02^{+0.11}_{-0.11}{}^{+0}_{-0.04}{}^{+0.54}_{-0.42}\pm0.13$%
           \\    
\hline $B^-\to \Sigma^{*0}\overline{p}$(**) 
          & $1.01^{+0.44}_{-0.39}{}^{+0}_{-0.31}{}^{+0.76}_{-0.55}\pm0.02$%
          & $\overline B{}^0\to \Sigma^{*+}\overline{p}$(***)
          & $2.15^{+0.84}_{-0.76}{}^{+0.73}_{-0}{}^{+1.39}_{-1.03}\pm0$%
           \\
$B^-\to \Sigma^{*-}\overline{n}$
          & $2.09^{+0.89}_{-0.80}{}^{+0}_{-0.00}{}^{+1.35}_{-1.02}{}^{+0.03}_{-0.02}$%
          & $\overline B{}^0\to \Sigma^{*0}\overline{n}$
          & $1.24^{+0.43}_{-0.39}{}^{+0.80}_{-0}{}^{+0.79}_{-0.56}\pm0$%
           \\ 
$B^-\to \Xi^{*0}\overline{\Sigma^{+}}$(**)
          & $1.54^{+0.68}_{-0.60}{}^{+0}_{-0.47}{}^{+1.17}_{-0.84}\pm0.03$%
          &$\overline B{}^0\to \Xi^{*0}\overline{\Sigma^{0}}$
          & $0.71^{+0.31}_{-0.28}{}^{+0}_{-0.22}{}^{+0.54}_{-0.39}\pm0$%
           \\            
$B^-\to \Xi^{*-}\overline{\Sigma^{0}}$(*)
          & $0.80^{+0.34}_{-0.31}{}^{+0}_{-0.00}{}^{+0.52}_{-0.39}\pm0.01$%
          & $\overline B{}^0\to \Xi^{*-}\overline{\Sigma^{-}}$
          & $1.48^{+0.63}_{-0.57}\pm0{}^{+0.95}_{-0.72}\pm0$%
           \\ 
$B^-\to \Omega^-\overline{\Xi^{0}}$(**) 
          & $3.85^{+1.64}_{-1.47}{}^{+0}_{-0.00}{}^{+2.48}_{-1.87}\pm0.05$%
          & $\overline B{}^0\to \Omega^-\overline{\Xi^{-}}$(***) 
          & $3.55^{+1.51}_{-1.36}\pm0{}^{+2.29}_{-1.72}\pm0$%
           \\ 
$B^-\to \Xi^{*-}\overline{\Lambda}$(**)
          & $2.53^{+1.08}_{-0.97}{}^{+0}_{-0.00}{}^{+1.63}_{-1.23}\pm0.03$%
          & $\overline B{}^0\to \Xi^{*0}\overline{\Lambda}$(***)
          & $2.60^{+1.02}_{-0.92}{}^{+0.88}_{-0}{}^{+1.68}_{-1.24}\pm0$%
           \\ 
$\overline B{}^0_s\to \Delta^+\overline{p}$
          & $0.00\pm0\pm0\pm0{}^{+0.0004}_{-0.00}$%
          & $\overline B{}^0_s\to \Sigma^{*+}\overline{\Sigma^{+}}$(**)
          & $2.06^{+0.78}_{-0.71}{}^{+0.67}_{-0}{}^{+1.37}_{-1.00}\pm0.01$%
           \\
$\overline B{}^0_s\to \Delta^0\overline{n}$
          & $0.00\pm0\pm0\pm0{}^{+0.0004}_{-0.00}$%
          & $\overline B{}^0_s\to \Sigma^{*0}\overline{\Sigma^{0}}$(*)
          & $1.95^{+0.77}_{-0.70}{}^{+0.32}_{-0}{}^{+1.27}_{-0.95}\pm0.01$%
           \\
$\overline B{}^0_s\to \Xi^{*0}\overline{\Xi^{0}}$(**)
          & $1.92^{+0.64}_{-0.59}{}^{+1.21}_{-0}{}^{+1.25}_{-0.88}\pm0.01$%
          & $\overline B{}^0_s\to \Sigma^{*-}\overline{\Sigma^{-}}$
          & $1.86^{+0.76}_{-0.69}\pm0{}^{+1.23}_{-0.92}\pm0$%
           \\ 
$\overline B{}^0_s\to \Xi^{*-}\overline{\Xi^{-}}$(**)
          & $1.51^{+0.62}_{-0.56}\pm0{}^{+1.00}_{-0.75}\pm0$%
          & $\overline B{}^0_s\to \Sigma^{*0}\overline{\Lambda}$
          & $0.04^{+0.00}_{-0.00}{}^{+0.05}_{-0}{}^{+0.02}_{-0.01}\pm0.01$%
           \\                                                                                                                                                                                                      
\end{tabular}
\end{ruledtabular}
\end{table}

\begin{table}[t!]
\caption{\label{tab: DDDS=0, -1} Same as Table~\ref{tab: BBDS=0, -1}, but for $\overline B_q\to\DD$ modes. 
}
\centering
\begin{ruledtabular}
\begin{tabular}{llll}
Mode
          & ${\mathcal B}(10^{-8})$
          & Mode
          & ${\mathcal B}(10^{-8})$
          \\
\hline $B^-\to \Delta^+ \overline{\Delta^{++}}$ (**) 
          & $14.78^{+1.63}_{-1.63}{}^{+0}_{-0.55}{}^{+7.84}_{-6.18}{}^{+1.96}_{-1.83}$%
          & $\overline B{}^0_s\to \Delta^{+} \overline{\Sigma^{*+}}$(**) 
          & $4.58^{+0.50}_{-0.50}{}^{+0}_{-0.18}{}^{+2.48}_{-1.94}\pm0$%
           \\
$B^-\to \Delta^0 \overline{\Delta^{+}}$ 
          & $5.91^{+0.82}_{-0.77}{}^{+0}_{-0.35}{}^{+3.87}_{-2.90}{}^{+1.36}_{-1.19}$%
          & $\overline B{}^0_s\to \Delta^{0} \overline{\Sigma^{*0}}$(**) 
          & $2.77^{+0.39}_{-0.36}{}^{+0}_{-0.17}{}^{+1.86}_{-1.39}\pm0$%
           \\ 
$B^-\to \Delta^- \overline{\Delta^{0}}$
          & $0.84^{+0.36}_{-0.32}{}^{+0.00}_{-0}{}^{+0.60}_{-0.41}{}^{+0.08}_{-0.00}$%
          &$\overline B{}^0_s\to \Delta^{-} \overline{\Sigma^{*-}}$
          & $0.83^{+0.34}_{-0.31}\pm0{}^{+0.61}_{-0.41}\pm0$%
           \\            
$B^-\to \Sigma^{*0} \overline{\Sigma^{*+}}$(**) 
          & $2.76^{+0.38}_{-0.36}{}^{+0}_{-0.16}{}^{+1.80}_{-1.35}{}^{+0.63}_{-0.55}$%
          & $\overline B{}^0_s\to \Sigma^{*0} \overline{\Xi^{*0}}$(**) 
          & $2.58^{+0.36}_{-0.34}{}^{+0}_{-0.15}{}^{+1.74}_{-1.29}\pm0$%
           \\ 
$B^-\to \Sigma^{*-} \overline{\Sigma^{*0}}$
          & $0.52^{+0.22}_{-0.20}{}^{+0.00}_{-0}{}^{+0.37}_{-0.25}{}^{+0.05}_{-0.00}$%
          & $\overline B{}^0_s\to \Sigma^{*-} \overline{\Xi^{*-}}$(**) 
          & $1.02^{+0.42}_{-0.38}\pm 0{}^{+0.76}_{-0.51}\pm0$%
           \\ 
$B^-\to \Xi^{*-} \overline{\Xi^{*0}}$
          & $0.24^{+0.10}_{-0.09}{}^{+0.00}_{-0}{}^{+0.17}_{-0.12}{}^{+0.02}_{-0.00}$%
          & $\overline B{}^0_s\to \Xi^{*-} \overline{\Omega^{-}}$(**) 
          & $0.71^{+0.29}_{-0.26}\pm0{}^{+0.53}_{-0.35}\pm0$%
           \\ 
$\overline B{}^0\to \Delta^{++} \overline{\Delta^{++}}$
          & $0.00\pm0\pm0\pm0{}^{+0.25}_{-0.00}$%
          & $\overline B{}^0\to \Sigma^{*+} \overline{\Sigma^{*+}}$
          & $0.00\pm0\pm0\pm0{}^{+0.12}_{-0.00}$%
           \\
$\overline B{}^0\to \Delta^{+} \overline{\Delta^{+}}$(*) 
          & $4.57^{+0.50}_{-0.50}{}^{+0}_{-0.17}{}^{+2.43}_{-1.91}{}^{+1.68}_{-1.42}$%
          & $\overline B{}^0\to \Sigma^{*0} \overline{\Sigma^{*0}}$(*) 
          & $1.28^{+0.18}_{-0.17}{}^{+0}_{-0.08}{}^{+0.84}_{-0.63}{}^{+0.54}_{-0.44}$%
           \\
$\overline B{}^0\to \Delta^{0} \overline{\Delta^{0}}$(***) 
          & $5.48^{+0.76}_{-0.72}{}^{+0}_{-0.32}{}^{+3.59}_{-2.69}{}^{+1.11}_{-1.00}$%
          & $\overline B{}^0\to \Sigma^{*-} \overline{\Sigma^{*-}}$(***) 
          & $0.96^{+0.41}_{-0.37}{}^{+0}_{-0.00}{}^{+0.69}_{-0.47}{}^{+0.17}_{-0.15}$%
           \\ 
$\overline B{}^0\to \Delta^{-} \overline{\Delta^{-}}$
          & $2.32^{+0.99}_{-0.89}{}^{+0}_{-0.00}{}^{+1.67}_{-1.13}{}^{+0.26}_{-0.24}$%
          & $\overline B{}^0\to \Xi^{*0} \overline{\Xi^{*0}}$
          & $0.00\pm0\pm0\pm0{}^{+0.04}_{-0.00}$%
           \\ 
$\overline B{}^0\to \Omega^{-} \overline{\Omega^{-}}$
          & $0.00\pm0\pm0\pm0{}^{+0.01}_{-0.00}$%
          & $\overline B{}^0\to \Xi^{*-} \overline{\Xi^{*-}}$
          & $0.22^{+0.09}_{-0.08}{}^{+0}_{-0.00}{}^{+0.16}_{-0.11}{}^{+0.08}_{-0.06}$%
           \\  
\hline $B^-\to \Sigma^{*+} \overline{\Delta^{++}}$(***) 
          & $20.53^{+8.03}_{-7.28}{}^{+6.93}_{-0}{}^{+13.28}_{-9.80}\pm0.15$%
          & $\overline B{}^0\to \Sigma^{*+} \overline{\Delta^{+}}$(**) 
          & $6.34^{+2.48}_{-2.25}{}^{+2.14}_{-0}{}^{+4.10}_{-3.03}\pm0$%
           \\
$B^-\to \Sigma^{*0} \overline{\Delta^{+}}$(*) 
          & $12.94^{+5.32}_{-4.79}{}^{+2.18}_{-0}{}^{+8.22}_{-6.18}{}^{+0.13}_{-0.12}$%
          & $\overline B{}^0\to \Sigma^{*0} \overline{\Delta^{0}}$(**) 
          & $11.99^{+4.93}_{-4.45}{}^{+2.03}_{-0}{}^{+7.62}_{-5.72}\pm0$%
           \\ 
$B^-\to \Sigma^{*-} \overline{\Delta^{0}}$(***) 
          & $6.19^{+2.64}_{-2.36}{}^{+0}_{-0.00}{}^{+3.99}_{-3.00}{}^{+0.08}_{-0.07}$%
          &$\overline B{}^0\to \Sigma^{*-} \overline{\Delta^{-}}$
          & $17.21^{+7.33}_{-6.58}\pm0{}^{+11.11}_{-8.36}\pm0$%
           \\            
$B^-\to \Xi^{*0} \overline{\Sigma^{*+}}$(***) 
          & $23.99^{+9.86}_{-8.89}{}^{+4.05}_{-0}{}^{+15.24}_{-11.46}\pm0.23$%
          & $\overline B{}^0\to \Xi^{*0} \overline{\Sigma^{*0}}$(**) 
          & $11.12^{+4.57}_{-4.12}{}^{+1.88}_{-0}{}^{+7.06}_{-5.31}\pm0$%
           \\ 
$B^-\to \Xi^{*-} \overline{\Sigma^{*0}}$(*) 
          & $11.47^{+4.89}_{-4.38}{}^{+0}_{-0.00}{}^{+7.41}_{-5.57}\pm0.14$%
          & $\overline B{}^0\to \Xi^{*-} \overline{\Sigma^{*-}}$(**) 
          & $21.25^{+9.06}_{-8.12}\pm0{}^{+13.72}_{-10.32}\pm0$%
           \\ 
$B^-\to \Omega^{-} \overline{\Xi^{*0}}$(***) 
          & $15.80^{+6.74}_{-6.04}{}^{+0}_{-0.01}{}^{+10.20}_{-7.67}\pm0.19$%
          & $\overline B{}^0\to \Omega^{-} \overline{\Xi^{*-}}$(**) 
          & $14.64^{+6.24}_{-5.60}\pm0{}^{+9.45}_{-7.11}\pm0$%
           \\ 
$\overline B{}^0_s\to \Delta^{++} \overline{\Delta^{++}}$
          & $0.00\pm0\pm0\pm0{}^{+0.20}_{-0.00}$%
          & $\overline B{}^0_s\to \Sigma^{*+} \overline{\Sigma^{*+}}$(***) 
          & $6.74^{+2.55}_{-2.33}{}^{+2.19}_{-0}{}^{+4.48}_{-3.29}{}^{+2.20}_{-1.87}$%
           \\
$\overline B{}^0_s\to \Delta^{+} \overline{\Delta^{+}}$
          & $0.00\pm0\pm0\pm0{}^{+0.18}_{-0.00}$%
          & $\overline B{}^0_s\to \Sigma^{*0} \overline{\Sigma^{*0}}$(*) 
          & $6.39^{+2.53}_{-2.30}{}^{+1.03}_{-0}{}^{+4.17}_{-3.12}{}^{+2.12}_{-1.81}$%
           \\
$\overline B{}^0_s\to \Delta^{0} \overline{\Delta^{0}}$
          & $0.00\pm0\pm0\pm0{}^{+0.17}_{-0.00}$%
          & $\overline B{}^0_s\to \Sigma^{*-} \overline{\Sigma^{*-}}$(***) 
          & $6.12^{+2.51}_{-2.27}{}^{+0}_{-0.01}{}^{+4.06}_{-3.03}{}^{+2.02}_{-1.73}$%
           \\ 
$\overline B{}^0_s\to \Delta^{-} \overline{\Delta^{-}}$
          & $0.00\pm0\pm0\pm0{}^{+0.15}_{-0.00}$%
          & $\overline B{}^0_s\to \Xi^{*0} \overline{\Xi^{*0}}$(***) 
          & $23.64^{+9.38}_{-8.51}{}^{+3.82}_{-0}{}^{+15.45}_{-11.53}{}^{+3.79}_{-3.50}$%
           \\ 
$\overline B{}^0_s\to \Omega^{-} \overline{\Omega^{-}}$(***) 
          & $46.75^{+19.22}_{-17.35}{}^{+0}_{-0.08}{}^{+31.03}_{-23.17}{}^{+4.91}_{-4.66}$%
          & $\overline B{}^0_s\to \Xi^{*-} \overline{\Xi^{*-}}$(*) 
          & $22.63^{+9.30}_{-8.39}{}^{+0}_{-0.04}{}^{+15.02}_{-11.21}{}^{+3.60}_{-3.33}$%
           \\                                                                                                                                                                                 
\end{tabular}
\end{ruledtabular}
\end{table}

\begin{table}[t!]
\caption{\label{tab: BcDS=0, -1} 
$B^-_c\to\bfBB'$ decay rates. The central values correspond to factorization contributions.
}
\begin{ruledtabular}
\centering
{\begin{tabular}{lllr}
Mode ($\Delta S=0$)
          & ${\mathcal B}(10^{-9})$
          & Mode ($\Delta S=-1$)
          & ${\mathcal B}(10^{-9})$
          \\
\hline $B^-_c\to n\overline{p}$
          & $0.00096^{+75.68}_{-0.00}$%
          & $B^-_c\to \Sigma^{0}\overline{p}$
          & $0.00032^{+0.09}_{-0.00}$%
           \\
$B^-_c\to \Sigma^{0}\overline{\Sigma^{+}}$
          & $0.00030^{+51.11}_{-0.00}$%
          & $B^-_c\to \Sigma^{-}\overline{n}$
          & $0.00063^{+0.19}_{-0.00}$%
           \\ 
$B^-_c\to \Sigma^{-}\overline{\Sigma^{0}}$
          & $0.00030^{+51.05}_{-0.00}$%
          & $B^-_c\to \Xi^{0}\overline{\Sigma^{+}}$
          & $0.01505^{+4.42}_{-0.00}$%
           \\            
$B^-_c\to \Sigma^{-}\overline{\Lambda}$
          & $0.00022^{+17.09}_{-0.00}$%
          & $B^-_c\to\Xi^{-}\overline{\Sigma^{0}}$
          & $0.00751^{+2.21}_{-0.00}$%
           \\ 
$B^-_c\to \Xi^{-}\overline{\Xi^{0}}$
          & $0.00004^{+2.75}_{-0.00}$%
          & $B^-_c\to \Xi^{-}\overline{\Lambda}$
          & $0.00091^{+1.28}_{-0.00}$%
           \\ 
$B^-_c\to \Lambda\overline{\Sigma^+}$
          & $0.00022^{+17.11}_{-0.00}$%
          & $B^-_c\to \Lambda\overline{p}$
          & $0.00860^{+3.68}_{-0.00}$
           \\  
\hline $B^-_c\to p\overline{\Delta^{++}}$
          & $0.00017^{+13.54}_{-0.00}$%
          & $B^-_c\to \Sigma^+\overline{\Delta^{++}}$
          & $0.00265^{+0.78}_{-0.00}$%
           \\
$B^-_c\to n\overline{\Delta^+}$
          & $0.00006^{+4.51}_{-0.00}$%
          & $B^-_c\to \Sigma^0\overline{\Delta^+}$
          & $0.00176^{+0.52}_{-0.00}$%
           \\ 
$B^-_c\to\Sigma^0\overline{\Sigma^{*+}}$
          & $0.00003^{+1.91}_{-0.00}$%
          & $B^-_c\to \Sigma^-\overline{\Delta^0}$
          & $0.00088^{+0.26}_{-0.00}$%
           \\            
$B^-_c\to\Sigma^-\overline{\Sigma^{*0}}$
          & $0.00003^{+1.91}_{-0.00}$%
          & $B^-_c\to \Xi^{0}\overline{\Sigma^{*+}}$
          & $0.00078^{+0.23}_{-0.00}$%
           \\ 
$B^-_c\to\Xi^{-}\overline{\Xi^{*0}}$
          & $0.00004^{+3.33}_{-0.00}$%
          & $B^-_c\to \Xi^{-}\overline{\Sigma^{*0}}$
          & $0.00039^{+0.11}_{-0.00}$%
           \\ 
$B^-_c\to\Lambda\overline{\Sigma^{*+}}$
          & $0.00008^{+5.93}_{-0.00}$%
          & $B^-_c\to \Lambda\overline{\Delta^{+}}$
          & $0$%
           \\ 
\hline $B^-_c\to \Delta^0\overline{p}$
          & $0.00006^{+4.51}_{-0.00}$%
          & $B^-_c\to \Sigma^{*0}\overline{p}$
          & $0.00045^{+0.13}_{-0.00}$%
           \\
$B^-_c\to \Delta^-\overline{n}$
          & $0.00017^{+13.53}_{-0.00}$%
          & $B^-_c\to \Sigma^{*-}\overline{n}$
          & $0.00090^{+0.27}_{-0.00}$%
           \\ 
$B^-_c\to \Sigma^{*0}\overline{\Sigma^{+}}$
          & $0.00003^{+1.91}_{-0.00}$%
          & $B^-_c\to \Xi^{*0}\overline{\Sigma^{+}}$
          & $0.00076^{+0.22}_{-0.00}$%
           \\            
$B^-_c\to \Sigma^{*-}\overline{\Sigma^{0}}$
          & $0.00003^{+1.91}_{-0.00}$%
          & $B^-_c\to \Xi^{*-}\overline{\Sigma^{0}}$
          & $0.00038^{+0.11}_{-0.00}$%
           \\ 
$B^-_c\to \Xi^{*-}\overline{\Xi^{0}}$
          & $0.00004^{+3.33}_{-0.00}$%
          & $B^-_c\to \Omega^-\overline{\Xi^{0}}$ 
          & $0.00198^{+0.58}_{-0.00}$%
           \\ 
$B^-_c\to \Sigma^{*-}\overline{\Lambda}$
          & $0.00008^{+5.91}_{-0.06}$%
          & $B^-_c\to \Xi^{*-}\overline{\Lambda}$
          & $0.00118^{+0.35}_{-0.00}$%
           \\ 
\hline $B^-_c\to \Delta^+ \overline{\Delta^{++}}$
          & $0.00044^{+33.83}_{-0.00}$%
          & $B^-_c\to \Sigma^{*+} \overline{\Delta^{++}}$
          & $0.00711^{+2.09}_{-0.00}$%
           \\
$B^-_c\to \Delta^0 \overline{\Delta^{+}}$
          & $0.00059^{+45.11}_{-0.00}$%
          & $B^-_c\to \Sigma^{*0} \overline{\Delta^{+}}$
          & $0.00474^{+1.39}_{-0.00}$%
           \\ 
$B^-_c\to \Delta^- \overline{\Delta^{0}}$
          & $0.00044^{+33.83}_{-0.00}$%
          & $B^-_c\to \Sigma^{*-} \overline{\Delta^{0}}$
          & $0.00237^{+0.70}_{-0.00}$%
           \\            
$B^-_c\to \Sigma^{*0} \overline{\Sigma^{*+}}$
          & $0.00029^{+21.54}_{-0.00}$%
          & $B^-_c\to \Xi^{*0} \overline{\Sigma^{*+}}$
          & $0.00903^{+2.65}_{-0.00}$%
           \\ 
$B^-_c\to \Sigma^{*-} \overline{\Sigma^{*0}}$
          & $0.00029^{+21.52}_{-0.00}$%
          & $B^-_c\to \Xi^{*-} \overline{\Sigma^{*0}}$
          & $0.00451^{+1.32}_{-0.00}$%
           \\ 
$B^-_c\to \Xi^{*-} \overline{\Xi^{*0}}$
          & $0.00014^{+10.21}_{-0.00}$%
          & $B^-_c\to \Omega^{-} \overline{\Xi^{*0}}$
          & $0.00642^{+1.88}_{-0.00}$%
           \\ 
\end{tabular}
}
\\
\end{ruledtabular}
\end{table}

Predictions on branching ratios of $\overline B_q\to \bfBB$ decays with inputs using ${\cal B}(\overline B{}^0\to pp)$ and ${\cal B}(B^-\to \Lambda\bar p)$ data are given in Tables \ref{tab: BBDS=0, -1}, \ref{tab: BDDS=0, -1}, \ref{tab: DBDS=0, -1} and \ref{tab: DDDS=0, -1}.
There are four uncertainties, the first one is from the uncertainties of $\chi$ and $\kappa$ as shown in Eq. (\ref{eq:chikappa}), reflecting the experimental uncertainties in $\overline B{}^0\to p\bar p$ and $B^-\to \Lambda\bar p$ rates;
the second one is from varying the penguin strong phase $\phi$, where we use a common strong phase for $P^{(\prime)}$, $P^{(\prime)}_{PE}$ and $P^{(\prime)}_{EW}$ for simplicity;
the third one is by relaxing the asymptotic relations by varying $r_{t,i}$, $r_{p,i}$, $r_{ewp,i}$ and $r_{pe, i}$ in Eqs. (\ref{eq:correction0}) and 
(\ref{eq:correction2}); 
the last one is from the uncertainties in sub-leading contributions from $r_{a, i}$, $r_{e.i}$, $r_{pa, i}$, $\eta_{a, i}$, $\eta_{e.i}$ and $\eta_{pa, i}$ in Eq. (\ref{eq:correction3}).

Modes are ranked according to decay rates and detectability. Those with three asterisks are the most favorable ones. They have relatively large rates and the baryons can decay to all charged final states. Those with two asterisks are the second ranked ones. They need a $\pi^0$ or $\gamma$ for detection. Those with one asterisk are the third ranked ones, where $\pi^0\pi^0$, $\pi^0\gamma$ or $\gamma\gamma$ are needed for detection.

As shown in Table \ref{tab: BBDS=0, -1}, $\overline B{}^0\to pp$ and $B^-\to \Lambda\bar p$ decays are modes ranked as $***$. They have large rates and very good detectability. It is natural that they are the first two modes observed. In fact, they are the only two modes being detected so far.
Note that the second uncertainties of the rates of these two modes from varying the relative phase of $T^{(\prime)}$ and $P^{(\prime)}+PE^{(\prime)}$ are comparably small. 

There are other modes, such as $\overline B_s^0\to\Xi^-\overline{\Xi^-}$ decay, ranked as $***$, which are predicted to have sizable rates and good detectability.
The $\overline B_s^0\to p\bar p$ rate from factorization contribution is predicted to be very rare, being $10^{-9}$ or $10^{-10}$ of the $\overline B^0\to p\bar p$ or $B^-\to\Lambda \bar p$ decay rate. 
The smallness of the factorization contribution to this decay rate can be understood using Eqs.
(\ref{eq:P/T}) and (\ref{eq: A' E' PA' per PE'}).
The $\bar B_s\to p\bar p$ rate shown in Table~\ref{tab: BBDS=0, -1} comes mainly from non-factorizable contributions estimated using Eq. (\ref{eq:correction3}).
Therefore, once the $\bar B_s\to p\bar p$ rate is measured, one can use it to give valuable informations on non-factorizable contributions.
We will discuss the consequences of an enhanced $\bar B_s\to p\bar p$ decay rate saturating the present experimental bound in next section.

In Table \ref{tab: BcDS=0, -1}, we show the predictions of $B^-_c\to\bfBB'$ branching ratios. The central values correspond to decay rates from factorization contributions, which are very rare, ranging from $10^{-14}$ to $10^{-11}$. 
These modes are all governed by annihilation amplitudes $A^{(\prime) c}_{\bf BB'}$. Although, as shown in Eq. (\ref{eq: Ac per A}), the factorizable $A^{(\prime) c}_{B_c}$ is about 10 times larger than $A^{(\prime)}_{B_u}$, the latter is very small.
Hence, it is natural to have $B^-_c\to \bfBB'$ rates from factorization contributions be suppressed than a typical $B^-\to\bfBB'$ rate by several orders of magnitude.
Nevertheless with non-factorization contributions, the $B^-_c\to \bfBB'$ branching ratios can be enhanced to $10^{-9}$ or even $10^{-8}$ level.
Measuring these modes can provide valuable information on non-factorization contributions to annihilation amplitudes.

\subsection{Numerical Results on Direct $CP$ Asymmetries}

\begin{table}[t!]
\caption{\label{tab: AcpBBDS=0, -1} Direct $CP$ asymmetries ($\A$ in $\%$) for $\overline B_q\to\BB$ modes
for $\phi=0$, $\pm\pi/4$ and $\pm\pi/2$.
}
\begin{ruledtabular}
\centering
{
\footnotesize
\begin{tabular}{lccclccc}
Mode
          & $\phi=0$
          & $\phi=\pm\pi/4$
          & $\phi=\pm\pi/2$
          & Mode
          & $\phi=0$
          & $\phi=\pm\pi/4$
          & $\phi=\pm\pi/2$
          \\
\hline $B^-\to n\overline{p}$
          & $0\pm76.0$%
          & $\mp(66.9_{-71.5}^{+33.1})$%
          & $\mp(97.0_{-74.7}^{+3.0})$%
          & $\overline B{}^0_s\to p\overline{\Sigma^{+}}$ 
          & $0\pm53.0$%
          & $\mp(37.0_{-39.7}^{+52.4})$%
          & $\mp(53.1_{-26.6}^{+45.5})$%
           \\
$B^-\to \Sigma^{0}\overline{\Sigma^{+}}$
          & $0\pm66.1$%
          & $\mp(57.1_{-59.0}^{+42.6})$%
          & $\mp(82.5_{-42.2}^{+17.5})$%
          & $\overline B{}^0_s\to n\overline{\Sigma^{0}}$ 
          & $0\pm28.9$%
          & $\pm(30.9^{+27.8}_{-24.9})$%
          & $\pm(43.3^{+25.6}_{-20.3})$%
          \\ 
$B^-\to \Sigma^{-}\overline{\Sigma^{0}}$
          & $0\pm 100$%
          & $\pm(0.1^{+ 99.9}_{-100.1})$%
          & $\pm(0.1^{+99.9}_{-100.1})$%
          &$\overline B{}^0_s\to n\overline{\Lambda}$
          & $0\pm49.6$%
          & $\mp(67.3_{-44.1}^{+30.6})$%
          & $\mp(97.6_{-28.2}^{+2.5})$%
           \\            
$B^-\to \Sigma^{-}\overline{\Lambda}$
          & $0\pm 89.0$%
          & $\pm(0.1^{+88.9}_{-89.0})$%
          & $\pm(0.1^{+88.9}_{-89.0})$%
          & $\overline B{}^0_s\to \Sigma^{0}\overline{\Xi^{0}}$ 
          & $0\pm28.1$%
          & $\mp(43.5_{-24.8}^{+26.4})$%
          & $\mp(62.6_{-22.1}^{+23.6})$%
           \\ 
$B^-\to \Xi^{-}\overline{\Xi^{0}}$
          & $0\pm 89.0$%
          & $\pm(0.1^{+88.9}_{-89.0})$%
          & $\pm(0.1^{+88.9}_{-89.0})$%
          & $\overline B{}^0_s\to \Sigma^{-}\overline{\Xi^{-}}$ 
          & $0\pm50.4$%
          & $0\pm50.4$%
          & $0\pm50.4$%
           \\ 
$B^-\to \Lambda\overline{\Sigma^+}$
          & $0\pm 100$%
          & $\pm(0.1^{+99.9}_{-100.1})$%
          & $\pm(0.1^{+99.9}_{-100.1})$%
          & $\overline B{}^0_s\to \Lambda\overline{\Xi^0}$
          & $0\pm100$%
          & $0\pm100$%
          & $0\pm100$%
           \\ 
$\overline B{}^0\to p\overline{p}$ 
          & $0\pm100$ %
          & $\mp(36.1_{-117.9}^{+63.9})$ %
          & $\mp(51.7_{-102.8}^{+48.3})$ %
          & $\overline B{}^0\to \Sigma^{+}\overline{\Sigma^{+}}$
          & $0\pm 29.6$%
          & $0\pm 29.6$%
          & $0\pm 29.6$%
           \\           
$\overline B{}^0\to n\overline{n}$
          & $0\pm 53.5$%
          & $\mp(57.1_{-48.9}^{+38.5})$%
          & $\mp(82.5_{-40.2}^{+17.5})$%
          & $\overline B{}^0\to \Sigma^{0}\overline{\Sigma^{0}}$ 
          & $0\pm72.1$%
          & $\mp(57.1_{-65.6}^{+42.9})$%
          & $\mp(82.5_{-49.0}^{+17.5})$%
           \\
$\overline B{}^0\to \Xi^{0}\overline{\Xi^{0}}$
          & $0\pm97.8$%
          & $0\pm97.8$%
          & $0\pm97.8$%
          & $\overline B{}^0\to \Sigma^{-}\overline{\Sigma^{-}}$ 
          & $0\pm81.9$%
          & $0\pm81.9$%
          & $0\pm81.9$%
           \\      
$\overline B{}^0\to \Xi^{-}\overline{\Xi^{-}}$
          & $0\pm78.3$%
          & $0\pm78.3$%
          & $0\pm78.2$%
          & $\overline B{}^0\to \Sigma^{0}\overline{\Lambda}$ 
          & $0\pm 27.4$%
          & $\mp(36.1_{-23.2}^{+27.8})$%
          & $\mp(51.7_{-20.2}^{+27.0})$%
           \\
$\overline B{}^0\to \Lambda\overline{\Lambda}$
          & $0\pm 100$%
          & $0\pm 100$%
          & $0\pm 100$%
          & $\overline B{}^0\to \Lambda\overline{\Sigma^{0}}$ 
          & $0\pm 100$%
          & $0.0\pm100.0$%
          & $0.0\pm100.0$%
          \\ 
\hline $B^-\to \Sigma^{0}\overline{p}$
          & $0\pm36.1$%
          & $\mp(23.7_{-32.3}^{+38.9})$%
          & $\mp(37.8_{-28.0}^{+41.5})$%
          & $\overline B{}^0\to \Sigma^{+}\overline{p}$ 
          & $0\pm31.8$%
          & $\pm(23.8^{+31.8}_{-25.0})$%
          & $\pm(30.2^{+28.8}_{-19.8})$%
           \\
$B^-\to \Sigma^{-}\overline{n}$
          & $0\pm5.4$%
          & $0.0\pm 5.4$%
          & $0.0\pm 5.4$%
          & $\overline B{}^0\to \Sigma^{0}\overline{n}$
          & $0\pm55.9$%
          & $\pm(43.8^{+42.4}_{-45.7})$%
          & $\pm(51.3^{+35.1}_{-31.8})$%
           \\ 
$B^-\to \Xi^{0}\overline{\Sigma^{+}}$ 
          & $0\pm7.0$%
          & $\pm(4.9^{+8.4}_{-5.3})$%
          & $\pm(6.8^{+9.1}_{-5.9})$%
          &$\overline B{}^0\to \Xi^{0}\overline{\Sigma^{0}}$ 
          & $0\pm3.7$%
          & $\pm(4.9^{+5.0}_{-2.9})$%
          & $\pm(6.8^{+5.8}_{-3.6})$%
           \\            
$B^-\to\Xi^{-}\overline{\Sigma^{0}}$ 
          & $0\pm5.4$%
          & $0.0\pm5.4$%
          & $0.0\pm5.4$%
          & $\overline B{}^0\to \Xi^{0}\overline{\Lambda}$ 
          & $0\pm47.6$%
          & $\pm(23.8^{+53.6}_{-26.2})$%
          & $\pm(30.2^{+49.0}_{-16.7})$%
           \\ 
$B^-\to \Xi^{-}\overline{\Lambda}$ 
          & $0\pm24.7$%
          & $\pm(0.0^{+24.6}_{-24.7})$%
          & $\pm(0.0^{+24.6}_{-24.7})$%
          & $\overline B{}^0\to  \Xi^{-}\overline{\Sigma^{-}}$ 
          & $0\pm2.3$%
          & $0\pm2.3$%
          & $0\pm2.3$%
           \\ 
$B^-\to \Lambda\overline{p}$ 
          & $0\pm11.3$%
          & $\pm(8.2^{+13.7}_{-8.3})$%
          & $\pm(11.1^{+14.3}_{-8.0})$%
          & $\overline B{}^0\to \Lambda\overline{n}$
          & $0\pm13.9$%
          & $\pm(16.2^{+16.4}_{-10.1})$%
          & $\pm(21.2^{+16.3}_{-9.0})$%
          \\                    
$\overline B{}^0_s\to p\overline{p}$
          & $0\pm 80.6$%
          & $0\pm 80.6$%
          & $0\pm 80.6$%
          & $\overline B{}^0_s\to \Sigma^{+}\overline{\Sigma^{+}}$ 
          & $0\pm60.0$%
          & $\pm(23.1^{+56.8}_{-48.4})$%
          & $\pm(29.4^{+51.3}_{-33.6})$%
          \\
$\overline B{}^0_s\to n\overline{n}$
          & $0\pm 19.2$%
          & $0\pm 19.2$%
          & $0\pm 19.2$%
          & $\overline B{}^0_s\to \Sigma^{0}\overline{\Sigma^{0}}$ 
          & $0\pm33.4$%
          & $\pm(11.8^{+36.7}_{-25.2})$%
          & $\pm(15.8^{+35.8}_{-20.0})$%
           \\
$\overline B{}^0_s\to \Xi^{0}\overline{\Xi^{0}}$ 
          & $0\pm15.7$%
          & $\pm(11.8^{+20.2}_{-10.0})$%
          & $\pm(15.8^{+21.0}_{-8.8})$%
          & $\overline B{}^0_s\to \Sigma^{-}\overline{\Sigma^{-}}$
          & $0\pm4.2$%
          & $0\pm4.2$%
          & $0\pm4.2$%
           \\ 
$\overline B{}^0_s\to \Xi^{-}\overline{\Xi^{-}}$ 
          & $0\pm4.7$%
          & $0\pm4.7$%
          & $0\pm4.7$%
          & $\overline B{}^0_s\to \Sigma^{0}\overline{\Lambda}$
          & $0\pm96.6$%
          & $\pm(78.7^{+21.3}_{-98.7})$%
          & $\pm(80.9^{+19.1}_{-41.1})$%
           \\    
$\overline B{}^0_s\to \Lambda\overline{\Lambda}$ 
          & $0\pm13.8$%
          & $\pm(11.8^{+17.0}_{-9.5})$%
          & $\pm(15.8^{+17.6}_{-8.7})$%
          & $\overline B{}^0_s\to \Lambda\overline{\Sigma^{0}}$
          & $0\pm74.1$%
          & $\pm(78.7^{+21.3}_{-58.1})$%
          & $\pm(80.9^{+19.1}_{-31.0})$%
           \\                                                                                                                                                                                                              
\end{tabular}
}
\\
\end{ruledtabular}
\end{table}

\begin{table}[t!]
\caption{\label{tab: AcpBDDS=0, -1} Same as Table~\ref{tab: AcpBBDS=0, -1}, but for $\overline B_q\to\BD$ modes.
}
\begin{ruledtabular}
\centering
{
\footnotesize
\begin{tabular}{lccclccc}
Mode
          & $\phi=0$
          & $\phi=\pm\pi/4$
          & $\phi=\pm\pi/2$
          & Mode
          & $\phi=0$
          & $\phi=\pm\pi/4$
          & $\phi=\pm\pi/2$
          \\
\hline $B^-\to p\overline{\Delta^{++}}$
          & $0\pm77.7$%
          & $\mp(49.5_{-66.2}^{+50.5})$%
          & $\mp(71.3_{-40.3}^{+28.7})$%
          & $\overline B{}^0_s\to p\overline{\Sigma^{*+}}$ 
          & $0\pm72.3$%
          & $\mp(51.5_{-62.0}^{+48.5})$%
          & $\mp(74.2_{-39.8}^{+25.8})$%
          \\
$B^-\to n\overline{\Delta^+}$
          & $0\pm53.8$%
          & $\pm(45.8^{+42.1}_{-47.5})$%
          & $\pm(63.7^{+32.3}_{-32.9})$%
          & $\overline B{}^0_s\to n\overline{\Sigma^{*0}}$
          & $0\pm51.2$%
          & $\pm(48.3^{+38.5}_{-46.2})$%
          & $\pm(67.0^{+28.5}_{-32.9})$%
           \\ 
$B^-\to\Sigma^0\overline{\Sigma^{*+}}$
          & $0\pm24.6$%
          & $\mp(29.9_{-21.2}^{+25.7})$%
          & $\mp(42.8_{-19.1}^{+25.9})$%
          &$\overline B{}^0_s\to \Sigma^{0}\overline{\Xi^{*0}}$ 
          & $0\pm24.2$%
          & $\mp(31.4_{-21.4}^{+24.8})$%
          & $\mp(44.9_{-19.7}^{+24.7})$%
           \\            
$B^-\to\Sigma^-\overline{\Sigma^{*0}}$
          & $0\pm88.5$%
          & $\pm(0.1^{88.5}_{-88.6})$%
          & $\pm(0.1^{+88.5}_{-88.6})$%
          & $\overline B{}^0_s\to \Sigma^{-}\overline{\Xi^{*-}}$ 
          & $0\pm 65.6$%
          & $0\pm 65.6$%
          & $0\pm 65.6$%
           \\ 
$B^-\to\Xi^{-}\overline{\Xi^{*0}}$
          & $0\pm 88.5$%
          & $\pm(0.1^{+88.5}_{-88.6})$%
          & $\pm(0.1^{+88.5}_{-88.6})$%
          & $\overline B{}^0_s\to \Xi^{-}\overline{\Omega^-}$
          & $0\pm 65.6$%
          & $0\pm 65.6$%
          & $0\pm 65.6$%
           \\ 
$B^-\to\Lambda\overline{\Sigma^{*+}}$
          & $0\pm 100$%
          & $0.0\pm 100.0$%
          & $\pm(0.1^{+99.9}_{-100.1})$%
          & $\overline B{}^0_s\to \Lambda\overline{\Xi^{*0}}$
          & $0\pm100$%
          & $0\pm100$%
          & $0\pm100$%
           \\ 
$\overline B{}^0\to p\overline{\Delta^+}$ 
          & $0\pm77.6$%
          & $\mp(49.5_{-66.2}^{+50.5})$%
          & $\mp(71.3_{-40.3}^{+28.7})$%
          & $\overline B{}^0\to \Sigma^{+}\overline{\Sigma^{*+}}$
          & $0$%
          & $0$%
          & $0$%
           \\
$\overline B{}^0\to n\overline{\Delta^0}$
          & $0\pm53.8$%
          & $\pm(45.8^{+42.1}_{-47.5})$%
          & $\pm(63.7^{+32.3}_{-32.9})$%
          & $\overline B{}^0\to \Sigma^{0}\overline{\Sigma^{*0}}$
          & $0\pm24.6$%
          & $\mp(29.9_{-21.2}^{+25.7})$%
          & $\mp(42.7_{-19.1}^{+25.9})$%
           \\
$\overline B{}^0\to \Xi^{0}\overline{\Xi^{*0}}$
          & $0$%
          & $0$%
          & $0$%
          & $\overline B{}^0\to \Sigma^{-}\overline{\Sigma^{*-}}$
          & $0\pm64.0$%
          & $0\pm64.0$%
          & $0\pm64.0$%
           \\      
$\overline B{}^0\to \Xi^{-}\overline{\Xi^{*-}}$
          & $0\pm64.0$%
          & $0\pm64.0$%
          & $0\pm64.0$%
          & $\overline B{}^0\to \Lambda\overline{\Sigma^{*0}}$
          & $0\pm100$%
          & $0.0\pm100.0$%
          & $0.0\pm100.0$%
           \\
\hline $B^-\to \Sigma^+\overline{\Delta^{++}}$ 
          & $0\pm26.8$%
          & $\pm(15.5^{+29.4}_{-20.0})$%
          & $\pm(20.4^{+28.3}_{-16.1})$%
          & $\overline B{}^0\to \Sigma^{+}\overline{\Delta^+}$
          & $0\pm24.7$%
          & $\pm(15.5^{+27.3}_{-18.2})$%
          & $\pm(20.4^{+26.4}_{-14.8})$%
          \\
$B^-\to \Sigma^0\overline{\Delta^+}$ 
          & $0\pm15.1$%
          & $\pm(7.8^{+17.6}_{-11.2})$%
          & $\pm(10.7^{+18.1}_{-10.4})$%
          & $\overline B{}^0\to \Sigma^{0}\overline{\Delta^0}$ 
          & $0\pm12.7$%
          & $\pm(7.8^{+15.3}_{-9.3})$%
          & $\pm(10.7^{+15.9}_{-8.8})$%
           \\ 
$B^-\to \Sigma^-\overline{\Delta^0}$
          & $0\pm5.3$%
          & $0.0\pm5.3$%
          & $0.0\pm5.3$%
          &$\overline B{}^0\to \Sigma^{-}\overline{\Delta^-}$ 
          & $0\pm3.1$%
          & $0\pm3.1$%
          & $0\pm3.1$%
           \\            
$B^-\to \Xi^{0}\overline{\Sigma^{*+}}$
          & $0\pm20.4$%
          & $\mp(15.5_{-16.2}^{+24.2})$%
          & $\mp(23.7_{-12.5}^{+27.1})$%
          & $\overline B{}^0\to \Xi^{0}\overline{\Sigma^{*0}}$
          & $0\pm18.5$%
          & $\mp(15.5_{-14.5}^{+22.0})$%
          & $\mp(23.7_{-11.2}^{+24.7})$%
           \\ 
$B^-\to \Xi^{-}\overline{\Sigma^{*0}}$
          & $0\pm5.3$%
          & $0.0\pm5.3$%
          & $0.0\pm5.3$%
          & $\overline B{}^0\to \Xi^{-}\overline{\Sigma^{*-}}$
          & $0\pm3.1$%
          & $0\pm3.1$%
          & $0\pm3.1$%
           \\ 
$B^-\to \Lambda\overline{\Delta^{+}}$
          & $0\pm58.7$%
          & $\pm(78.7^{+20.3}_{-42.9})$%
          & $\pm(80.9^{+17.0}_{-25.1})$%
          & $\overline B{}^0\to \Lambda\overline{\Delta^0}$
          & $0\pm58.7$%
          & $\pm(78.7^{+20.3}_{-42.9})$%
          & $\pm(80.9^{+17.0}_{-25.1})$%
          \\ 
$\overline B{}^0_s\to p\overline{\Delta^+}$
          & $0$%
          & $0$%
          & $0$%
          & $\overline B{}^0_s\to \Sigma^{+}\overline{\Sigma^{*+}}$
          & $0\pm25.7$%
          & $\pm(14.6^{+28.5}_{-19.0})$%
          & $\pm(19.3^{+27.6}_{-15.4})$%
           \\
$\overline B{}^0_s\to n\overline{\Delta^0}$
          & $0$%
          & $0$%
          & $0$%
          & $\overline B{}^0_s\to \Sigma^{0}\overline{\Sigma^{*0}}$
          & $0\pm13.3$%
          & $\pm(7.4^{+15.9}_{-9.7})$%
          & $\pm(10.1^{+16.6}_{-9.3})$%
           \\
$\overline B{}^0_s\to \Xi^{0}\overline{\Xi^{*0}}$ 
          & $0\pm19.7$%
          & $\mp(14.6_{-15.5}^{+23.4})$%
          & $\mp(22.2_{-11.8}^{+26.2})$%
          & $\overline B{}^0_s\to \Sigma^{-}\overline{\Sigma^{*-}}$ 
          & $0\pm3.2$%
          & $0\pm3.2$%
          & $0\pm3.2$%
           \\ 
$\overline B{}^0_s\to \Xi^{-}\overline{\Xi^{*-}}$
          & $0\pm3.2$%
          & $0\pm3.2$%
          & $0\pm3.2$%
          & $\overline B{}^0_s\to \Lambda\overline{\Sigma^{*0}}$
          & $0\pm66.6$%
          & $\pm(78.7^{+21.2}_{-50.1})$%
          & $\pm(80.9^{+18.4}_{-28.1})$%
           \\            
\end{tabular}
}
\end{ruledtabular}
\end{table}

\begin{table}[t!]
\caption{\label{tab: AcpDBDS=0, -1} Same as Table~\ref{tab: AcpBBDS=0, -1}, but for $\overline B_q\to\DB$ modes.
}
\begin{ruledtabular}
\centering
{
\footnotesize
\begin{tabular}{lccclccc}
Mode
          & $\phi=0$
          & $\phi=\pm\pi/4$
          & $\phi=\pm\pi/2$
          & Mode
          & $\phi=0$
          & $\phi=\pm\pi/4$
          & $\phi=\pm\pi/2$
          \\
\hline $B^-\to \Delta^0\overline{p}$ 
          & $0\pm30.3$%
          & $\pm(29.9^{+30.3}_{-24.9})$%
          & $\pm(41.8^{+28.3}_{-19.6})$%
          & $\overline B{}^0_s\to \Delta^+\overline{\Sigma^{+}}$
          & $0\pm 23.9$%
          & $\mp(37.0_{-21.1}^{+23.8})$%
          & $\mp(53.1_{-19.2}^{+23.0})$%
           \\
$B^-\to \Delta^-\overline{n}$
          & $0\pm89.0$%
          & $\pm(0.1^{+88.9}_{-89.0})$%
          & $\pm(0.1^{+88.9}_{-89.0})$%
          & $\overline B{}^0_s\to \Delta^0\overline{\Sigma^{0}}$ 
          & $0\pm37.0$%
          & $\mp(58.1_{-32.8}^{+29.0})$%
          & $\mp(84.0_{-26.0}^{+15.7})$%
           \\ 
$B^-\to \Sigma^{*0}\overline{\Sigma^{+}}$
          & $0\pm30.3$%
          & $\pm(29.9^{+30.3}_{-24.9})$%
          & $\pm(41.8^{+28.3}_{-19.6})$%
          &$\overline B{}^0_s\to \Delta^-\overline{\Sigma^{-}}$ 
          & $0\pm50.2$%
          & $0\pm50.2$%
          & $0\pm50.2$%
           \\            
$B^-\to \Sigma^{*-}\overline{\Sigma^{0}}$
          & $0\pm89.0$%
          & $\pm(0.1^{+88.9}_{-89.0})$%
          & $\pm(0.1^{+88.9}_{-89.0})$%
          & $\overline B{}^0_s\to \Sigma^{*0}\overline{\Xi^{0}}$ 
          & $0\pm14.0$%
          & $\mp(21.7_{-12.2}^{+15.1})$%
          & $\mp(30.9_{-11.4}^{+15.7})$%
           \\ 
$B^-\to \Xi^{*-}\overline{\Xi^{0}}$
          & $0\pm89.0$%
          & $\pm(0.1^{+88.9}_{-89.0})$%
          & $\pm(0.1^{+88.9}_{-89.0})$%
          & $\overline B{}^0_s\to \Sigma^{*-}\overline{\Xi^{-}}$ 
          & $0\pm50.2$%
          & $0\pm50.2$%
          & $0\pm50.2$%
           \\ 
$B^-\to \Sigma^{*-}\overline{\Lambda}$
          & $0\pm89.0$%
          & $\pm(0.1^{+88.9}_{-89.0})$%
          & $\pm(0.1^{+88.9}_{-89.0})$%
          & $\overline B{}^0_s\to \Delta^0\overline{\Lambda}$ 
          & $0\pm2.2$%
          & $\mp(4.2_{-1.9}^{+2.5})$%
          & $\mp(6.0_{-1.8}^{+2.6})$%
           \\ 
$\overline B{}^0\to \Delta^+\overline{p}$
          & $0\pm27.4$%
          & $\mp(36.1_{-23.2}^{+27.8})$%
          & $\mp(51.7_{-20.2}^{+27.0})$%
          & $\overline B{}^0\to \Sigma^{*+}\overline{\Sigma^{+}}$
          & $0$%
          & $0$%
          & $0$%
           \\
$\overline B{}^0\to \Delta^0\overline{n}$
          & $0\pm14.7$%
          & $\mp(21.1_{-12.6}^{+16.2})$%
          & $\mp(30.1_{-11.5}^{+16.9})$%
          & $\overline B{}^0\to \Sigma^{*0}\overline{\Sigma^{0}}$
          & $0\pm30.3$%
          & $\pm(29.9^{+30.3}_{-24.9})$%
          & $\pm(41.8^{+28.2}_{-19.6})$%
           \\
$\overline B{}^0\to \Xi^{*0}\overline{\Xi^{0}}$
          & $0$%
          & $0$%
          & $0$%
          & $\overline B{}^0\to \Sigma^{*-}\overline{\Sigma^{-}}$ 
          & $0\pm48.8$%
          & $0\pm48.8$%
          & $0\pm48.8$%
           \\      
$\overline B{}^0\to \Xi^{*-}\overline{\Xi^{-}}$
          & $0\pm48.8$%
          & $0\pm48.8$%
          & $0\pm48.8$%
          & $\overline B{}^0\to \Sigma^{*0}\overline{\Lambda}$ 
          & $0\pm27.4$%
          & $\mp(36.1_{-23.2}^{+27.8})$%
          & $\mp(51.7_{-20.2}^{+27.0})$%
          \\
\hline $B^-\to \Sigma^{*0}\overline{p}$ 
          & $0\pm23.2$%
          & $\mp(23.7_{-19.9}^{+26.1})$%
          & $\mp(37.8_{-17.2}^{+29.0})$%
          & $\overline B{}^0\to \Sigma^{*+}\overline{p}$ 
          & $0\pm17.1$%
          & $\pm(23.8^{+18.4}_{-12.9})$%
          & $\pm(30.2^{+17.1}_{-11.0})$%
           \\
$B^-\to \Sigma^{*-}\overline{n}$
          & $0\pm5.4$%
          & $0.0\pm5.4$%
          & $0.0\pm5.4$%
          & $\overline B{}^0\to \Sigma^{*0}\overline{n}$
          & $0\pm31.3$%
          & $\pm(43.8^{+27.0}_{-22.7})$%
          & $\pm(51.3^{+22.3}_{-16.9})$%
           \\ 
$B^-\to \Xi^{*0}\overline{\Sigma^{+}}$ 
          & $0\pm23.2$%
          & $\mp(23.7_{-19.9}^{+26.1})$%
          & $\mp(37.8_{-17.3}^{+29.0})$%
          &$\overline B{}^0\to \Xi^{*0}\overline{\Sigma^{0}}$
          & $0\pm20.6$%
          & $\mp(23.7_{-17.5}^{+23.3})$%
          & $\mp(37.8_{-15.2}^{+26.1})$%
           \\            
$B^-\to \Xi^{*-}\overline{\Sigma^{0}}$ 
          & $0\pm5.4$%
          & $0.0\pm5.4$%
          & $0.0\pm5.4$%
          & $\overline B{}^0\to \Xi^{*-}\overline{\Sigma^{-}}$ 
          & $0\pm2.3$%
          & $0\pm2.3$%
          & $0\pm2.3$%
           \\ 
$B^-\to \Omega^-\overline{\Xi^{0}}$ 
          & $0\pm5.4$%
          & $0.0\pm5.4$%
          & $0.0\pm5.4$%
          & $\overline B{}^0\to \Omega^-\overline{\Xi^{-}}$ 
          & $0\pm2.3$%
          & $0\pm2.3$%
          & $0\pm2.3$%
           \\ 
$B^-\to \Xi^{*-}\overline{\Lambda}$ 
          & $0\pm5.4$%
          & $0.0\pm5.4$%
          & $0.0\pm5.4$%
          & $\overline B{}^0\to \Xi^{*0}\overline{\Lambda}$ 
          & $0\pm17.1$%
          & $\pm(23.8^{+18.4}_{-12.9})$%
          & $\pm(30.2^{+17.1}_{-11.0})$%
           \\ 
$\overline B{}^0_s\to \Delta^+\overline{p}$
          & $0$%
          & $0$%
          & $0$%
          & $\overline B{}^0_s\to \Sigma^{*+}\overline{\Sigma^{+}}$ 
          & $0\pm19.7$%
          & $\pm(23.1^{+21.0}_{-14.9})$%
          & $\pm(29.4^{+19.5}_{-12.5})$%
           \\
$\overline B{}^0_s\to \Delta^0\overline{n}$
          & $0$%
          & $0$%
          & $0$%
          & $\overline B{}^0_s\to \Sigma^{*0}\overline{\Sigma^{0}}$
          & $0\pm10.1$%
          & $\pm(11.8^{+11.9}_{-7.8})$%
          & $\pm(15.8^{+12.2}_{-7.4})$%
           \\
$\overline B{}^0_s\to \Xi^{*0}\overline{\Xi^{0}}$ 
          & $0\pm33.5$%
          & $\pm(42.7^{+29.1}_{-24.5})$%
          & $\pm(50.1^{+24.1}_{-18.2})$%
          & $\overline B{}^0_s\to \Sigma^{*-}\overline{\Sigma^{-}}$
          & $0\pm2.3$%
          & $0\pm2.3$%
          & $0\pm2.3$%
           \\ 
$\overline B{}^0_s\to \Xi^{*-}\overline{\Xi^{-}}$ 
          & $0\pm2.3$%
          & $0\pm2.3$%
          & $0\pm2.3$%
          & $\overline B{}^0_s\to \Sigma^{*0}\overline{\Lambda}$
          & $0\pm54.4$
          & $\pm(78.7^{+20.0}_{-36.1})$%
          & $\pm(80.9^{+16.6}_{-20.9})$%
           \\                           
\end{tabular}
}
\end{ruledtabular}
\end{table}

\begin{table}[t!]
\caption{\label{tab: AcpDDDS=0, -1} Same as Table~\ref{tab: AcpBBDS=0, -1}, but for
$\overline B_q\to\DD$ modes. 
}
\begin{ruledtabular}
\centering
{
\footnotesize
\begin{tabular}{lccclccc}
Mode
          & $\phi=0$
          & $\phi=\pm\pi/4$
          & $\phi=\pm\pi/2$
          & Mode
          & $\phi=0$
          & $\phi=\pm\pi/4$
          & $\phi=\pm\pi/2$
          \\
\hline $B^-\to \Delta^+ \overline{\Delta^{++}}$ 
          & $0\pm27.4$%
          & $\mp(36.1_{-23.2}^{+27.9})$%
          & $\mp(51.7_{-20.2}^{+27.0})$%
          & $\overline B{}^0_s\to \Delta^{+} \overline{\Sigma^{*+}}$ 
          & $0\pm23.9$%
          & $\mp(37.0_{-21.1}^{+23.8})$%
          & $\mp(53.1_{-19.2}^{+23.0})$%
           \\
$B^-\to \Delta^0 \overline{\Delta^{+}}$ 
          & $0\pm48.3$%
          & $\mp(57.1_{-41.8}^{+36.3})$%
          & $\mp(82.5_{-30.3}^{+17.5})$%
          & $\overline B{}^0_s\to \Delta^{0} \overline{\Sigma^{*0}}$ 
          & $0\pm37.0$%
          & $\mp(58.1_{-32.8}^{+29.0})$%
          & $\mp(84.0_{-26.0}^{+15.7})$%
           \\ 
$B^-\to \Delta^- \overline{\Delta^{0}}$
          & $0\pm89.0$%
          & $\pm(0.1^{+88.9}_{-89.0})$%
          & $\pm(0.1^{+88.9}_{-89.0})$%
          & $\overline B{}^0_s\to \Delta^{-} \overline{\Sigma^{*-}}$ 
          & $0\pm50.2$%
          & $0\pm50.2$%
          & $0\pm50.2$%
           \\            
$B^-\to \Sigma^{*0} \overline{\Sigma^{*+}}$ 
          & $0\pm48.3$%
          & $\mp(57.1_{-41.8}^{+36.3})$%
          & $\mp(82.5_{-30.3}^{+17.5})$%
          & $\overline B{}^0_s\to \Sigma^{*0} \overline{\Xi^{*0}}$ 
          & $0\pm37.0$%
          & $\mp(58.1_{-32.8}^{+29.0})$%
          & $\mp(84.0_{-26.0}^{+15.7})$%
           \\ 
$B^-\to \Sigma^{*-} \overline{\Sigma^{*0}}$
          & $0\pm89.0$%
          & $\pm(0.1^{+88.9}_{-89.0})$%
          & $\pm(0.1^{+88.9}_{-89.0})$%
          & $\overline B{}^0_s\to \Sigma^{*-} \overline{\Xi^{*-}}$
          & $0\pm50.2$%
          & $0\pm50.2$%
          & $0\pm50.2$%
           \\ 
$B^-\to \Xi^{*-} \overline{\Xi^{*0}}$
          & $0\pm89.0$%
          & $\pm(0.1^{+88.9}_{-89.0})$%
          & $\pm(0.1^{+88.9}_{-89.0})$%
          & $\overline B{}^0_s\to \Xi^{*-} \overline{\Omega^{-}}$
          & $0\pm50.2$%
          & $0\pm50.2$%
          & $0\pm50.2$%
           \\ 
$\overline B{}^0\to \Delta^{++} \overline{\Delta^{++}}$
          & $0\pm 39.6$%
          & $0\pm 39.6$%
          & $0\pm 39.6$%
          & $\overline B{}^0\to \Sigma^{*+} \overline{\Sigma^{*+}}$
          & $0\pm 59.3$%
          & $0\pm 59.3$%
          & $0\pm 59.3$%
           \\
$\overline B{}^0\to \Delta^{+} \overline{\Delta^{+}}$
          & $0\pm40.4$%
          & $\mp(36.1_{-33.6}^{+40.3})$%
          & $\mp(51.8_{-27.6}^{+37.8})$%
          & $\overline B{}^0\to \Sigma^{*0} \overline{\Sigma^{*0}}$ 
          & $0\pm60.5$%
          & $\mp(57.1_{-54.2}^{+41.3})$%
          & $\mp(82.5_{-41.0}^{+17.5})$%
           \\
$\overline B{}^0\to \Delta^{0} \overline{\Delta^{0}}$ 
          & $0\pm48.0$%
          & $\mp(57.1_{-42.6}^{+36.1})$%
          & $\mp(82.5_{-33.3}^{+17.5})$%
          & $\overline B{}^0\to \Sigma^{*-} \overline{\Sigma^{*-}}$
          & $0\pm63.8$%
          & $0\pm63.8$%
          & $0\pm63.8$%
           \\ 
$\overline B{}^0\to \Delta^{-} \overline{\Delta^{-}}$
          & $0\pm58.8$%
          & $0\pm58.8$%
          & $0\pm58.8$%
          & $\overline B{}^0\to \Xi^{*0} \overline{\Xi^{*0}}$
          & $0\pm 97.7$%
          & $0\pm 97.7$%
          & $0\pm 97.7$%
           \\ 
$\overline B{}^0\to \Omega^{-} \overline{\Omega^{-}}$
          & $0\pm 63.8$%
          & $0\pm 63.8$%
          & $0\pm 63.8$%
          & $\overline B{}^0\to \Xi^{*-} \overline{\Xi^{*-}}$ 
          & $0\pm 78.1$%
          & $0\pm 78.1$%
          & $0\pm 78.2$%
           \\
\hline $B^-\to \Sigma^{*+} \overline{\Delta^{++}}$ 
          & $0\pm20.0$%
          & $\pm(23.8^{+21.2}_{-15.2})$%
          & $\pm(30.2^{+19.6}_{-12.8})$%
          & $\overline B{}^0\to \Sigma^{*+} \overline{\Delta^{+}}$ 
          & $0\pm17.1$%
          & $\pm(23.8^{+18.4}_{-12.9})$%
          & $\pm(30.2^{+17.1}_{-11.0})$%
           \\          
$B^-\to \Sigma^{*0} \overline{\Delta^{+}}$
          & $0\pm11.8$%
          & $\pm(12.2^{+13.6}_{-9.1})$%
          & $\pm(16.3^{+13.7}_{-8.5})$%
          & $\overline B{}^0\to \Sigma^{*0} \overline{\Delta^{0}}$ 
          & $0\pm8.7$%
          & $\pm(12.2^{+10.4}_{-6.8})$%
          & $\pm(16.3^{+10.8}_{-6.6})$%
           \\ 
$B^-\to \Sigma^{*-} \overline{\Delta^{0}}$ 
          & $0\pm5.4$%
          & $0.0\pm5.4$%
          & $0.0\pm5.4$%
          &$\overline B{}^0\to \Sigma^{*-} \overline{\Delta^{-}}$
          & $0\pm2.3$%
          & $0\pm2.3$%
          & $0\pm2.3$%
           \\            
$B^-\to \Xi^{*0} \overline{\Sigma^{*+}}$
          & $0\pm11.8$%
          & $\pm(12.2^{+13.6}_{-9.1})$%
          & $\pm(16.3^{+13.7}_{-8.5})$%
          & $\overline B{}^0\to \Xi^{*0} \overline{\Sigma^{*0}}$ 
          & $0\pm8.7$%
          & $\pm(12.2^{+10.4}_{-6.8})$%
          & $\pm(16.3^{+10.8}_{-6.6})$%
           \\ 
$B^-\to \Xi^{*-} \overline{\Sigma^{*0}}$
          & $0\pm5.4$%
          & $0.0\pm5.4$%
          & $0.0\pm5.4$%
          & $\overline B{}^0\to \Xi^{*-} \overline{\Sigma^{*-}}$
          & $0\pm2.3$%
          & $0\pm2.3$%
          & $0\pm2.3$%
           \\ 
$B^-\to \Omega^{-} \overline{\Xi^{*0}}$ 
          & $0\pm5.4$%
          & $0.0\pm5.4$%
          & $0.0\pm5.4$%
          & $\overline B{}^0\to \Omega^{-} \overline{\Xi^{*-}}$
          & $0\pm2.3$%
          & $0\pm2.3$%
          & $0\pm2.3$%
           \\ 
$\overline B{}^0_s\to \Delta^{++} \overline{\Delta^{++}}$
          & $0\pm 48.0$%
          & $0\pm 48.0$%
          & $0\pm 48.0$%
          & $\overline B{}^0_s\to \Sigma^{*+} \overline{\Sigma^{*+}}$ 
          & $0\pm31.6$%
          & $\pm(23.1^{+34.5}_{-21.4})$%
          & $\pm(29.4^{+31.9}_{-16.3})$%
           \\
$\overline B{}^0_s\to \Delta^{+} \overline{\Delta^{+}}$
          & $0\pm 34.1$%
          & $\pm(0^{+34.0}_{-34.1})$%
          & $0\pm 34.1$%
          & $\overline B{}^0_s\to \Sigma^{*0} \overline{\Sigma^{*0}}$ 
          & $0\pm16.7$%
          & $\pm(11.8^{+20.6}_{-11.2})$%
          & $\pm(15.8^{+21.2}_{-9.8})$%
           \\
$\overline B{}^0_s\to \Delta^{0} \overline{\Delta^{0}}$
          & $0\pm 19.2$%
          & $\pm(0^{+19.1}_{-19.2})$%
          & $0\pm 19.2$%
          & $\overline B{}^0_s\to \Sigma^{*-} \overline{\Sigma^{*-}}$
          & $0\pm 4.2$%
          & $0\pm 4.2$%
          & $0\pm 4.2$%
           \\ 
$\overline B{}^0_s\to \Delta^{-} \overline{\Delta^{-}}$
          & $0\pm3.7$%
          & $0\pm3.7$%
          & $0\pm3.7$%
          & $\overline B{}^0_s\to \Xi^{*0} \overline{\Xi^{*0}}$ 
          & $0\pm12.3$%
          & $\pm(11.8^{+15.0}_{-8.8})$%
          & $\pm(15.8^{+15.5}_{-8.2})$%
           \\ 
$\overline B{}^0_s\to \Omega^{-} \overline{\Omega^{-}}$
          & $0\pm2.9$%
          & $0\pm2.9$%
          & $0\pm2.9$%
          & $\overline B{}^0_s\to \Xi^{*-} \overline{\Xi^{*-}}$ 
          & $0\pm 3.2$%
          & $0\pm 3.2$%
          & $0\pm 3.2$%
           \\            
\end{tabular}
}
\end{ruledtabular}
\end{table}

We show the predictions on direct $CP$ violations in all $\overline B_q\to \bfBB'$ modes in Tables~\ref{tab: AcpBBDS=0, -1}, \ref{tab: AcpBDDS=0, -1}, \ref{tab: AcpDBDS=0, -1} and \ref{tab: AcpDDDS=0, -1}.
Results are given with $\phi=0$, $\pm\pi/4$ and $\pm\pi/2$,
where $\phi$ is the penguin strong phase and we use a common strong phase for $P^{(\prime)}$, $P^{(\prime)}_{PE}$ and $P^{(\prime)}_{EW}$ for simplicity.
Uncertainties are obtained by varying all other strong phases in Eqs. (\ref{eq:correction0}),  (\ref{eq:correction2}) and (\ref{eq:correction3}).

\begin{table}[t!]
\caption{\label{tab: Acp pure penguin} 
Direct $CP$ asymmetries ($\A$ in $\%$) of $\Delta S=-1$ pure penguin modes.
These are robust predictions of the SM.}
\begin{ruledtabular}
\centering
{
\footnotesize
\begin{tabular}{lccclccc}
Mode
          & $\A(\%)$
          & Mode
          & $\A(\%)$
          \\
\hline 
$\overline B{}^0\to  \Xi^{-}\overline{\Sigma^{-}}$ 
          & $0\pm2.3$%
          & $\overline B{}^0_s\to \Sigma^{-}\overline{\Sigma^{-}}$
          & $0\pm4.2$%
           \\ 
$\overline B{}^0_s\to \Xi^{-}\overline{\Xi^{-}}$ 
          & $0\pm4.7$%
           \\                                                                                                                                                      
\hline 
          $\overline B{}^0\to \Sigma^{-}\overline{\Delta^-}$ 
          & $0\pm3.1$%
          & $\overline B{}^0\to \Xi^{-}\overline{\Sigma^{*-}}$ 
          & $0\pm3.1$%
           \\ 
$\overline B{}^0_s\to \Sigma^{-}\overline{\Sigma^{*-}}$ 
          & $0\pm3.2$%
          & $\overline B{}^0_s\to \Xi^{-}\overline{\Xi^{*-}}$ 
          & $0\pm3.2$%
           \\            
\hline 
$\overline B{}^0\to \Xi^{*-}\overline{\Sigma^{-}}$ 
          & $0\pm2.3$%
          & $\overline B{}^0\to \Omega^-\overline{\Xi^{-}}$ 
          & $0\pm2.3$%
           \\ 
$\overline B{}^0_s\to \Sigma^{*-}\overline{\Sigma^{-}}$ 
          & $0\pm2.3$%
          & $\overline B{}^0_s\to \Xi^{*-}\overline{\Xi^{-}}$ 
          & $0\pm2.3$%
           \\                           
\hline 
$\overline B{}^0\to \Sigma^{*-} \overline{\Delta^{-}}$ 
          & $0\pm2.3$%
          & $\overline B{}^0_s\to \Sigma^{*-} \overline{\Sigma^{*-}}$ 
          & $0\pm 4.2$%
           \\            
$\overline B{}^0\to \Xi^{*-} \overline{\Sigma^{*-}}$ 
          & $0\pm2.3$%
          & $\overline B{}^0_s\to \Delta^{-} \overline{\Delta^{-}}$
          & $0\pm3.7$%
           \\ 
$\overline B{}^0\to \Omega^{-} \overline{\Xi^{*-}}$ 
          & $0\pm2.3$%
          & $\overline B{}^0_s\to \Omega^{-} \overline{\Omega^{-}}$ 
          & $0\pm2.9$%
           \\            
\end{tabular}
}
\end{ruledtabular}
\end{table}

\begin{table}[t!]
\caption{\label{tab: Acp null test} 
Vanishing direct CP violations of pure exchange modes are null tests of the SM.
}
\begin{ruledtabular}
\centering
{
\footnotesize
\begin{tabular}{lclc}
Mode
          & $\A$
          & Mode
          & $\A$
          \\
\hline 
$\overline B{}^0\to \Sigma^{+}\overline{\Sigma^{*+}}$
          & $0$
          & $\overline B{}^0\to \Xi^{0}\overline{\Xi^{*0}}$
          & $0$
           \\      
$\overline B{}^0_s\to p\overline{\Delta^+}$
          & $0$
          & $\overline B{}^0_s\to n\overline{\Delta^0}$
          & $0$
           \\
 \hline          
$\overline B{}^0\to \Sigma^{*+}\overline{\Sigma^{+}}$
          & $0$
          & $\overline B{}^0\to \Xi^{*0}\overline{\Xi^{0}}$
          & $0$
           \\      
$\overline B{}^0_s\to \Delta^+\overline{p}$
          & $0$
          & $\overline B{}^0_s\to \Delta^0\overline{n}$
          & $0$
           \\
\end{tabular}
}
\end{ruledtabular}
\end{table}

\begin{table}[t!]
\caption{\label{tab: AcpBcDS=0, -1} 
$B^-_c\to\bfBB'$ direct CP violations. 
These vanishing $\A(B^-_c\to\bfBB')$ are null tests of the standard model.
}
\begin{ruledtabular}
\centering
{\begin{tabular}{lclc}
Mode ($\Delta S=0$)
          & ${\mathcal A}$
          & Mode ($\Delta S=-1$)
          & ${\mathcal A}$
          \\
\hline $B^-_c\to n\overline{p}$
          & 0
          & $B^-_c\to \Sigma^{0}\overline{p}$
          & 0
           \\
$B^-_c\to \Sigma^{0}\overline{\Sigma^{+}}$
          & 0
          & $B^-_c\to \Sigma^{-}\overline{n}$
          & 0
           \\ 
$B^-_c\to \Sigma^{-}\overline{\Sigma^{0}}$
          & 0
          & $B^-_c\to \Xi^{0}\overline{\Sigma^{+}}$
          & 0
           \\            
$B^-_c\to \Sigma^{-}\overline{\Lambda}$
          & 0
          & $B^-_c\to\Xi^{-}\overline{\Sigma^{0}}$
          & 0
           \\ 
$B^-_c\to \Xi^{-}\overline{\Xi^{0}}$
          & 0
          & $B^-_c\to \Xi^{-}\overline{\Lambda}$
          & 0
           \\ 
$B^-_c\to \Lambda\overline{\Sigma^+}$
          & 0
          & $B^-_c\to \Lambda\overline{p}$
          & 0
           \\  
\hline $B^-_c\to p\overline{\Delta^{++}}$
          & 0
          & $B^-_c\to \Sigma^+\overline{\Delta^{++}}$
          & 0
           \\
$B^-_c\to n\overline{\Delta^+}$
          & 0
          & $B^-_c\to \Sigma^0\overline{\Delta^+}$
          & 0
           \\ 
$B^-_c\to\Sigma^0\overline{\Sigma^{*+}}$
          & 0
          & $B^-_c\to \Sigma^-\overline{\Delta^0}$
          & 0
           \\            
$B^-_c\to\Sigma^-\overline{\Sigma^{*0}}$
          & 0
          & $B^-_c\to \Xi^{0}\overline{\Sigma^{*+}}$
          & 0
           \\ 
$B^-_c\to\Xi^{-}\overline{\Xi^{*0}}$
          & 0
          & $B^-_c\to \Xi^{-}\overline{\Sigma^{*0}}$
          & 0
           \\ 
$B^-_c\to\Lambda\overline{\Sigma^{*+}}$
          & 0
           \\ 
\hline $B^-_c\to \Delta^0\overline{p}$
          & 0
          & $B^-_c\to \Sigma^{*0}\overline{p}$
          & 0
           \\
$B^-_c\to \Delta^-\overline{n}$
          & 0
          & $B^-_c\to \Sigma^{*-}\overline{n}$
          & 0
           \\ 
$B^-_c\to \Sigma^{*0}\overline{\Sigma^{+}}$
          & 0
          & $B^-_c\to \Xi^{*0}\overline{\Sigma^{+}}$
          & 0
           \\            
$B^-_c\to \Sigma^{*-}\overline{\Sigma^{0}}$
          & 0
          & $B^-_c\to \Xi^{*-}\overline{\Sigma^{0}}$
          & 0
           \\ 
$B^-_c\to \Xi^{*-}\overline{\Xi^{0}}$
          & 0
          & $B^-_c\to \Omega^-\overline{\Xi^{0}}$ 
          & 0
           \\ 
$B^-_c\to \Sigma^{*-}\overline{\Lambda}$
          & 0
          & $B^-_c\to \Xi^{*-}\overline{\Lambda}$
          & 0
           \\ 
\hline $B^-_c\to \Delta^+ \overline{\Delta^{++}}$
          & 0
          & $B^-_c\to \Sigma^{*+} \overline{\Delta^{++}}$
          & 0
           \\
$B^-_c\to \Delta^0 \overline{\Delta^{+}}$
          & 0
          & $B^-_c\to \Sigma^{*0} \overline{\Delta^{+}}$
          & 0
           \\ 
$B^-_c\to \Delta^- \overline{\Delta^{0}}$
          & 0
          & $B^-_c\to \Sigma^{*-} \overline{\Delta^{0}}$
          & 0
           \\            
$B^-_c\to \Sigma^{*0} \overline{\Sigma^{*+}}$
          & 0
          & $B^-_c\to \Xi^{*0} \overline{\Sigma^{*+}}$
          & 0
           \\ 
$B^-_c\to \Sigma^{*-} \overline{\Sigma^{*0}}$
          & 0
          & $B^-_c\to \Xi^{*-} \overline{\Sigma^{*0}}$
          & 0
           \\ 
$B^-_c\to \Xi^{*-} \overline{\Xi^{*0}}$
          & 0
          & $B^-_c\to \Omega^{-} \overline{\Xi^{*0}}$
          & 0
           \\ 
\end{tabular}
}
\\
\end{ruledtabular}
\end{table}

Note that for $\Delta S=-1$ transition, the amplitudes of $\bar B_q\to\bfBB'$ and its conjugated modes are given by
\be
A&=&V_{ub} V^*_{us} |A_u| e^{i\delta_u}+V_{cb} V^*_{cs} |A_c| e^{i\delta_c},
\non\\
\bar A&=&V^*_{ub} V_{us} |A_u| e^{i\delta_u}+V^*_{cb} V_{cs} |A_c| e^{i\delta_c}.
\en
Since we have $|V_{ub} V^*_{us}|/|V_{cb} V^*_{cs}|\simeq 0.02$, for 
\be
\frac{|A_u|}{|A_c|}\lesssim {\cal O}(1),
\label{eq: Au Ac}
\en
we should have the following estimation on the direct CP violation $\A$,
\be
|\A|\simeq 
2\left|\frac{V_{ub}V^*_{us}}{V_{cb}V^*_{cs}}\right|\frac{|A_u|}{|A_c|} |\sin(\delta_u-\delta_c)|
\sin\gamma
\lesssim \frac{|A_u|}{|A_c|}\times 3.7\%.
\label{eq: A DeltaS penguin}
\en
Indeed, Eq. (\ref{eq: Au Ac}) can be satisfied in the case of pure penguin modes, where we expect
\be
\frac{|A_u|}{|A_c|}={\cal O}(1),
\en
and, consequently, from Eq. (\ref{eq: A DeltaS penguin}), we should have,
\be
|\A|\lesssim \frac{|A_u|}{|A_c|}\times 3.7\%\simeq  {\cal O}(1)\times 3.7\%,
\label{eq: A DeltaS penguin1}
\en
for direct CP violations of $\Delta S=-1$ pure penguin modes.
In Table \ref{tab: Acp pure penguin} we collect the predictions of direct CP violation of these modes. 
We see that the sizes of the predicted direct CP violations agree with the above expectation.

In Table~\ref{tab: Acp null test}, we collect results of vanishing direct CP violations from pure exchange modes. Since there is no any penguin contribution, the direct CP violations of these modes are predicted to be vanishing. These are null tests of the SM.

In Table~\ref{tab: AcpBcDS=0, -1}, we give the predictions of direct CP violation of $B^-_c\to\bfBB'$ decays. As the decays are from annihilation diagrams, there is no any penguin contribution, and the direct CP violations are all predicted to be vanishing. These are also null tests of the SM.

\section{Discussions and Conclusion}

\begin{table}[t!]
\caption{\label{tab: comparison} 
Comparisons of data and theoretical results on the branching ratios (in the unit of $10^{-8}$) of some 
$\bar B_{u,d,s}\to\bfBB'$ decays. 
}
\begin{ruledtabular}
\begin{tabular}{lcccc}
Mode
          & Expt
          & This work 
          & Ref. \cite{Jin:2021onb}
          & Ref. \cite{Cheng:2001tr} 
          \\
\hline $B^-\to\Lambda\bar p$
          & $24^{+10}_{-9}$ \cite{LHCb:2016nbc, PDG} 
          & $24.00^{+10.00}_{-9.00}{}^{+2.69}_{-0}{}^{+19.37}_{-13.65}{}^{+0.31}_{-0.30}$%
          & $24\pm 9$
          & 22
          \\
$B^-\to\Sigma^{*0}\bar p$
          & $<47$ \cite{Belle:2007lbz}
          & $1.01^{+0.44}_{-0.39}{}^{+0}_{-0.31}{}^{+0.76}_{-0.55}\pm0.02$%
           \\
$B^-\to\Lambda\overline{\Delta^+}$
          & $<82$ \cite{Belle:2007lbz}
          & $0.10^{+0.01}_{-0.01}{}^{+0.13}_{-0}{}^{+0.04}_{-0.03}\pm0$%
          \\
$B^-\to\Delta^0\bar p$
          & $<138$ \cite{Belle:2007oni}
          & $1.68^{+0.19}_{-0.19}{}^{+0.05}_{-0}{}^{+0.82}_{-0.62}{}^{+0.23}_{-0.22}$%
          \\
$B^-\to p \overline{\Delta^{++}}$
          & $<14$ \cite{Belle:2007oni}
          &  $5.88^{+0.61}_{-0.61}{}^{+0}_{-0.30}{}^{+9.26}_{-4.88}{}^{+0.71}_{-0.67}$%
          &
          & $140$
          \\   
$B^-\to \Sigma^+ \overline{\Delta^{++}}$
          & 
          &  $15.68^{+3.92}_{-3.81}{}^{+3.39}_{-0}{}^{+14.09}_{-9.39}\pm0.09$%
          &
          & $20$
          \\   
$B^-\to \Sigma^- \overline{\Delta^{0}}$
          & 
          &  $4.91^{+1.29}_{-1.25}{}^{+0}_{-0.00}{}^{+4.18}_{-2.90}\pm0.04$%
          &
          & $8.7$
          \\                              
\hline $\bar B^0\to p\bar p$
          &  $1.27\pm 0.13\pm 0.05\times 0.03$ \cite{LHCb:2022lff} 
          &  $1.27^{+0.14}_{-0.14}{}^{+0}_{-0.05}{}^{+1.85}_{-1.02}{}^{+1.32}_{-0.84}$%
          &  $1.2\pm 0.3$
          &  $11$
          \\
$\bar B^0\to\Sigma^{*+}\bar p$
          & $<26$ \cite{Belle:2007lbz}
         & $2.15^{+0.84}_{-0.76}{}^{+0.73}_{-0}{}^{+1.39}_{-1.03}\pm0$%
          \\
$\bar B^0\to p\overline{\Delta^+}, \Delta^+\bar p$
          & $<160$ \cite{Belle:2019abe}
          & $3.48^{+0.37}_{-0.37}{}^{+0}_{-0.15}{}^{+3.74}_{-2.20}{}^{+0.44}_{-0.41}$%
          & 
          \\ 
$\bar B^0\to p\overline{\Delta^+}$
          & 
          & $1.82^{+0.19}_{-0.19}{}^{+0}_{-0.09}{}^{+2.86}_{-1.51}{}^{+0.22}_{-0.21}$%
          &
          & $43$
          \\                         
$\bar B^0\to \Lambda\overline{\Delta^0}$
           & $<93$ \cite{Belle:2007lbz}
          & $0.09^{+0.01}_{-0.01}{}^{+0.12}_{-0}{}^{+0.04}_{-0.03}\pm0$%
          \\
$\bar B^0\to \Lambda\overline{\Lambda}$
          & $<32$ \cite{Belle:2007gob}
          & $0.00\pm0\pm0{}^{+0.24}_{-0}{}^{+0.03}_{-0.00}$%
          & $0.4$
          & 0
          \\
$\bar B^0\to \Delta^0\overline{\Delta^0}$
          & $<1.5\times 10^{5}$ \cite{CLEO:1989xsn}
          & $5.48^{+0.76}_{-0.72}{}^{+0}_{-0.32}{}^{+3.59}_{-2.69}{}^{+1.11}_{-1.00}$%
          \\ 
$\bar B^0\to \Delta^{++}\overline{\Delta^{++}}$
          & $<1.1\times 10^{4}$ \cite{CLEO:1989xsn}
          & $0.00\pm0\pm0\pm0{}^{+0.25}_{-0.00}$%
          \\ 
$\bar B^0\to \Sigma^+\overline{\Delta^+}$
          & 
          & $4.85^{+1.21}_{-1.18}{}^{+1.05}_{-0}{}^{+4.35}_{-2.90}\pm0$%
          &
          & $6.3$
          \\                                          
\hline $\bar B^0_s\to p\bar p$
          & $<0.44$  \cite{LHCb:2022lff}
          & $0.00\pm0\pm0{}^{+0.07}_{-0.00}$%
          & $<0.01$
          \\ 
$\bar B^0_s\to\Xi^0\overline{\Xi^0}$
          &
          & $26.38^{+10.47}_{-9.50}{}^{+4.27}_{-0}{}^{+28.15}_{-17.95}{}^{+2.08}_{-2.00}$%
          & $193\pm 27$
          \\    
$\bar B^0_s\to\Xi^-\overline{\Xi^-}$
          &
          & $25.23^{+10.27}_{-9.35}{}^{+0}_{-0.04}{}^{+26.94}_{-17.26}\pm 0.00$
          & $194\pm 27$
          \\                                                                       
\end{tabular}
\end{ruledtabular}
\end{table}

In Table~\ref{tab: comparison} we compare our results on some of the $\bar B_{u,d,s}\to\bfBB'$ decay rates to data and other theoretical predictions.
It is encouraging that using $\bar B^0\to p\bar p$ and $B^-\to\Lambda \bar p$ rates as inputs, our results satisfy all existing experimental bounds.
This by itself is a non-trivial test.  
From the table we see that in the $B^-$ decay modes, our results agree with those in refs. \cite{Jin:2021onb, Cheng:2001tr}, except the $B^-\to p\overline{\Delta^{++}}$ rate, where the prediction of ref. \cite{Cheng:2001tr} is much larger than ours and exceeds the experimental bound \cite{Belle:2007oni} by one order of magnitude, while the agreement with ref. \cite{Jin:2021onb} on $\bar B^0\to p\bar p$ rate rests on the fact that we all use the measured rate as an input.
For the $\bar B^0$ decays, again the agreement with ref. \cite{Jin:2021onb} on $B^-\to \Lambda \bar p$ rate is simply reflecting that we are using the same data as input. The predictions on $B^-\to \Lambda\bar \Lambda$ rate from ref. \cite{Jin:2021onb} and ours are of the same order.
On the other hand our results on the $\bar B^0$ decays differ from those in ref.~\cite{Cheng:2001tr}, except for the $\bar B^0\to \Sigma^+\overline{\Delta^+}$ decay.
For the $\bar B^0_s$ decays, our prediction on $\bar B^0_s\to p\bar p$ rate is below the present experimental bound \cite{LHCb:2022lff}. 
Our predictions on $\bar B^0_s\to \Xi^0\overline{\Xi^0}$ and $\bar B^0_s\to\Xi^-\overline{\Xi^-}$ rates are below those from ref. \cite{Jin:2021onb} by roughly one order of magnitude.

Note that the experimental limits on $B^-\to\Sigma^{*0}\bar p$, $\Lambda\overline{\Delta^+}$, $\Delta^0\bar p$, $p \overline{\Delta^{++}}$, $\bar B^0\to\Sigma^{*+}\bar p$
$\bar B^0\to \Lambda\overline{\Delta^0}$
were reported in 2007 \cite{Belle:2007lbz, Belle:2007oni}, 
while those on $\bar B^0\to \Delta^0\overline{\Delta^0}$ and $\Delta^{++}\overline{\Delta^{++}}$
were reported in 1989 \cite{CLEO:1989xsn}.
It will be interesting to see the updated results on these modes.
In particular, as one can see from Table~\ref{tab: comparison}, the experimental upper limit on $B^-\to p \overline{\Delta^{++}}$ rate reported in ref. \cite{Belle:2007oni}
is only a factor of two larger than our predicted rate.
It will be interesting to see the updated search on this mode.

\begin{table}[t!]
\caption{\label{tab: BBEEPA} Branching ratios (in unit of $10^{-8}$) of $\overline B_q\to\BB$ decays with $E^{(\prime)}_{1\BB}$, $E^{(\prime)}_{2\BB}$ or $PA^{(\prime)}_\BB$ enlarged. For illustration we assume $\B(\bar B^0_s\to p\bar p)=0.44\times 10^{-8}$ saturating the present experimental bound \cite{LHCb:2022lff}. 
Note that values in parentheses are rates that are unaffected by the enlargements and we only show the central values of these modes, 
see Table~\ref{tab: BBDS=0, -1} for detail numbers. 
}
\begin{ruledtabular}
\centering
{\begin{tabular}{lccclccc}
Mode
          & 
          & ${\mathcal B}(10^{-8})$
          &
          & Mode
          &
          & ${\mathcal B}(10^{-8})$
          &
          \\
          & $E^{(\prime)}_{1\BB}$
          & $E^{(\prime)}_{2\BB}$
          & $PA^{(\prime)}_{\BB}$
          &
          & $E^{(\prime)}_{1\BB}$
          & $E^{(\prime)}_{2\BB}$
          & $PA^{(\prime)}_{\BB}$
          \\          
\hline $B^-\to n\overline{p}$
          & $(3.4)$%
          & $(3.4)$%
          & $(3.4)$%
          & $\overline B{}^0_s\to p\overline{\Sigma^{+}}$
          & $(1.3)$%
          & $(1.3)$%
          & $(1.3)$%
           \\
$B^-\to \Sigma^{0}\overline{\Sigma^{+}}$
          & $(3.0)$ %
          & $(3.0)$%
          & $(3.0)$%
          & $\overline B{}^0_s\to n\overline{\Sigma^{0}}$
          & $(0.6)$ %
          & $(0.6)$%
          & $(0.6)$%
           \\ 
$B^-\to \Sigma^{-}\overline{\Sigma^{0}}$
          & $(0.6)$%
          & $(0.6)$%
          & $(0.6)$%
          & $\overline B{}^0_s\to n\overline{\Lambda}$
          & $(2.9)$ %
          & $(2.9)$%
          & $(2.9)$%
           \\            
$B^-\to \Sigma^{-}\overline{\Lambda}$
          & $(0.4)$%
          & $(0.4)$%
          & $(0.4)$%
          & $\overline B{}^0_s\to \Sigma^{0}\overline{\Xi^{0}}$
          & $(9.8)$%
          & $(9.8)$%
          & $(9.8)$%
           \\ 
$B^-\to \Xi^{-}\overline{\Xi^{0}}$
          & $(0.1)$%
          & $(0.1)$%
          & $(0.1)$%
          & $\overline B{}^0_s\to \Sigma^{-}\overline{\Xi^{-}}$ 
          & $(1.8)$ %
          & $(1.8)$%
          & $(1.8)$%
           \\ 
$B^-\to \Lambda\overline{\Sigma^+}$
          & $(0.4)$%
          & $(0.4)$%
          & $(0.4)$%
          & $\overline B{}^0_s\to \Lambda\overline{\Xi^0}$
          & $(0.1)$%
          & $(0.1)$%
          & $(0.1)$%
           \\ 
$\overline B{}^0\to p\overline{p}$
          & $3.2^{+12.3}_{-0.0}$%
          & $15.5^{+0.0}_{-12.3}$%
          & $1.2^{+0.4}_{-0.2}$ %
          & $\overline B{}^0\to \Sigma^{+}\overline{\Sigma^{+}}$
          & $7.4\pm0.0$%
          & $7.4\pm0.0$%
          & $0.02^{+0.01}_{-0.00}$ %
           \\
$\overline B{}^0\to n\overline{n}$
          & $(6.1)$%
          & $1.7^{+24.9}_{-0.0}$%
          & $6.4^{+0.5}_{-1.1}$ %
          & $\overline B{}^0\to \Sigma^{0}\overline{\Sigma^{0}}$
          & $6.1^{+0.0}_{-5.7}$%
          & $0.4^{+5.7}_{-0.0}$%
          & $1.6^{+0.2}_{-0.5}$ %
           \\
$\overline B{}^0\to \Xi^{0}\overline{\Xi^{0}}$
          & $(0.0)$ %
          & $7.0\pm0.0$ %
          & $0.02^{+0.01}_{-0.00}$ 
          & $\overline B{}^0\to \Sigma^{-}\overline{\Sigma^{-}}$
          & $(1.0)$ %
          & $(1.0)$ %
          & $1.3^{+0.0}_{-0.5}$ %
           \\      
$\overline B{}^0\to \Xi^{-}\overline{\Xi^{-}}$
          & $(0.06)$ %
          & $(0.06)$ %
          & $0.01^{+0.14}_{-0.00}$%
          & $\overline B{}^0\to \Sigma^{0}\overline{\Lambda}$
          & $1.3^{+5.7}_{-0.0}$ %
          & $1.3^{+5.7}_{-0.0}$%
          & $(3.5)$ %
           \\
$\overline B{}^0\to \Lambda\overline{\Lambda}$
          & $0.2\pm0.0$%
          & $5.3\pm 0.0$%
          & $0.02^{+0.01}_{-0.00}$%
          & $\overline B{}^0\to \Lambda\overline{\Sigma^{0}}$
          & $0.8^{+0.1}_{-0.0}$ %
          & $0.8^{+0.1}_{-0.0}$ %
          & $(0.2)$ %
           \\ 
\hline $B^-\to \Sigma^{0}\overline{p}$
          & $(0.8)$ %
          & $(0.8)$%
          & $(0.8)$%
          & $\overline B{}^0\to \Sigma^{+}\overline{p}$
          & $(1.7)$ %
          & $(1.7)$%
          & $(1.7)$%
          \\
$B^-\to \Sigma^{-}\overline{n}$
          & $(1.7)$ %
          & $(1.7)$%
          & $(1.7)$%
          & $\overline B{}^0\to \Sigma^{0}\overline{n}$
          & $(1.0)$ %
          & $(1.0)$%
          & $(1.0)$%
          \\ 
$B^-\to \Xi^{0}\overline{\Sigma^{+}}$
          & $(40.3)$ %
          & $(40.3)$%
          & $(40.3)$%
          & $\overline B{}^0\to \Xi^{0}\overline{\Sigma^{0}}$
          & $(18.7)$ %
          & $(18.7)$%
          & $(18.7)$%
          \\            
$B^-\to\Xi^{-}\overline{\Sigma^{0}}$
          & $(19.8)$ %
          & $(19.8)$%
          & $(19.8)$%
          & $\overline B{}^0\to \Xi^{0}\overline{\Lambda}$
          & $(2.5)$ %
          & $(2.5)$%
          & $(2.5)$%
          \\ 
$B^-\to \Xi^{-}\overline{\Lambda}$
          & $(2.4)$ %
          & $(2.4)$%
          & $(2.4)$%
          & $\overline B{}^0\to  \Xi^{-}\overline{\Sigma^{-}}$
          & $(36.7)$ %
          & $(36.7)$%
          & $(36.7)$%
          \\ 
$B^-\to \Lambda\overline{p}$
          & $(24.0)$ %
          & $(24.0)$%
          & $(24.0)$%
          & $\overline B{}^0\to \Lambda\overline{n}$
          & $(23.0)$ %
          & $(23.0)$ %
          & $(23.0)$ %
          \\ 
$\overline B{}^0_s\to p\overline{p}$
          & $0.44\pm0.00$ %
          & $0.44\pm0.00$%
          & $0.44\pm0.00$%
          & $\overline B{}^0_s\to \Sigma^{+}\overline{\Sigma^{+}}$
          & $2.7^{+0.0}_{-1.0}$ %
          & $1.8^{+1.0}_{-0.0}$ %
          & $0.5^{+3.3}_{-0.0}$ %
           \\
$\overline B{}^0_s\to n\overline{n}$
          & $(0.0)$ %
          & $0.4\pm0.0$ %
          & $0.4\pm0.0$ %
          & $\overline B{}^0_s\to \Sigma^{0}\overline{\Sigma^{0}}$
          & $2.1^{+0.0}_{-0.6}$%
          & $1.6^{+0.6}_{-0.0}$ %
          & $0.5^{+3.3}_{-0.0}$ %
           \\
$\overline B{}^0_s\to \Xi^{0}\overline{\Xi^{0}}$
          & $(26.4)$ %
          & $29.0^{+0.0}_{-4.5}$ %
          & $33.1^{+0.0}_{-12.7}$ %
          & $\overline B{}^0_s\to \Sigma^{-}\overline{\Sigma^{-}}$
          & $(1.7)$ %
          & $(1.7)$ %
          & $0.4^{+3.3}_{-0.0}$ %
           \\ 
$\overline B{}^0_s\to \Xi^{-}\overline{\Xi^{-}}$
          & $(25.2)$ %
          & $(25.2)$ %
          & $31.8^{+2.1}_{-12.4}$ %
          & $\overline B{}^0_s\to \Sigma^{0}\overline{\Lambda}$
          & $0.13^{+0.00}_{-0.11}$ %
          & $0.13^{+0.00}_{-0.11}$ %
          & $(0.04)$ %
           \\    
$\overline B{}^0_s\to \Lambda\overline{\Lambda}$
          & $15.8^{+0.6}_{-0.0}$ %
          & $17.9^{+0.0}_{-3.1}$ %
          & $21.6^{+0.0}_{-10.3}$ %
          & $\overline B{}^0_s\to \Lambda\overline{\Sigma^{0}}$
          & $0.01^{+0.11}_{-0.00}$ %
          & $0.01^{+0.11}_{-0.00}$ %
          & $(0.04)$ %
           \\                                                                                                                                                                                                             
\end{tabular}
}
\\
\end{ruledtabular}
\end{table}

From Eq.~(\ref{eq: BBBs, DS=-1}), we see that the amplitude of $\overline B_s^0\to p\bar p$ decay is given by,
$A(\bar B^0_s\to p\overline{p})=-5E'_{1\BB}+ E'_{2\BB}-3 PA'_\BB$. 
The enhancement in $\overline B_s^0\to p\bar p$ decay rate can be achieved through the enhancement of $E'_{1\BB}$, $E'_{2\BB}$ or $PA'_\BB$.
For illustration, we assume $\B(\bar B^0_s\to p\bar p)=0.44\times 10^{-8}$ saturating the present experimental bound \cite{LHCb:2022lff}
through the enhancement in these topological amplitudes. 
Note that $E_{1\BB}$, $E_{2\BB}$ or $PA_\BB$ will also be enlarged as they are related to $E'_{1\BB}$, $E'_{2\BB}$ and $PA'_\BB$, respectively, through CKM factors. 
To fit the $\bar B^0_s\to p\bar p$ rate,
we enhance the non-factorization contributions by adjusting the values of $\eta^{(\prime)}_{E,1}$, $\eta^{(\prime)}_{E,2}$ and $\eta^{(\prime)}_{PA}$ separately, see Eq. (\ref{eq:correction3}).
In enhancing $E'_{1\BB}$, $E'_{2\BB}$ or $PA'_\BB$, we need 
$|\eta^{(\prime)}_{E,1}|=7.6$,
$|\eta^{(\prime)}_{E,2}|=37.9$
or
$|\eta^{(\prime)}_{PA}|=3.2$,
respectively.
It seems that enhancing $PA'$ is the most effective choice.
 
Given that decay rates of other modes may be affected by the enhancement, 
we show in Table \ref{tab: BBEEPA} branching ratios of $\overline B_q\to\BB$ decays with $E^{(\prime)}_{1\BB}$, $E^{(\prime)}_{2\BB}$ or $PA^{(\prime)}_\BB$ enlarged. The uncertainties in rates are from the strong phases of these topological amplitudes. 
Note that in $\Delta S=0$ transition, the $B^-\to \BB'$ and $\bar B^0_s\to \BB'$ decays are unaffected, while in $\Delta S=-1$ transition, 
the $B^-\to \BB'$ and $\bar B^0\to \BB'$ decays are unaffected as well.
In particular, the $B^-\to\Lambda \bar p$ decay is not affected as it does not have exchange and penguin-annihilation diagrams.
Our finding are as following.
(i) By enlarging $E^{(\prime)}_{1\BB}$ or $E^{(\prime)}_{2\BB}$, we see that the $\bar B^0_s\to p\bar p$ rate is enlarged, but $\bar B^0\to p\bar p$ rate is also enlarged and is in tension with data. 
(ii) By enlarging $PA^{(\prime)}_\BB$, $\bar B^0_s\to p\bar p$ rate is enlarged, while the $\bar B^0\to p\bar p$ rate agree with data,
and $\bar B^0_s\to \Xi^0\overline{\Xi^0}$,  $\Xi^-\overline{\Xi^-}$
and $\bar B_s^0\to\Lambda\overline\Lambda$ rates are slightly enlarged.
It seems that enlarging $PA'$ is a possible way to enhance $\bar B_s\to p\bar p$ rate without having significant impacts on other modes.

For $\Delta S=-1$ pure penguin modes, the direct CP violation of these modes are predicted to be at most at few percent, see Eqs. (\ref{eq: A DeltaS penguin}), (\ref{eq: A DeltaS penguin1}) and Table~\ref{tab: Acp pure penguin}. 
It is possible that rare decay modes are sensitive to other effects such as final state interaction (FSI) \cite{FSI, Chua:2018ikx}.
Nevertheless, even in the presence of final state interaction the topological amplitude formalism is still applicable \cite{Chiang:2004nm, Chua:2018ikx}.
Indeed, it is possible to have final states with charmed flavor to rescatter into charmless final states \cite{FSI}. 
These FSI can enhance the charming penguin contribution, $|A_c|$, \cite{Ciuchini:1997rj}, giving $|A_u|/|A_c|<1$, and, consequently, further reduces the sizes of $\A$ of these pure penguin modes, as one can see by using Eq. (\ref{eq: A DeltaS penguin}). 
Hence $|\A|$ at most at the level of few \%, as shown in Table \ref{tab: Acp pure penguin}, is a robust prediction of the SM.

In this work, we study the rates and direct CP violations of $\bar B_{u,d,s}\to {{\rm\bf B}\overline{\rm\bf B}}'$ and $B^-_c\to{{\rm\bf B}\overline{\rm\bf B}}'$ decays. We incorporate topological amplitude formalism and the factorization approach. Asymptotic relations at large $m_b$ are used to simplify decay amplitudes. Using the most up-to-date data on $\bar B^0\to p\bar p$ and $B^-\to\Lambda\bar p$ decay rates as inputs, rates and direct CP violations of $\bar B_{u,d,s}\to \bfBB'$ decays are revised and predicted.
It is interesting that our results satisfy all existing experimental bounds and some predicted rates are close to the bounds.
In particular, the experimental limit on $B^-\to p \overline{\Delta^{++}}$ rate reported in ref. \cite{Belle:2007oni}
is only a factor of two larger than our predicted rate.
It will be interesting to see the updated search on this decay mode.
Factorization diagrams contribute to penguin-exchange, exchange, annihilation and penguin-annihilation amplitudes. Although the resulting penguin-exchange amplitudes are sizable, 
the factorization contributions to exchange, annihilation and penguin-annihilation amplitudes suffer from chiral suppression. 
Therefore the non-factorizable contributions on these topological amplitudes are important and cannot be ignored.
The factorizable contributions to $\bar B_s\to p\bar p$ rate is predicted to be several orders of magnitudes below the present bound, but it can be enhanced by including non-factorizable contributions, as it is governed by exchange and penguin-annihilation diagrams. 
The case where the rate can be enhanced through the enhancement on exchange or penguin annihilation amplitudes is discussed.
We find that 
by enlarging $E^{(\prime)}_{1\BB}$ or $E^{(\prime)}_{2\BB}$, we see that the $\bar B^0\to p\bar p$ rate is also enlarged and is in tension with data. 
On the other hand, by enlarging $PA^{(\prime)}_\BB$, we see that the $\bar B^0\to p\bar p$ rate agree with data,
while $\bar B^0_s\to \Xi^0\overline{\Xi^0}$,  $\Xi^-\overline{\Xi^-}$
and $\bar B_s^0\to\Lambda\overline\Lambda$ rates are slightly enlarged.
The measurement of $\bar B_s^0\to p\bar p$ rate can clarify the role of these topological amplitudes and provide valuable information on non-factorization contributions.  
The $B^-_c\to\bfBB'$ decays are annihilation modes. Their rates from factorization calculation are found to be very rare but can be enlarged to $10^{-9}$ or even $10^{-8}$ via non-factorizable contributions.
Small direct CP violations of pure penguin modes in $\Delta S=-1$ $\bar B_{u,d,s}\to{{\rm\bf B}\overline{\rm\bf B}}'$ decays (see Table \ref{tab: Acp pure penguin}) are robust predictions of the SM, while vanishing direct CP violations of exchange modes in $\bar B_{u,d,s}\to{{\rm\bf B}\overline{\rm\bf B}}'$ decays (see Table \ref{tab: Acp null test}) and in all $B^-_c\to{{\rm\bf B}\overline{\rm\bf B}}'$ decay modes (see Table \ref{tab: AcpBcDS=0, -1}) are null tests of the SM.
They can be used to test the SM in rare $B_{u,d,s,c}$ decays in the baryonic sector.

\begin{acknowledgments}
This work is supported in part by
the Ministry of Science and Technology of R.O.C.
under Grant No MOST-110-2112-M-033 -003.
\end{acknowledgments}

\appendix

\section{Topological amplitudes of two-body charmless baryonic $B$ decays}\label{App: Topological Amplitudes}

We collect all $\overline B_{u,d,s,c}\to\BB$, $\BD$, $\DB$, $\DD$ decay amplitudes using Eqs. (\ref{eq: DD}), (\ref{eq: DB}), (\ref{eq: BD}), (\ref{eq: BB}),  (\ref{eq: Bc DD}), (\ref{eq: Bc DB}), (\ref{eq: Bc BD}) and (\ref{eq: Bc BB}), in this appendix.

\subsection{$\overline B$ to octet-anti-octet baryonic decays}

The full $\bar B_{u,d,s,c}\to\BB$ decay amplitudes for $\Delta S=0$ processes are given by
\be
A(B^-\to n\overline{p})
   &=&-T_{1\BB}-5 P_{1\BB}-5 PE_{1\BB}
           +\frac{2}{3}(P_{1EW\BB}
           -P_{3EW\BB}+P_{4EW\BB})
           -5 A_{1\BB}, %
   \non\\
A(B^-\to\Sigma^{0}\overline{\Sigma^{+}})
   &=&\sq2T_{3\BB}
         +\frac{1}{\sq2}(5P_{1\BB}-P_{2\BB})
         +\frac{1}{\sq2}(5PE_{1\BB}-PE_{2\BB})
          +\frac{1}{3\sq2}(P_{1EW\BB}
   \non\\
   &&        +P_{2EW\BB}+2P_{3EW\BB}
-2P_{4EW\BB})
         +\frac{1}{\sq2}(5 A_{1\BB}-A_{2\BB}),%
   \non\\   
A(B^-\to\Sigma^{-}\overline{\Sigma^{0}})
   &=&-\frac{1}{\sq2}(5P_{1\BB}-P_{2\BB})
         -\frac{1}{\sq2}(5PE_{1\BB}-PE_{2\BB})
           -\frac{1}{3\sq2}(P_{1EW\BB}+P_{2EW\BB}
   \non\\
   &&-4P_{3EW\BB}-2P_{4EW\BB})-\frac{1}{\sq2}(5 A_{1\BB}- A_{2\BB}),%
   \non\\
A(B^-\to\Sigma^{-}\overline{\Lambda})
   &=&-\frac{1}{\sq6}(5P_{1\BB}+P_{2\BB})
          -\frac{1}{\sq6}(5PE_{1\BB}+PE_{2\BB})
           -\frac{1}{3\sq6}(P_{1EW\BB}-P_{2EW\BB}
   \non\\
   &&-4P_{3EW\BB}-2P_{4EW\BB})-\frac{1}{\sq6}(5 A_{1\BB}+ A_{2\BB}),%
   \non\\  
A(B^-\to\Xi^{-}\overline{\Xi^{0}})
   &=&-P_{2\BB}-PE_{2\BB}+\frac{1}{3}P_{2EW\BB}- A_{2\BB},%
   \non\\
A(B^-\to\Lambda\overline{\Sigma^+})
   &=&-\sq{\frac{2}{3}}(T_{1\BB}-T_{3\BB})
          -\frac{1}{\sq6}(5P_{1\BB}+P_{2\BB})
          -\frac{1}{\sq6}(5PE_{1\BB}+PE_{2\BB})
   \non\\
   &&   +\frac{1}{3\sq6}(5P_{1EW\BB}+P_{2EW\BB}-4P_{3EW\BB}+2P_{4EW\BB})
           -\frac{1}{\sq6}(5 A_{1\BB}+A_{2\BB}),%
 \non\\
\label{eq: BBBm, DS=0}           
\en
\be
A(\bar B^0\to p\overline{p})
   &=&-T_{2\BB}+2T_{4\BB}
          +P_{2\BB}
          +PE_{2\BB}
           +\frac{2}{3}P_{2EW\BB}
          -5 E_{1\BB}+ E_{2\BB}
          -3PA_\BB,%
   \non\\
A(\bar B^0\to n\overline{n})
   &=&-(T_{1\BB}+T_{2\BB})
          -(5P_{1\BB}-P_{2\BB})
          -(5PE_{1\BB}-PE_{2\BB})
   \non\\
   &&+\frac{2}{3}(P_{1EW\BB}+P_{2EW\BB}
   -P_{3EW\BB}-2P_{4EW\BB})
         + E_{2\BB}
         -3 PA_\BB,%
   \non\\
A(\bar B^0\to\Sigma^{+}\overline{\Sigma^{+}})
   &=&-5 E_{1\BB}+ E_{2\BB}
          -3 PA_\BB,%
   \non\\
A(\bar B^0\to\Sigma^{0}\overline{\Sigma^{0}})
   &=&-T_{3\BB}
          -\frac{1}{2}(5P_{1\BB}-P_{2\BB})
          -\frac{1}{2}(5PE_{1\BB}-PE_{2\BB})
          -\frac{1}{6}(P_{1EW\BB}
   \non\\
   &&+P_{2EW\BB}+2P_{3EW\BB}-2P_{4EW\BB})
   -\frac{1}{2}(5 E_{1\BB}- E_{2\BB})
          -3 PA_\BB,%
   \non\\
A(\bar B^0\to\Sigma^{0}\overline{\Lambda})
   &=&\frac{1}{\sq3}(T_{3\BB}+2T_{4\BB})
          +\frac{1}{2\sq3}(5P_{1\BB}+P_{2\BB})
          +\frac{1}{2\sq3}(5PE_{1\BB}+PE_{2\BB})
   \non\\
   && +\frac{1}{6\sq3}(P_{1EW\BB}-P_{2EW\BB}+2P_{3EW\BB}+10P_{4EW\BB})
   \non\\
   &&     -\frac{1}{2\sqrt3}(5 E_{1\BB}+ E_{2\BB}),%
   \non\\
A(\bar B^0\to\Sigma^{-}\overline{\Sigma^{-}})
   &=&-(5P_{1\BB}-P_{2\BB})
          -(5PE_{1\BB}-PE_{2\BB})
           -\frac{1}{3}(P_{1EW\BB}+P_{2EW\BB}
   \non\\
   &&-4P_{3EW\BB}-2P_{4EW\BB})-3 PA_\BB,%
   \non\\   
A(\bar B^0\to\Xi^{0}\overline{\Xi^{0}})
   &=& E_{2\BB}
          -3 PA_\BB,%
   \non\\
A(\bar B^0\to\Xi^{-}\overline{\Xi^{-}})
   &=&P_{2\BB}+PE_{2\BB}
           -\frac{1}{3}P_{2EW\BB}
           -3PA_\BB,%
   \non\\
A(\bar B^0\to\Lambda\overline{\Sigma^{0}})
   &=&\frac{1}{\sq3}(T_{1\BB}-T_{3\BB})
          +\frac{1}{2\sq3}(5P_{1\BB}+P_{2\BB})
          +\frac{1}{2\sq3}(5PE_{1\BB}+PE_{2\BB})
   \non\\
   && -\frac{1}{6\sq3}(5P_{1EW\BB}+P_{2EW\BB}-2P_{3EW\BB}+2P_{4EW\BB})
   \non\\
    &&   -\frac{1}{2\sq3}(5 E_{1\BB}+E_{2\BB}), %
   \non\\
A(\bar B^0\to\Lambda\overline{\Lambda})
   &=&-\frac{1}{3}(T_{1\BB}+2T_{2\BB}-T_{3\BB}-2T_{4\BB})
          -\frac{5}{6}(P_{1\BB}-P_{2\BB}) 
          -\frac{5}{6}(PE_{1\BB}       
   \non\\
   &&-PE_{2\BB})   +\frac{1}{18}(5P_{1EW\BB}+7P_{2EW\BB}-2P_{3EW\BB}-10P_{4EW\BB})
   \non\\ 
   && -\frac{5}{6}(E_{1\BB}-E_{2\BB})-3PA_\BB,  %
\label{eq: BBB0, DS=0}   
\en
\be
A(\bar B^0_s\to p\overline{\Sigma^{+}})
   &=&T_{2\BB}-2T_{4\BB}
          -P_{2\BB} -PE_{2\BB}
          -\frac{2}{3}P_{2EW\BB},%
   \non\\
A(\bar B^0_s\to n\overline{\Sigma^{0}})
   &=&-\frac{1}{\sq2}T_{2\BB}
          +\frac{1}{\sq2}P_{2\BB}     
          +\frac{1}{\sq2}PE_{2\BB}          
          +\frac{\sq2}{3}(P_{2EW\BB}-3P_{4EW\BB}),%
   \non\\
A(\bar B^0_s\to n\overline{\Lambda})
   &=&\frac{1}{\sq6}(2T_{1\BB}+T_{2\BB})
          +\frac{1}{\sq6}(10P_{1\BB}-P_{2\BB})      
          +\frac{1}{\sq6}(10PE_{1\BB}-PE_{2\BB})          
   \non\\
   &&-\frac{1}{3}\sqrt{\frac{2}{3}}(2P_{1EW\BB}+P_{2EW\BB}-2P_{3EW\BB}-P_{4EW\BB}),%
   \non\\   
A(\bar B^0_s\to\Sigma^{0}\overline{\Xi^{0}})
   &=&\sq2(T_{3\BB}+T_{4\BB})
          +\frac{5}{\sq2}P_{1\BB}+\frac{5}{\sq2}PE_{1\BB}
          +\frac{1}{3\sq2}(P_{1EW\BB}+2P_{3EW\BB}
    \non\\      
          &&+4P_{4EW\BB}),%
   \non\\
A(\bar B^0_s\to\Sigma^{-}\overline{\Xi^{-}})
   &=&-5P_{1\BB} -5PE_{1\BB}          
           +\frac{1}{3}(-P_{1EW\BB}+4P_{3EW\BB}+2P_{4EW\BB}),%
   \non\\   
A(\bar B^0_s\to\Lambda\overline{\Xi^0})
   &=&-\sq{\frac{2}{3}}(T_{1\BB}+T_{2\BB}-T_{3\BB}-T_{4\BB})
          -\frac{1}{\sq6}(5P_{1\BB}-2P_{2\BB})   
           -\frac{1}{\sq6}(5PE_{1\BB}      
   \non\\
   &&-2PE_{2\BB})    +\frac{1}{3\sq6}(5P_{1EW\BB}+4P_{2EW\BB}-2P_{3EW\BB}-4P_{4EW\BB}),%
\label{eq: BBBs, DS=0}
\en
and
\be
A(B_c^-\to n\overline{p})
   &=& -5 A^c_{1\BB}, %
   \non\\
A(B_c^-\to\Sigma^{0}\overline{\Sigma^{+}})
   &=&\frac{1}{\sq2}(5 A^c_{1\BB}-A^c_{2\BB}),%
   \non\\   
A(B_c^-\to\Sigma^{-}\overline{\Sigma^{0}})
   &=& -\frac{1}{\sq2}(5 A^c_{1\BB}- A^c_{2\BB}),%
   \non\\
A(B_c^-\to\Sigma^{-}\overline{\Lambda})
   &=&-\frac{1}{\sq6}(5 A^c_{1\BB}+ A^c_{2\BB}),%
   \non\\  
A(B_c^-\to\Xi^{-}\overline{\Xi^{0}})
   &=&- A^c_{2\BB},%
   \non\\
A(B_c^-\to\Lambda\overline{\Sigma^+})
   &=& -\frac{1}{\sq6}(5 A^c_{1\BB}+A^c_{2\BB}),
\label{eq: BBBc, DS=0}           
\en
while those for $\Delta S=1$ transitions are given by
\be
A(B^-\to\Sigma^{0}\overline{p})
   &=&-\frac{1}{\sq2}(T'_{1\BB}-2T'_{3\BB})
           -\frac{1}{\sq2} P'_{2\BB}
            -\frac{1}{\sq2} PE'_{2\BB}
           +\frac{1}{3\sq2}(3P'_{1EW\BB}
   \non\\        
           &&+P'_{2EW\BB})-\frac{1}{\sq2}A'_{2\BB},%
   \non\\
A(B^-\to\Sigma^{-}\overline{n})
   &=&-P'_{2\BB}
          -PE'_{2\BB}
          +\frac{1}{3}P'_{2EW\BB}
          - A'_{2\BB},%
   \non\\
A(B^-\to\Xi^{0}\overline{\Sigma^{+}})
   &=&-T'_{1\BB}
           -5 P'_{1\BB}
           -5 PE'_{1\BB}
           +\frac{2}{3}(P'_{1EW\BB}-P'_{3EW\BB}+P'_{4EW\BB})
           -5 A'_{1\BB},%
   \non\\
A(B^-\to\Xi^{-}\overline{\Sigma^{0}})
   &=&-\frac{5}{\sq2}P'_{1\BB}
          -\frac{5}{\sq2}PE'_{1\BB}
           -\frac{1}{3\sq2}(P'_{1EW\BB}-4P'_{3EW\BB}-2P'_{4EW\BB})
    \non\\
     &&      -\frac{5}{\sq2}A'_{1\BB},%
   \non\\  
A(B^-\to\Xi^{-}\overline{\Lambda})
   &=&-\frac{1}{\sqrt6}(5P'_{1\BB}-2P'_{2\BB})
   -\frac{1}{\sqrt6}(5PE'_{1\BB}-2PE'_{2\BB})
           -\frac{1}{3\sq6}(P'_{1EW\BB}
   \non\\
   &&+2P'_{2EW\BB}-4P'_{3EW\BB}-2P'_{4EW\BB})-\frac{1}{\sq6}(5 A'_{1\BB}-2 A'_{2\BB}),%
   \non\\      
A(B^-\to\Lambda\overline{p})
   &=&\frac{1}{\sq6}(T'_{1\BB}+2T'_{3\BB})
           +\frac{1}{\sqrt6}(10P'_{1\BB}-P'_{2\BB})
            +\frac{1}{\sqrt6}(10PE'_{1\BB}-PE'_{2\BB})
            \non\\
   &&  -\frac{1}{3\sq6}(P'_{1EW\BB}-P'_{2EW\BB}-4P'_{3EW\BB}+4P'_{4EW\BB})
         \non\\
      &&   +\frac{1}{\sq6}(10 A'_{1\BB}- A'_{2\BB}), %
\label{eq: BBBm, DS=-1}                
\en
\be
A(\bar B^0\to \Sigma^{+}\overline{p})
   &=& T'_{2\BB}-2T'_{4\BB}
           -P'_{2\BB}
            -PE'_{2\BB}
           -\frac{2}{3}P'_{2EW\BB},%
   \non\\
A(\bar B^0\to\Sigma^{0}\overline{n})
   &=&-\frac{1}{\sq2}(T'_{1\BB}+T'_{2\BB}-2T'_{3\BB}-2T'_{4\BB})
          +\frac{1}{\sq2}P'_{2\BB}
          +\frac{1}{\sq2}PE'_{2\BB}
    \non\\
    && +\frac{1}{3\sq2}(3P'_{1EW\BB}+2P'_{2EW\BB}),%
   \non\\
A(\bar B^0\to\Xi^{0}\overline{\Sigma^{0}})
   &=&\frac{1}{\sq2}T'_{1\BB}
           +\frac{5}{\sq2}P'_{1\BB}
            +\frac{5}{\sq2}PE'_{1\BB}
           -\frac{\sq2}{3}(P'_{1EW\BB}-P'_{3EW\BB}+P'_{4EW\BB}),%
   \non\\
A(\bar B^0\to\Xi^{0}\overline{\Lambda})
   &=&-\frac{1}{\sq6}(T'_{1\BB}+2T'_{2\BB})
           -\frac{1}{\sq6}(5P'_{1\BB}-2P'_{2\BB})
           -\frac{1}{\sq6}(5PE'_{1\BB}-2PE'_{2\BB})
    \non\\
    &&+\frac{1}{3}\sq{\frac{2}{3}}(P'_{1EW\BB}+2P'_{2EW\BB}-P'_{3EW\BB}-5P'_{4EW\BB}),%
   \non\\ 
A(\bar B^0\to\Xi^{-}\overline{\Sigma^{-}})
   &=&-5P'_{1\BB}-5PE'_{1\BB}
           -\frac{1}{3}(P'_{1EW\BB}-4P'_{3EW\BB}-2P'_{4EW\BB}),%
   \non\\   
A(\bar B^0\to\Lambda\overline{n})
   &=&\frac{1}{\sq6}\bigg[(T'_{1\BB}+T'_{2\BB}+2T'_{3\BB}+2T'_{4\BB})
           +(10P'_{1\BB}-P'_{2\BB})
            +(10PE'_{1\BB}
    \non\\
    &&-PE'_{2\BB})-\frac{1}{3}(P'_{1EW\BB}+2P'_{2EW\BB}-4P'_{3EW\BB}-8P'_{4EW\BB})\bigg],%
\label{eq: BBB0, DS=-1} 
\en
\be
A(\bar B^0_s\to p\overline{p})
   &=&-5E'_{1\BB}+ E'_{2\BB}-3 PA'_\BB,%
   \non\\
A(\bar B^0_s\to n\overline{n})
   &=& E'_{2\BB}-3PA'_\BB,%
   \non\\
A(\bar B^0_s\to\Sigma^{+}\overline{\Sigma^{+}})
   &=&-T'_{2\BB}+2T'_{4\BB}
           +P'_{2\BB}
            +PE'_{2\BB}
           +\frac{2}{3}P'_{2EW\BB}
           -5 E'_{1\BB}+ E'_{2\BB}
    \non\\       
           &&-3PA'_\BB,%
   \non\\
A(\bar B^0_s\to\Sigma^{0}\overline{\Sigma^{0}})
   &=&-\frac{1}{2}(T'_{2\BB}-2T'_{4\BB})
           +P'_{2\BB}
           +PE'_{2\BB}
           +\frac{1}{6}P'_{2EW\BB}
           -\frac{1}{2}(5 E'_{1\BB}- E'_{2\BB})
   \non\\        
          && -3 PA'_\BB,%
   \non\\
A(\bar B^0_s\to\Sigma^{0}\overline{\Lambda})
   &=&\frac{1}{2\sq3}(2T'_{1\BB}+T'_{2\BB}-4T'_{3\BB}-2T'_{4\BB})
           -\frac{1}{2\sq3}(2P'_{1EW\BB}+P'_{2EW\BB})
    \non\\
    &&-\frac{1}{2\sq3}(5 E'_{1\BB}+ E'_{2\BB}),%
     \non\\     
A(\bar B^0_s\to\Sigma^{-}\overline{\Sigma^{-}})
   &=&P'_{2\BB}
          +PE'_{2\BB}
           -\frac{1}{3}P'_{2EW\BB}
           -3 PA'_\BB,%
   \non\\   
A(\bar B^0_s\to\Xi^{0}\overline{\Xi^{0}})
   &=&-T'_{1\BB}-T'_{2\BB}
           -(5P'_{1\BB}-P'_{2\BB})
            -(5PE'_{1\BB}-PE'_{2\BB})
           +\frac{2}{3}(P'_{1EW\BB}
    \non\\
    &&+P'_{2EW\BB}-P'_{3EW\BB}-2P'_{4EW\BB})
           + E'_{2\BB}
           -3 PA'_\BB,%
   \non\\
A(\bar B^0_s\to\Xi^{-}\overline{\Xi^{-}})
   &=&-(5P'_{1\BB}-P'_{2\BB})
          -(5PE'_{1\BB}-PE'_{2\BB})
          -\frac{1}{3}(P'_{1EW\BB}+P'_{2EW\BB}
   \non\\
    &&-4P'_{3EW\BB}-2P'_{4EW\BB})-3 PA'_\BB,%
    \non\\     
A(\bar B^0_s\to\Lambda\overline{\Sigma^{0}})
   &=&\frac{1}{2\sq3}(T'_{2\BB}+2T'_{4\BB})
           +\frac{1}{2\sq3}(-P'_{2EW\BB}+4P'_{4EW\BB})
    \non\\       
    &&-\frac{1}{2\sq3}(5 E'_{1\BB}+E'_{2\BB}),%
   \non\\
A(\bar B^0_s\to\Lambda\overline{\Lambda})
   &=&-\frac{1}{6}(2T'_{1\BB}+T'_{2\BB}+4T'_{3\BB}+2T'_{4\BB})
           -\frac{1}{3}(10P'_{1\BB}-P'_{2\BB})
           -\frac{1}{3}(10PE'_{1\BB}
   \non\\       
   &&-PE'_{2\BB})+\frac{1}{18}(2P'_{1EW\BB}+P'_{2EW\BB}-8P'_{3EW\BB}-4P'_{4EW\BB})
   \non\\      
   &&-\frac{5}{6}(E'_{1\BB}-E'_{2\BB})
         -3PA'_\BB,%
\label{eq: BBBs, DS=-1}
\en
and
\be
A(B^-_c\to\Sigma^{0}\overline{p})
   &=& -\frac{1}{\sq2}A^{\prime c}_{2\BB},
   \non\\
A(B^-_c\to\Sigma^{-}\overline{n})
   &=&
          - A^{\prime c}_{2\BB},
   \non\\
A(B^-_c\to\Xi^{0}\overline{\Sigma^{+}})
   &=&
           -5 A^{\prime c}_{1\BB},
   \non\\
A(B^-_c\to\Xi^{-}\overline{\Sigma^{0}})
   &=&
           -\frac{5}{\sq2}A^{\prime c}_{1\BB},
   \non\\  
A(B^-_c\to\Xi^{-}\overline{\Lambda})
   &=&-\frac{1}{\sq6}(5 A^{\prime c}_{1\BB}-2 A^{\prime c}_{2\BB}),
   \non\\      
A(B^-_c\to\Lambda\overline{p})
   &=&
         \frac{1}{\sq6}(10 A^{\prime c}_{1\BB}- A^{\prime c}_{2\BB}).
\label{eq: BBBc, DS=-1}                
\en

\subsection{$\overline B$ to octet-anti-decuplet baryonic decays}

The full $\bar B_{u,d,s,c}\to\BD$ decay amplitudes for $\Delta S=0$ processes are given by
\be
A(B^-\to p\overline{\Delta^{++}})
   &=&-\sq6 (T_{1\BD}-2T_{2\BD})+\sq6 P_\BD+\sq6 PE_\BD+2\sq{\frac{2}{3}}P_{1EW\BD}+\sq6 A_\BD,%
   \non\\
A(B^-\to n\overline{\Delta^+})
   &=&-\sq2T_{1\BD}+\sq2 P_\BD+\sq2 PE_\BD+\frac{2\sq2}{3}(P_{1EW\BD}-3P_{2EW\BD})+\sq2A_\BD,%
   \non\\
A(B^-\to\Sigma^0\overline{\Sigma^{*+}})
   &=&-2T_{2\BD}-P_\BD-PE_\BD+\frac{1}{3}(P_{1EW\BD}-6P_{2EW\BD})- A_\BD,%
   \non\\ 
A(B^-\to\Sigma^-\overline{\Sigma^{*0}})
   &=&-P_\BD-PE_\BD+\frac{1}{3}P_{1EW\BD}- A_\BD,%
   \non\\      
A(B^-\to\Xi^{-}\overline{\Xi^{*0}})
   &=&-\sq2 P_\BD-\sq2 PE_\BD+\frac{\sq2}{3}P_{1EW\BD}-\sq2A_\BD,%
   \non\\
A(B^-\to\Lambda\overline{\Sigma^{*+}})
   &=&\frac{2}{\sq3}(T_{1\BD}-T_{2\BD})-\sq3 P_\BD-\sq3 PE_\BD-\frac{1}{\sq3}(P_{1EW\BD}-2P_{2EW\BD})
   \non\\
   && -\sq3 A_\BD,%
\label{eq: BDBm, DS=0} 
\en
\be
A(\bar B^0\to p\overline{\Delta^+})
   &=&-\sq2(T_{1\BD}-2T_{2\BD})+\sq2 P_\BD+\sq2 PE_\BD+\frac{2\sq2}{3}P_{1EW\BD}-\sq2 E_\BD,%
   \non\\
A(\bar B^0\to n\overline{\Delta^0})
   &=&-\sq2 T_{1\BD}+\sq2 P_\BD+\sq2 PE_\BD+\frac{2\sq2}{3}(P_{1EW\BD}-3P_{2EW\BD})
   \non\\
   &&-\sq2 E_\BD,%
   \non\\
A(\bar B^0\to\Sigma^{+}\overline{\Sigma^{*+}})
   &=&\sq2 E_\BD,%
   \non\\
A(\bar B^0\to\Sigma^{0}\overline{\Sigma^{*0}})
   &=&-\sq2 T_{2\BD}-\frac{1}{\sq2}P_\BD-\frac{1}{\sq2}PE_\BD+\frac{1}{3\sq2}(P_{1EW\BD}-6P_ {2EW\BD})
   \non\\
   && -\frac{1}{\sq2} E_\BD,%
   \non\\
A(\bar B^0\to\Sigma^{-}\overline{\Sigma^{*-}})
   &=&-\sq2 P_\BD-\sq2 PE_\BD+\frac{\sq2}{3} P_{1EW\BD},
   \non\\
A(\bar B^0\to\Xi^{0}\overline{\Xi^{*0}})
   &=&\sq2 E_{\BD},%
   \non\\
A(\bar B^0\to\Xi^{-}\overline{\Xi^{*-}})
   &=&-\sq2 P_\BD-\sq2 PE_\BD+\frac{\sq2}{3}P_{1EW\BD},%
   \non\\
A(\bar B^0\to\Lambda\overline{\Sigma^{*0}})
   &=&\sq{\frac{2}{3}}(T_{1\BD}-T_{2\BD})-\sq{\frac{3}{2}} P_\BD-\sq{\frac{3}{2}} PE_\BD
   -\frac{1}{\sq6}(P_{1EW\BD}-2P_{2EW\BD})
   \non\\   
   &&+\sq{\frac{3}{2}}E_\BD,%
\label{eq: BDB0, DS=0} 
\en
\be
A(\bar B^0_s\to p\overline{\Sigma^{*+}})
   &=&-\sq2 (T_{1\BD}-2T_{2\BD})+\sq2 P_\BD+\sq2 PE_\BD+\frac{2\sq2}{3}P_{1EW\BD},%
   \non\\
A(\bar B^0_s\to n\overline{\Sigma^{*0}})
   &=&-T_{1\BD}+P_\BD+PE_\BD+\frac{2}{3}(P_{1EW\BD}-3P_{2EW\BD}),%
   \non\\
A(\bar B^0_s\to\Sigma^{0}\overline{\Xi^{*0}})
   &=&-2T_{2\BD}-P_\BD-PE_\BD+\frac{1}{3}(P_{1EW\BD}-6P_{2EW\BD}),%
   \non\\
A(\bar B^0_s\to\Sigma^{-}\overline{\Xi^{*-}})
   &=&-\sq2 P_\BD-\sq2 PE_\BD+\frac{\sq2}{3}P_{1EW\BD},%
   \non\\   
A(\bar B^0_s\to\Xi^{-}\overline{\Omega^-})
   &=&-\sq6 P_\BD-\sq6 PE_\BD+\sq{\frac{2}{3}}P_{1EW\BD},%
   \non\\
A(\bar B^0_s\to\Lambda\overline{\Xi^{*0}})
   &=&\frac{2}{\sq3}(T_{1\BD}-T_{2\BD})
         -\sq3 P_\BD
         -\sq3 PE_\BD
         -\frac{1}{\sq3}(P_{1EW\BD}-2P_{2EW\BD}),%
   \non\\
\label{eq: BDBs, DS=0} 
\en
and
\be
A(B^-_c\to p\overline{\Delta^{++}})
   &=&\sq6 A^c_\BD,
   \non\\
A(B^-_c\to n\overline{\Delta^+})
   &=&\sq2A^c_\BD,
   \non\\
A(B^-_c\to\Sigma^0\overline{\Sigma^{*+}})
   &=&- A^c_\BD,
   \non\\ 
A(B^-_c\to\Sigma^-\overline{\Sigma^{*0}})
   &=&- A^c_\BD,
   \non\\      
A(B^-_c\to\Xi^{-}\overline{\Xi^{*0}})
   &=&-\sq2A^c_\BD,
   \non\\
A(B^-_c\to\Lambda\overline{\Sigma^{*+}})
   &=&-\sq3 A^c_\BD,
\label{eq: BDBc, DS=0} 
\en
while those for $\Delta S=1$ transitions are given by
\be
A(B^-\to \Sigma^+\overline{\Delta^{++}})
   &=&\sq6 (T'_{1\BD}-2T'_{2\BD})
          -\sq6P'_\BD
          -\sq6PE'_\BD
          -2\sq{\frac{2}{3}}P'_{1EW\BD}
          -\sq6 A'_\BD,%
   \non\\
A(B^-\to\Sigma^0\overline{\Delta^+})
   &=&-T'_{1\BD}+2T'_{2\BD}
   +2P'_\BD
   +2PE'_\BD
   +\frac{1}{3}P'_{1EW\BD}
   +2 A'_\BD,
   \non\\
A(B^-\to\Sigma^-\overline{\Delta^0})
   &=&\sq2 P'_\BD
         +\sq2 PE'_\BD
         -\frac{\sq2}{3}P'_{1EW\BD}
         +\sq2 A'_\BD,
   \non\\
A(B^-\to\Xi^{0}\overline{\Sigma^{*+}})
   &=&\sq2T'_{1\BD}
         -\sq2 P'_\BD
           -\sq2 PE'_\BD
         -\frac{2\sq2}{3}(P'_{1EW\BD}-3P'_{2EW\BD})
         -\sq2 A'_\BD,
   \non\\   
A(B^-\to\Xi^{-}\overline{\Sigma^{*0}})
   &=&P'_\BD+PE'_\BD-\frac{1}{3}P'_{1EW\BD}+ A'_\BD,
   \non\\
A(B^-\to\Lambda\overline{\Delta^{+}})
   &=&\frac{1}{\sq3}(T'_{1\BD}+2T'_{2\BD})
         -\frac{1}{\sq3}(P'_{1EW\BD}-4P'_{2EW\BD}),
\label{eq: BDBm, DS=-1}
\en
\be
A(\bar B^0\to \Sigma^{+}\overline{\Delta^+})
   &=&\sq2 (T'_{1\BD}-2 T'_{2\BD})
          -\sq2 P'_\BD
           -\sq2 PE'_\BD
          -\frac{2\sq2}{3}P'_{1EW\BD},%
   \non\\
A(\bar B^0\to\Sigma^{0}\overline{\Delta^0})
   &=&-T'_{1\BD}+2T'_{2\BD}
         +2P'_\BD
         +2PE'_\BD
         +\frac{1}{3}P'_{1EW\BD},%
   \non\\
A(\bar B^0\to\Sigma^{-}\overline{\Delta^-})
   &=&\sq6 P'_\BD
         +\sq6 PE'_\BD
         -\sq{\frac{2}{3}}P'_{1EW\BD},%
   \non\\
A(\bar B^0\to\Xi^{0}\overline{\Sigma^{*0}})
   &=&T'_{1\BD}
         -P'_\BD
         -PE'_\BD
         -\frac{2}{3}(P'_{1EW\BD}-3P'_{2EW\BD}),%
   \non\\
A(\bar B^0\to\Xi^{-}\overline{\Sigma^{*-}})
   &=&\sq2P'_\BD
         +\sq2PE'_\BD
         -\frac{\sq2}{3}P'_{1EW\BD},%
   \non\\   
A(\bar B^0\to\Lambda\overline{\Delta^0})
   &=&\frac{1}{\sq3}(T'_{1\BD}+2T'_{2\BD})
         -\frac{1}{\sq3}(P'_{1EW\BD}-4P'_{2EW\BD}),%
\label{eq: BDB0, DS=-1}
\en
\be
A(\bar B^0_s\to p\overline{\Delta^+})
   &=&-\sq2E'_\BD,%
   \non\\
A(\bar B^0_s\to n\overline{\Delta^0})
   &=&-\sq2E'_\BD,%
   \non\\
A(\bar B^0_s\to\Sigma^{+}\overline{\Sigma^{*+}})
   &=&\sq2(T'_{1\BD}-2T'_{2\BD})
         -\sq2P'_\BD
         -\sq2PE'_\BD
         -\frac{2\sq2}{3}P'_{1EW\BD}
        +\sq2E'_\BD,%
   \non\\
A(\bar B^0_s\to\Sigma^{0}\overline{\Sigma^{*0}})
   &=&-\frac{1}{\sq2}(T'_{1\BD}-2T'_{2\BD})
         +\sq2 P'_\BD
         +\sq2 PE'_\BD
         +\frac{1}{3\sq2}P'_{1EW\BD}
         -\frac{1}{\sq2} E'_\BD,
   \non\\
A(\bar B^0_s\to\Sigma^{-}\overline{\Sigma^{*-}})
   &=&\sq2P'_\BD
          +\sq2PE'_\BD
         -\frac{\sq2}{3}P'_{1EW\BD},%
   \non\\   
A(\bar B^0_s\to\Xi^{0}\overline{\Xi^{*0}})
   &=&\sq2T'_{1\BD}
         -\sq2P'_\BD
         -\sq2PE'_\BD
         -\frac{2\sq2}{3}(P'_{1EW\BD}-3P'_{2EW\BD})
         +\sq2 E'_\BD,%
   \non\\
A(\bar B^0_s\to\Xi^{-}\overline{\Xi^{*-}})
   &=&\sq2P'_\BD
        +\sq2PE'_\BD
         -\frac{\sq2}{3}P'_{1EW\BD},%
   \non\\
A(\bar B^0_s\to\Lambda\overline{\Sigma^{*0}})
   &=&\frac{1}{\sq6}(T'_{1\BD}+2T'_{2\BD})
          -\frac{1}{\sq6}(P'_{1EW\BD}-4P'_{2EW\BD})
          +\sq{\frac{3}{2}} E'_\BD,  %
\label{eq: BDBs, DS=-1}
\en
and
\be
A(B^-_c\to \Sigma^+\overline{\Delta^{++}})
   &=&
          -\sq6 A^{\prime c}_\BD,
   \non\\
A(B^-_c\to\Sigma^0\overline{\Delta^+})
   &=&2 A^{\prime c}_\BD,
   \non\\
A(B^-_c\to\Sigma^-\overline{\Delta^0})
   &=&\sq2 A^{\prime c}_\BD,
   \non\\
A(B^-_c\to\Xi^{0}\overline{\Sigma^{*+}})
   &=&
         -\sq2 A^{\prime c}_\BD,
   \non\\   
A(B^-_c\to\Xi^{-}\overline{\Sigma^{*0}})
   &=& A^{\prime c}_\BD,
   \non\\
A(B^-_c\to\Lambda\overline{\Delta^{+}})
   &=&0.
\label{eq: BDBc, DS=-1}
\en

\subsection{$\overline B$ to decuplet-anti-octet baryonic decays}

The full $\bar B_{u,d,s,c}\to\DB$ decay amplitudes for $\Delta S=0$ processes are given by
\be
A(B^-\to\Delta^0\overline{p})
   &=&\sq2T_{1\DB}
          -\sq2 P_\DB
          -\sq2 PE_\DB
          +\frac{\sq2}{3}(3P_{1EW\DB}+P_{2EW\DB})
          -\sq2 A_\DB,%
   \non\\
A(B^-\to\Delta^-\overline{n})
   &=&-\sq6 P_\DB
          -\sq6 PE_\DB
         +\sq{\frac{2}{3}}P_{2EW\DB}
         -\sq6 A_\DB,%
   \non\\
A(B^-\to\Sigma^{*0}\overline{\Sigma^{+}})
   &=&-T_{1\DB}
          +P_\DB
          +PE_\DB
          -\frac{1}{3}(3P_{1EW\DB}+P_{2EW\DB})
          +A_\DB,%
   \non\\   
A(B^-\to\Sigma^{*-}\overline{\Sigma^{0}})
   &=&-P_\DB
          -PE_\DB
          +\frac{1}{3}P_{2EW\DB}
          - A_\DB,%
   \non\\
A(B^-\to\Xi^{*-}\overline{\Xi^{0}})
   &=&\sq2 P_\DB
         + \sq2 PE_\DB
         -\frac{\sq2}{3}P_{2EW\DB}
         +\sq2A_\DB,%
   \non\\
A(B^-\to\Sigma^{*-}\overline{\Lambda})
   &=&\sq3 P_\DB
          +\sq3 PE_\DB
          -\frac{1}{\sq3}P_{2EW\DB}
          +\sq3 A_\DB,%
\label{eq: DBBm, DS=0}
\en
\be
A(\bar B^0\to\Delta^+\overline{p})
   &=&\sq2T_{2\DB}
          +\sq2 P_\DB
          +\sq2 PE_\DB
          +\frac{2\sq2}{3}P_{2EW\DB}
          -\sq2E_\DB,%
   \non\\
A(\bar B^0\to\Delta^0\overline{n})
   &=&\sq2(T_{1\DB}+T_{2\DB})
         +\sq2 P_\DB
         +\sq2 PE_\DB
         +\frac{\sq2}{3}(3P_{1EW\DB}+2P_{2EW\DB})
   \non\\      
         &&-\sq2E_\DB,%
   \non\\
A(\bar B^0\to\Sigma^{*+}\overline{\Sigma^{+}})
   &=&\sq2 E_\DB,%
   \non\\
A(\bar B^0\to\Sigma^{*0}\overline{\Sigma^{0}})
   &=&\frac{1}{\sq2}T_{1\DB}
         -\frac{1}{\sq2}P_\DB
         -\frac{1}{\sq2}PE_\DB
         +\frac{1}{3\sq2}(3P_{1EW\DB}+P_{2EW\DB})
   \non\\
         &&-\frac{1}{\sq2} E_\DB,%
   \non\\
A(\bar B^0\to\Sigma^{*-}\overline{\Sigma^{-}})
   &=&-\sq2P_\DB
          -\sq2PE_\DB
          +\frac{\sq2}{3}P_{2EW\DB},%
   \non\\   
A(\bar B^0\to\Xi^{*0}\overline{\Xi^{0}})
   &=&\sq2E_{\DB},%
   \non\\
A(\bar B^0\to\Xi^{*-}\overline{\Xi^{-}})
   &=&-\sq2 P_\DB
          -\sq2 PE_\DB
          +\frac{\sq2}{3}P_{2EW\DB},%
   \non\\
A(\bar B^0\to\Sigma^{*0}\overline{\Lambda})
   &=&-\frac{1}{\sq6}(T_{1\DB}+2T_{2\DB})
          -\sq{\frac{3}{2}}P_\DB
          -\sq{\frac{3}{2}}PE_\DB
          -\frac{1}{\sq6}(P_{1EW\DB}+P_{2EW\DB})
  \non\\
      &&+\sq{\frac{3}{2}} E_\DB,  %
\label{eq: DBB0, DS=0}
\en
\be
A(\bar B^0_s\to \Delta^+\overline{\Sigma^{+}})
   &=&-\sq2 T_{2\DB}
          -\sq2 P_\DB
          -\sq2 PE_\DB
          -\frac{2\sq2}{3}P_{2EW\DB},%
   \non\\
A(\bar B^0_s\to\Delta^0\overline{\Sigma^{0}})
   &=&T_{2\DB}
         +2 P_\DB
         +2 PE_\DB
         +\frac{1}{3}P_{2EW\DB},%
   \non\\
A(\bar B^0_s\to\Delta^-\overline{\Sigma^{-}})
   &=&\sq6 P_\DB
         +\sq6 PE_\DB
          -\sq{\frac{2}{3}}P_{2EW\DB},%
   \non\\
A(\bar B^0_s\to\Sigma^{*0}\overline{\Xi^{0}})
   &=&-(T_{1\DB}+T_{2\DB})
          -P_\DB
          -PE_\DB
          -\frac{1}{3}(3P_{1EW\DB}+2P_{2EW\DB}),%
   \non\\
A(\bar B^0_s\to\Sigma^{*-}\overline{\Xi^{-}})
   &=&\sq2 P_\DB
        +\sq2 PE_\DB
          -\frac{\sq2}{3}P_{2EW\DB},%
   \non\\   
A(\bar B^0_s\to\Delta^0\overline{\Lambda})
   &=&-\frac{1}{\sq3}(2T_{1\DB}+T_{2\DB})
          -\frac{1}{\sq3}(2P_{1EW\DB}+P_{2EW\DB}),%
\label{eq: DBBs, DS=0}
\en
and
\be
A(B^-_c\to\Delta^0\overline{p})
   &=&
          -\sq2 A^c_\DB,
   \non\\
A(B^-_c\to\Delta^-\overline{n})
   &=&
         -\sq6 A^c_\DB,
   \non\\
A(B^-_c\to\Sigma^{*0}\overline{\Sigma^{+}})
   &=&
          A^c_\DB,
   \non\\   
A(B^-_c\to\Sigma^{*-}\overline{\Sigma^{0}})
   &=&
          - A^c_\DB,
   \non\\
A(B^-_c\to\Xi^{*-}\overline{\Xi^{0}})
   &=&
         \sq2A^c_\DB,
   \non\\
A(B^-_c\to\Sigma^{*-}\overline{\Lambda})
   &=&
          \sq3 A^c_\DB,
\label{eq: DBBc, DS=0}
\en
while those for $\Delta S=1$ transitions are given by
\be
A(B^-\to\Sigma^{*0}\overline{p})
   &=&T'_{1\DB}- P'_\DB- PE'_\DB
          +\frac{1}{3}(3P'_{1EW\DB}+P'_{2EW\DB})
          -A'_\DB,%
   \non\\
A(B^-\to\Sigma^{*-}\overline{n})
   &=&-\sq2 P'_\DB-\sq2 PE'_\DB
          +\frac{\sq2}{3}P'_{2EW\DB}
          -\sq2A'_\DB,%
   \non\\
A(B^-\to\Xi^{*0}\overline{\Sigma^{+}})
   &=&-\sq2T'_{1\DB}
         +\sq2 P'_\DB+\sq2 PE'_\DB
         -\frac{\sq2}{3}(3P'_{1EW\DB}+P'_{2EW\DB})
         +\sq2 A'_\DB,%
   \non\\
A(B^-\to\Xi^{*-}\overline{\Sigma^{0}})
   &=&-P'_\DB-PE'_\DB
          +\frac{1}{3}P'_{2EW\DB}
          -A'_\DB,%
   \non\\   
A(B^-\to\Omega^-\overline{\Xi^{0}})
   &=&\sq6 P'_\DB+\sq6 PE'_\DB
          -\sq{\frac{2}{3}}P'_{2EW\DB}
          +\sq6 A'_\DB,%
   \non\\
A(B^-\to\Xi^{*-}\overline{\Lambda})
   &=&\sq3 P'_\DB+\sq3 PE'_\DB
          -\frac{1}{\sq3}P'_{2EW\DB}
          +\sq3A'_\DB, %
\label{eq: DBBm, DS=-1}
\en
\be
A(\bar B^0\to \Sigma^{*+}\overline{p})
   &=&\sq2 T'_{2\DB}
         +\sq2 P'_\DB+\sq2 PE'_\DB
         +\frac{2\sq2}{3}P'_{2EW\DB},%
   \non\\
A(\bar B^0\to\Sigma^{*0}\overline{n})
   &=&T'_{1\DB}+T'_{2\DB}
          +P'_\DB+PE'_\DB
          +\frac{1}{3}(3P'_{1EW\DB}+2P'_{2EW\DB}),%
   \non\\
A(\bar B^0\to\Xi^{*0}\overline{\Sigma^{0}})
   &=&T'_{1\DB}
          -P'_\DB-PE'_\DB
          +\frac{1}{3}(3P'_{1EW\DB}+P'_{2EW\DB}),%
   \non\\
A(\bar B^0\to\Xi^{*-}\overline{\Sigma^{-}})
   &=&-\sq2 P'_\DB-\sq2 PE'_\DB
          +\frac{\sq2}{3}P'_{2EW\DB},%
   \non\\   
A(\bar B^0\to\Omega^-\overline{\Xi^{-}})
   &=&-\sq6 P'_\DB-\sq6 PE'_\DB
          +\sq{\frac{2}{3}}P_{2EW\DB},%
    \non\\
A(\bar B^0\to\Xi^{*0}\overline{\Lambda})
   &=&-\frac{1}{\sq3}(T'_{1\DB}+2T'_{2\DB})
          -\sq3 P'_\DB-\sq3 PE'_\DB
          -\frac{1}{\sq3}(P'_{1EW\DB}+P'_{2EW\DB}),%
    \non\\
\label{eq: DBB0, DS=-1}
\en
\be
A(\bar B^0_s\to\Delta^+\overline{p})
   &=&-\sq2 E'_\DB, %
   \non\\
A(\bar B^0_s\to\Delta^0\overline{n})
   &=&-\sq2E'_\DB, %
   \non\\
A(\bar B^0_s\to\Sigma^{*+}\overline{\Sigma^{+}})
   &=&-\sq2T'_{2\DB}
          -\sq2P'_\DB -\sq2PE'_\DB
          -\frac{2\sq2}{3}P'_{2EW\DB}
          +\sq2 E'_\DB, %
   \non\\
A(\bar B^0_s\to\Sigma^{*0}\overline{\Sigma^{0}})
   &=& \frac{1}{\sq2}T'_{2\DB}
         +\sq2P'_\DB+\sq2PE'_\DB
         +\frac{1}{3\sq2}P'_{2EW\DB}
         -\frac{1}{\sq2} E'_\DB, %
   \non\\
A(\bar B^0_s\to\Sigma^{*-}\overline{\Sigma^{-}})
   &=&\sq2P'_\DB+\sq2PE'_\DB
          -\frac{\sq2}{3}P'_{2EW\DB}, %
   \non\\   
A(\bar B^0_s\to\Xi^{*0}\overline{\Xi^{0}})
   &=&-\sq2(T'_{1\DB}+T'_{2\DB})
          -\sq2P'_\DB  -\sq2PE'_\DB
          -\frac{\sq2}{3}(3P'_{1EW\DB}+2P'_{2EW\DB})
    \non\\      
        &&+\sq2 E'_\DB, %
   \non\\
A(\bar B^0_s\to\Xi^{*-}\overline{\Xi^{-}})
   &=&\sq2 P'_\DB+\sq2 PE'_\DB
         -\frac{\sq2}{3}P'_{2EW\DB}, %
   \non\\
A(\bar B^0_s\to\Sigma^{*0}\overline{\Lambda})
   &=&-\frac{1}{\sq6}(2T'_{1\DB}+T'_{2\DB})
          -\frac{1}{\sq6}(2P'_{1EW\DB}+P'_{2EW\DB})
          +\sq{\frac{3}{2}} E'_\DB, %
\label{eq: DBBs, DS=-1}
\en
and
\be
A(B^-_c\to\Sigma^{*0}\overline{p})
   &=&          -A^{\prime c}_\DB,
   \non\\
A(B^-_c\to\Sigma^{*-}\overline{n})
   &=&          -\sq2A^{\prime c}_\DB,
   \non\\
A(B^-_c\to\Xi^{*0}\overline{\Sigma^{+}})
   &=&         +\sq2 A^{\prime c}_\DB,
   \non\\
A(B^-_c\to\Xi^{*-}\overline{\Sigma^{0}})
   &=&          -A^{\prime c}_\DB,
   \non\\   
A(B^-_c\to\Omega^-\overline{\Xi^{0}})
   &=&          \sq6 A^{\prime c}_\DB,
   \non\\
A(B^-_c\to\Xi^{*-}\overline{\Lambda})
   &=&          \sq3A^{\prime c}_\DB.   
\label{eq: DBBc, DS=-1}
\en

\subsection{$\overline B$ to decuplet-anti-decuplet baryonic decays}

The full $\bar B_{u,d,s,c}\to\DD$ decay amplitudes for $\Delta S=0$ processes are given by
\be
A(B^-\to \Delta^+\overline{\Delta^{++}})
   &=&2\sq3 T_\DD+2\sq3 P_\DD+2\sq3 PE_\DD+\frac{4}{\sq3}P_{EW\DD}+2\sq3A_\DD,%
   \non\\
A(B^-\to\Delta^0\overline{\Delta^+})
   &=&2T_\DD+4 P_\DD+4 PE_\DD+\frac{2}{3}P_{EW\DD}+4A_\DD,%
   \non\\
A(B^-\to\Delta^-\overline{\Delta^0})
   &=&2\sq3 P_\DD+2\sq3 PE_\DD-\frac{2}{\sq3}P_{EW\DD}+2\sq3A_\DD,%
   \non\\
A(B^-\to\Sigma^{*0}\overline{\Sigma^{*+}})
   &=&\sq2T_\DD+2\sq2 P_\DD+2\sq2 PE_\DD+\frac{\sq2}{3}P_{EW\DD}+2\sq2A_\DD,%
   \non\\   
A(B^-\to\Sigma^{*-}\overline{\Sigma^{*0}})
   &=&2\sq2 P_\DD+2\sq2 PE_\DD-\frac{2\sq2}{3}P_{EW\DD}+2\sq2A_\DD,%
   \non\\
A(B^-\to\Xi^{*-}\overline{\Xi^{*0}})
   &=&2 P_\DD+2 PE_\DD-\frac{2}{3}P_{EW\DD}+2A_\DD,%
\label{eq: DDBm, DS=0}
\en
\be
A(\bar B^0\to \Delta^{++}\overline{\Delta^{++}})
   &=&6E_\DD+6PA_\DD,%
   \non\\
A(\bar B^0\to\Delta^+\overline{\Delta^+})
   &=&2T_\DD+2 P_\DD+2 PE_\DD+\frac{4}{3}P_{EW\DD}+4E_\DD+6PA_\DD,%
   \non\\
A(\bar B^0\to\Delta^0\overline{\Delta^0})
   &=&2T_\DD+4 P_\DD+4 PE_\DD+\frac{2}{3}P_{EW\DD}+2E_\DD+6PA_\DD,%
   \non\\
A(\bar B^0\to\Delta^-\overline{\Delta^-})
   &=&6P_\DD+6PE_\DD-2P_{EW\DD}+6PA_\DD,%
   \non\\
A(\bar B^0\to\Sigma^{*+}\overline{\Sigma^{*+}})
   &=&4E_\DD+6PA_\DD,%
   \non\\
A(\bar B^0\to\Sigma^{*0}\overline{\Sigma^{*0}})
   &=&T_\DD+2P_\DD+2PE_\DD+\frac{1}{3}P_{EW\DD}+2E_\DD+6PA_\DD,%
   \non\\
A(\bar B^0\to\Sigma^{*-}\overline{\Sigma^{*-}})
   &=&4P_\DD+4PE_\DD-\frac{4}{3}P_{EW\DD}+6PA_\DD,%
   \non\\   
A(\bar B^0\to\Xi^{*0}\overline{\Xi^{*0}})
   &=&2E_{\DD}+6PA_\DD,%
   \non\\
A(\bar B^0\to\Xi^{*-}\overline{\Xi^{*-}})
   &=&2 P_\DD+2 PE_\DD-\frac{2}{3}P_{EW\DD}+6PA_\DD,%
   \non\\
A(\bar B^0\to\Omega^{-}\overline{\Omega^{-}})
   &=&6PA_\DD,   %
\label{eq: DDB0, DS=0}
\en
\be
A(\bar B^0_s\to \Delta^+\overline{\Sigma^{*+}})
   &=&2 T_\DD+2 P_\DD+2 PE_\DD+\frac{4}{3}P_{EW\DD},%
   \non\\
A(\bar B^0_s\to\Delta^0\overline{\Sigma^{*0}})
   &=&\sq2T_\DD+2\sq2 P_\DD+2\sq2 PE_\DD+\frac{\sq2}{3}P_{EW\DD},%
   \non\\
A(\bar B^0_s\to\Delta^-\overline{\Sigma^{*-}})
   &=&2\sq3 P_\DD+2\sq3 PE_\DD-\frac{2}{\sq3}P_{EW\DD},%
   \non\\
A(\bar B^0_s\to\Sigma^{*0}\overline{\Xi^{*0}})
   &=&\sq2T_\DD+2\sq2 P_\DD+2\sq2 PE_\DD+\frac{\sq2}{3}P_{EW\DD},%
   \non\\
A(\bar B^0_s\to\Sigma^{*-}\overline{\Xi^{*-}})
   &=&4P_\DD+4PE_\DD-\frac{4}{3}P_{EW\DD},%
   \non\\   
A(\bar B^0_s\to\Xi^{*-}\overline{\Omega^-})
   &=&2\sq3 P_\DD+2\sq3 PE_\DD-\frac{2}{\sq3}P_{EW\DD},%
\label{eq: DDBs, DS=0}
\en
and
\be
A(B^-_c\to \Delta^+\overline{\Delta^{++}})
   &=&2\sq3A^c_\DD,
   \non\\
A(B^-_c\to\Delta^0\overline{\Delta^+})
   &=&4A^c_\DD,
   \non\\
A(B^-_c\to\Delta^-\overline{\Delta^0})
   &=&2\sq3A^c_\DD,
   \non\\
A(B^-_c\to\Sigma^{*0}\overline{\Sigma^{*+}})
   &=&2\sq2A^c_\DD,
   \non\\   
A(B^-_c\to\Sigma^{*-}\overline{\Sigma^{*0}})
   &=&2\sq2A^c_\DD,
   \non\\
A(B^-_c\to\Xi^{*-}\overline{\Xi^{*0}})
   &=&2A^c_\DD,
\label{eq: DDBc, DS=0}
\en
while those for $\Delta S=1$ transitions are given by
\be
A(B^-\to \Sigma^{*+}\overline{\Delta^{++}})
   &=&2\sq3 T'_\DD+2\sq3 P'_\DD+2\sq3 PE'_\DD+\frac{4}{\sq3}P'_{EW\DD}+2{\sq3}A'_\DD,%
   \non\\
A(B^-\to\Sigma^{*0}\overline{\Delta^+})
   &=&\sq2T'_\DD+2\sq2 P'_\DD+2\sq2 PE'_\DD+\frac{\sq2}{3}P'_{EW\DD}+2{\sq2}A'_\DD,%
   \non\\
A(B^-\to\Sigma^{*-}\overline{\Delta^0})
   &=&2 P'_\DD+2 PE'_\DD-\frac{2}{3}P'_{EW\DD}+2A'_\DD,%
   \non\\
A(B^-\to\Xi^{*0}\overline{\Sigma^{*+}})
   &=&2T'_\DD+4 P'_\DD+4 PE'_\DD+\frac{2}{3}P'_{EW\DD}+4A'_\DD,%
   \non\\
A(B^-\to\Xi^{*-}\overline{\Sigma^{*0}})
   &=&2\sq2 P'_\DD+2\sq2 PE'_\DD-\frac{2\sq2}{3}P'_{EW\DD}+2{\sq2}A'_\DD,%
   \non\\   
A(B^-\to\Omega^-\overline{\Xi^{*0}})
   &=&2\sq3 P'_\DD+2\sq3 PE'_\DD-\frac{2}{\sq3}P'_{EW\DD}+2{\sq3}A'_\DD,%
\label{eq: DDBm, DS=-1}
\en
\be
A(\bar B^0\to \Sigma^{*+}\overline{\Delta^+})
   &=&2 T'_\DD+2 P'_\DD+2 PE'_\DD+\frac{4}{3}P'_{EW\DD},%
   \non\\
A(\bar B^0\to\Sigma^{*0}\overline{\Delta^0})
   &=&\sq2T'_\DD+2\sq2 P'_\DD+2\sq2 PE'_\DD+\frac{\sq2}{3}P'_{EW\DD},%
   \non\\
A(\bar B^0\to\Sigma^{*-}\overline{\Delta^-})
   &=&2\sq3 P'_\DD+2\sq3 PE'_\DD-\frac{2}{\sq3}P'_{EW\DD},%
   \non\\
A(\bar B^0\to\Xi^{*0}\overline{\Sigma^{*0}})
   &=&\sq2T'_\DD+2\sq2 P'_\DD+2\sq2 PE'_\DD+\frac{\sq2}{3}P'_{EW\DD},%
   \non\\
A(\bar B^0\to\Xi^{*-}\overline{\Sigma^{*-}})
   &=&4P'_\DD+4PE'_\DD-\frac{4}{3}P'_{EW\DD},%
   \non\\   
A(\bar B^0\to\Omega^-\overline{\Xi^{*-}})
   &=&2\sq3 P'_\DD+2\sq3 PE'_\DD-\frac{2}{\sq3}P'_{EW\DD},%
\label{eq: DDB0, DS=-1}
\en
\be
A(\bar B^0_s\to \Delta^{++}\overline{\Delta^{++}})
   &=&6E'_\DD+6PA'_\DD, %
   \non\\
A(\bar B^0_s\to\Delta^+\overline{\Delta^+})
   &=&4E'_\DD+6PA'_\DD, %
   \non\\
A(\bar B^0_s\to\Delta^0\overline{\Delta^0})
   &=&2E'_\DD+6PA'_\DD, %
   \non\\
A(\bar B^0_s\to\Delta^-\overline{\Delta^-})
   &=&6PA'_\DD, %
   \non\\
A(\bar B^0_s\to\Sigma^{*+}\overline{\Sigma^{*+}})
   &=&2T'_\DD+2P'_\DD+2PE'_\DD+\frac{4}{3}P'_{EW\DD}+4E'_\DD+6PA'_\DD,%
   \non\\
A(\bar B^0_s\to\Sigma^{*0}\overline{\Sigma^{*0}})
   &=&T'_\DD+2P'_\DD+2PE'_\DD+\frac{1}{3}P'_{EW\DD}+2E'_\DD+6PA'_\DD,%
   \non\\
A(\bar B^0_s\to\Sigma^{*-}\overline{\Sigma^{*-}})
   &=&2P'_\DD+2PE'_\DD-\frac{2}{3}P'_{EW\DD}+6PA'_\DD,%
   \non\\   
A(\bar B^0_s\to\Xi^{*0}\overline{\Xi^{*0}})
   &=&2T'_\DD+4P'_\DD+4PE'_\DD+\frac{2}{3}P'_{EW\DD}+2E'_{\DD}+6PA'_\DD,%
   \non\\
A(\bar B^0_s\to\Xi^{*-}\overline{\Xi^{*-}})
   &=&4 P'_\DD+4 PE'_\DD-\frac{4}{3}P'_{EW\DD}+6PA'_\DD,%
   \non\\
A(\bar B^0_s\to\Omega^{-}\overline{\Omega^{-}})
   &=&6 P_\DD+6 PE_\DD-2P_{EW\DD}+6PA'_\DD,%
\label{eq: DDBs, DS=-1}
\en
and
\be
A(B^-_c\to \Sigma^{*+}\overline{\Delta^{++}})
   &=&2{\sq3}A^{\prime c}_\DD,
   \non\\
A(B^-_c\to\Sigma^{*0}\overline{\Delta^+})
   &=&2{\sq2}A^{\prime c}_\DD,
   \non\\
A(B^-_c\to\Sigma^{*-}\overline{\Delta^0})
   &=&2A^{\prime c}_\DD,
   \non\\
A(B^-_c\to\Xi^{*0}\overline{\Sigma^{*+}})
   &=&4A^{\prime c}_\DD,
   \non\\
A(B^-_c\to\Xi^{*-}\overline{\Sigma^{*0}})
   &=&2{\sq2}A^{\prime c}_\DD,
   \non\\   
A(B^-_c\to\Omega^-\overline{\Xi^{*0}})
   &=&2{\sq3}A^{\prime c}_\DD.
\label{eq: DDBc, DS=-1}
\en


\section{Formulas for decay rates and asymptotic relations for $\la \bfBB'|(\bar q q')_{S,P}|0\ra$}\label{App: asym}

The decay $\bar B\to \BB, \BD, \DB$ and $\DD$ decay have the following forms~\cite{Jarfi:1990ej}
\begin{eqnarray}
A(\overline B\to {\cal B}_1 \overline {\cal B}_2)&=&\bar
u_1(A_{{\cal B}\overline {\cal B}}+\gamma_5 B_{{\cal B}\overline
{\cal B}}) v_2,
\nonumber\\
A(\overline B\to {\cal D}_1 \overline {\cal B}_2)&=&i
\frac{q^\mu}{m_B} \bar u^\mu_1(A_{{\cal D}\overline {\cal
B}}+\gamma_5 B_{{\cal D}\overline {\cal B}}) v_2,
\nonumber\\
A(\overline B\to {\cal B}_1 \overline {\cal D}_2)&=&i
\frac{q^\mu}{m_B}\bar u_1(A_{{\cal B}\overline {\cal D}}+\gamma_5
B_{{\cal B}\overline {\cal D}}) v^\mu_2,
\nonumber\\
A(\overline B\to {\cal D}_1 \overline {\cal D}_2)&=&\bar
u^\mu_1(A_{{\cal D}\overline {\cal D}}+\gamma_5 B_{{\cal
D}\overline {\cal D}}) v_{2\mu}+\frac{q^\mu q^\nu}{m^2_B}\bar
u^\mu_1(C_{{\cal D}\overline {\cal D}}+\gamma_5 D_{{\cal
D}\overline {\cal D}}) v_{2\nu},
\label{eq: A1}
\end{eqnarray}
where $q=p_1-p_2$ is the difference of the momenta of the baryons and $u^\mu,\,v^\mu$ are the Rarita-Schwinger vector spinors,~\cite{Moroi:1995fs}
\be
u_\mu(\pm\frac{3}{2})&=&\epsilon_\mu(\pm1) u(\pm\frac{1}{2})
\non\\
u_\mu(\pm\frac{1}{2})&=&(\epsilon_\mu(\pm1)
u(\mp\frac{1}{2})+\sqrt{2}\,\epsilon_\mu(0)
u(\pm\frac{1}{2}))/\sqrt3,
\en 
with $\epsilon_\mu(\lambda)$ the polarization vector.
Using
\be
q\cdot\epsilon(\lambda)_{1,2}&=&\mp\,\delta_{\lambda,0}\,m_B
p_c/m_{1,2},
\non\\
\epsilon^*_1(0)\cdot\epsilon_2(0)&=&(m_B^2-m^2_1-m^2_2)/2m_1 m_2,
\en
with $p_c$ the baryon momentum in the center of mass frame
and the fact that $\epsilon^*_1(0)\cdot\epsilon_2(0)$
is the largest product among the scalar products of $\epsilon^*_1(\lambda_1)$ and 
$\epsilon_2(\lambda_2)$, the last three amplitudes in Eq. (\ref{eq: A1}) can be expressed or approximated as
\begin{eqnarray}
A(\overline B\to {\cal D}_1 \overline {\cal B}_2)&=&-i
\sqrt{\frac{2}{3}}\frac{p_{cm}}{m_1} \bar u_1(A_{{\cal D}\overline
{\cal B}}+\gamma_5 B_{{\cal D}\overline {\cal B}}) v_2,
\nonumber\\
A(\overline B\to {\cal B}_1 \overline {\cal D}_2)&=&i
\sqrt{\frac{2}{3}}\frac{p_{cm}}{m_2}\bar u_1(A_{{\cal B}\overline
{\cal D}}+\gamma_5 B_{{\cal B}\overline {\cal D}}) v_2,
\nonumber\\
A(\overline B\to {\cal D}_1 \overline {\cal
D}_2)&\simeq&\frac{m_B^2}{3m_1m_2}\bar u_1(A^\prime_{{\cal
D}\overline {\cal D}}+\gamma_5 B^\prime_{{\cal D}\overline {\cal
D}})v_2, \label{eq:largemB}
\end{eqnarray}
where 
\be
A^\prime_{{\cal D}\overline {\cal D}}=A_{{\cal D}\overline
{\cal D}}-2(p_{cm}/m_B)^2 C_{{\cal D}\overline {\cal D}},
\quad
B^\prime_{{\cal D}\overline {\cal D}}=B_{{\cal D}\overline {\cal
D}}-2(p_{cm}/m_B)^2 D_{{\cal D}\overline {\cal D}}.
\en
Hence decay modes with decuplets (or anti-decuplets) are only in or dominantly in the $\pm\frac{1}{2}$-helicity states. 

All $\overline B\to {\cal B} \overline {\cal B}$,
${\cal D} \overline {\cal B}$, ${\cal B} \overline {\cal D}$,
${\cal D} \overline {\cal D}$ decay amplitudes can be effectively expressed as
\begin{equation}
A(\overline B\to {\mathbf B}_1\overline {\mathbf B}_2)
 =\bar u_1({\bf A}+\gamma_5 {\bf B})v_2,
 \label{eq:asymptoticform}
\end{equation}
and it is straightforward to obtain the decay rates giving
\be
\Gamma(\overline B\to {\mathbf B}_1\overline {\mathbf B}_2)
=\frac{p_{cm}}{8 \pi m_B^2} [(2 m_B^2 - 2 (m_{{\mathbf B}_1}+ m_{{\mathbf B}_1})^2) {\bf A}^2+
(2 m_B^2 - 2 (m_{{\mathbf B}_1}- m_{{\mathbf B}_1})^2) {\bf B}^2].
\en

We now change to the discussion of finding the asymptotic relations for form factors of scalar and pseudo-scalar density matrix elements, 
$\la \bfBB'|(\bar q q')_{S,P}|0\ra$.
We follow ref.~\cite{Brodsky:1980sx} to obtain the asymptotic relations.
The wave function of a octet or decuplet baryon with helicity $\lambda=-1/2$ can be expressed as 
\begin{equation}
|{\mathbf B}\,;\downarrow\rangle\sim
\frac{1}{\sqrt3}(|{\mathbf B}\,;\downarrow\uparrow\downarrow\rangle
                +|{\mathbf B}\,;\downarrow\downarrow\uparrow\rangle
                +|{\mathbf B}\,;\uparrow\downarrow\downarrow\rangle), 
\end{equation}
which are composed of 13-, 12- and 23-symmetric terms,
respectively.
For octet baryons, 
we have
\begin{eqnarray}
|p\,;\downarrow\uparrow\downarrow\rangle&=&
\left[\frac{d(1)u(3)+u(1)d(3)}{\sqrt6} u(2)
 -\sqrt{\frac{2}{3}} u(1)d(2)u(3)\right]
|\downarrow\uparrow\downarrow\rangle,
\nonumber\\
|n\,;\downarrow\uparrow\downarrow\rangle&=&
(-|p\,;\downarrow\uparrow\downarrow\rangle
\,\,{\rm with}\,\,\,u \leftrightarrow d),
\nonumber\\
|\Sigma^+\,;\downarrow\uparrow\downarrow\rangle&=&
(-|p\,;\downarrow\uparrow\downarrow\rangle
\,\,{\rm with}\,\,d \rightarrow s),
\nonumber\\
|\Sigma^0\,;\downarrow\uparrow\downarrow\rangle&=&
\bigg[-\frac{u(1)d(3)+d(1)u(3)}{\sqrt3}\,s(2)
      +\frac{u(2)d(3)+d(2)u(3)}{2\sqrt3}\,s(1)
\nonumber\\
      &&\,\,+\frac{u(1)d(2)+d(1)u(2)}{2\sqrt3}\,s(3)\bigg]
|\downarrow\uparrow\downarrow\rangle,
\nonumber\\
|\Sigma^-\,;\downarrow\uparrow\downarrow\rangle&=&
(|p\,;\downarrow\uparrow\downarrow\rangle
\,\,{\rm with}\,\,u, d\rightarrow d,s),
\nonumber\\
|\Lambda\,;\downarrow\uparrow\downarrow\rangle&=&
\bigg[\frac{d(2)u(3)-u(2)d(3)}{2}\,s(1)
      +\frac{u(1)d(2)-d(1)u(2)}{2}\,s(3)\bigg]
|\downarrow\uparrow\downarrow\rangle,
\non\\
|\Xi^0\,;\downarrow\uparrow\downarrow\rangle&=&
(|p\,;\downarrow\uparrow\downarrow\rangle
\,\,{\rm with}\,\,u, d\rightarrow s, u),
\nonumber\\
|\Xi^-\,;\downarrow\uparrow\downarrow\rangle&=&
(-|p\,;\downarrow\uparrow\downarrow\rangle
\,\,{\rm with}\,\,u\rightarrow s),
\en
and for decuplet baryons, we have
\be
|\Delta^{++};\downarrow\uparrow\downarrow\rangle&=&u(1)u(2)u(3)|\downarrow\uparrow\downarrow\rangle,\qquad\qquad
|\Delta^{-};\downarrow\uparrow\downarrow\rangle=d(1)d(2)d(3)|\downarrow\uparrow\downarrow\rangle,
\nonumber\\
|\Delta^{+};\downarrow\uparrow\downarrow\rangle&=&
\frac{1}{\sqrt3}[u(1)u(2)d(3)+u(1)d(2)u(3)+d(1)u(2)u(3)]|\downarrow\uparrow\downarrow\rangle,
\nonumber\\
|\Delta^{0};\downarrow\uparrow\downarrow\rangle&=&(|\Delta^{+};\downarrow\uparrow\downarrow\rangle\,\,{\rm
with}\,\,u \leftrightarrow d),\qquad
|\Sigma^{*+};\downarrow\uparrow\downarrow\rangle=(|\Delta^{+};\downarrow\uparrow\downarrow\rangle\,\,{\rm
with}\,\,d \leftrightarrow s),
\nonumber\\
|\Sigma^{*0};\downarrow\uparrow\downarrow\rangle&=&\frac{1}{\sqrt6}[u(1)d(2)s(3)+{\rm
permutation}]|\downarrow\uparrow\downarrow\rangle,
\non\\
|\Omega^-;\downarrow\uparrow\downarrow\rangle&=&(|\Delta^{++};\downarrow\uparrow\downarrow\rangle
\,\,{\rm with}\,\,u\rightarrow s),
\end{eqnarray}
for the $|{\mathbf B}\,;\downarrow\uparrow\downarrow\rangle$ parts.
while the 12- and 23-symmetric parts can be easily obtained by suitable permutation.

\begin{table}[t!]
\caption{\label{tab: eAeP} 
The coefficients $e^{-}_{\parallel}$ for various $\la {\mathbf B}|\bar q_R q'_L|{\mathbf B}'\ra$ in $A^{(\prime)}$ of $\bar B_{u}\to\bfBB'$ decays and $E^{(\prime)}$ of $\bar B_{d(s)}\to\bfBB'$ decays.
}
\begin{ruledtabular}
\begin{tabular}{cccc}
$\la {\mathbf B}|\bar q_R q'_L|{\mathbf B}'\ra$
          & $e^{-}_{\parallel}$ in $A^{(\prime)}$
          & $\la {\mathbf B}|\bar q_R q'_L|{\mathbf B}'\ra$
          & $e^{-}_{\parallel}$ in $E^{(\prime)}$
          \\
\hline $\la\Delta^0|\bar d_R u_L|\Delta^+\ra$
          & $4$%
          & $\la\Sigma^{*+}|\bar u_R u_L|\Sigma^{*+}\ra$
          & $4$ %
          \\
$\la\Sigma^{*0}|\bar d_R u_L|\Sigma^{*+}\ra$
          & $2\sqrt 2$
          & $\la\Sigma^{*0}|\bar u_R u_L|\Sigma^{*0}\ra$
          & $2$%
          \\
$\la\Sigma^{*0}|\bar s_R u_L|\Delta^{+}\ra$
         & $2\sqrt 2$%
         & $\la\Xi^{*0}|\bar u_R u_L|\Xi^{*0}\ra$
         & 2%
          \\          
\hline $\la p|\bar d_R u_L|\Delta^{++}\ra$
          & $\sqrt{6}$%
          & $\la \Lambda|\bar u_R u_L|\Sigma^{*0}\ra$
          & $\sqrt{\frac{3}{2}}$%
          \\
$\la n|\bar d_R u_L|\Delta^{+}\ra$
          & $\sqrt2$%
          & $\la p|\bar u_R u_L|\Delta^+\ra$
          & $-\sqrt2$%
          \\
$\la\Sigma^{0}|\bar s_R u_L|\Delta^{+}\ra$
            & 2%
            & $\la \Sigma^0|\bar u_R u_L|\Sigma^{*0}\ra$
            & $-\frac{1}{\sqrt2}$%
             \\          
\hline $\la\Delta^0|\bar d_R u_L|p\ra$
          & $-\sqrt2$%
          & $\la \Delta^+|\bar u_R u_L|p\ra$
          & $-\sqrt2$%
          \\
$\la\Delta^-|\bar d_R u_L|n\ra$
          & $-\sqrt 6$%
          & $\la\Sigma^{*0}|\bar u_R u_L|\Lambda\ra$
          & $\sqrt{\frac{3}{2}}$%
          \\
$\la\Sigma^{*0}|\bar s_R u_L|p\ra$
          & $-1$%
          & $\la\Xi^{*0}|\bar u_R u_L|\Xi^0\ra$
          & $\sqrt2$%
          \\
\hline $\la n|\bar d_R u_L|p\ra$
          & $-5$%
          & $\la p|\bar u_R u_L|p\ra$
          & $-4$%
          \\
$\la\Lambda|\bar s_R u_L|p\ra$
          & $3\sqrt{\frac{3}{2}}$%
          & $\la \Lambda|\bar u_R u_L|\Sigma^0\ra$
          & $-\sqrt3$%
          \\
$\la\Sigma^0|\bar s_R u_L|p\ra$
          & $-\frac{1}{\sqrt2}$
          & $\la \Lambda|\bar u_R u_L|\Lambda\ra$
          & $0$
          \\
\end{tabular}
\end{ruledtabular}
\end{table}

\begin{table}[t!]
\caption{\label{tab: ePE} The coefficients $e^{-}_{\parallel}$ for various $\la {\mathbf B}|\bar q_R q'_L|{\mathbf B}'\ra$ in $PE^{(\prime)}$ of $\bar B_{q'}\to \bfBB'$ decays.
}
\begin{ruledtabular}
\begin{tabular}{cccccccc}
$\la {\mathbf B}|\bar q_R q'_L|{\mathbf B}'\ra$
          & $e^{-}_{\parallel}$ in $PE^{(\prime)}$
          & $\la {\mathbf B}|\bar q_R q'_L|{\mathbf B}'\ra$
          & $e^{-}_{\parallel}$ in $PE^{(\prime)}$
          \\
\hline  $\la\Delta^-|\bar d_R u_L|\Delta^0\ra$
          & $2\sqrt3$
          & $\la \Sigma^{*0}|\bar d_R d_L|\Sigma^{*0}\ra$
          & $2$
          \\
$\la\Delta^0|\bar d_R u_L|\Delta^+\ra$
          & $4$
          & $\la\Sigma^{*0}|\bar s_R u_L|\Delta^{+}\ra$
          & $2\sqrt2$
          \\
\hline $\la p|\bar d_R u_L|\Delta^{++}\ra$
          & $\sqrt{6}$
          & $\la n|\bar d_R u_L|\Delta^{+}\ra$
          & $\sqrt2$
          \\
$\la p|\bar d_R d_L|\Delta^+\ra$
          & $\sqrt2$
          & $\la\Lambda|\bar d_R d_L|\Sigma^{*0}\ra$
          & $-\sqrt{\frac{3}{2}}$
          \\
$\la p|\bar d_L s_R|\Sigma^{*+}\ra$
          & $\sqrt2$
          & $\la\Sigma^{0}|\bar d_R s_L|\Xi^{*0}\ra$
          & $-1$
          \\
\hline $\la\Delta^0|\bar d_R u_L|p\ra$
          & $-\sqrt2$
          & $\la\Delta^-|\bar d_R u_L|n\ra$
          & $-\sqrt{6}$
           \\
$\la\Delta^+|\bar d_R d_L|p\ra$
          & $\sqrt2$
          & $\la\Sigma^{*0}|\bar d_R d_L|\Lambda\ra$
          & $-\sqrt{\frac{3}{2}}$
          \\
$\la\Sigma^{*0}|\bar s_R u_L|p\ra$
          & $-1$
          & $\la\Sigma^{*0}|\bar s_R d_L|n\ra$
          & $1$
          \\
\hline $\la p|\bar d_R d_L|p\ra$
          & $1$
          & $\la n|\bar d_R u_L|p\ra$
          & $-5$
          \\
$\la n|\bar d_R d_L|n\ra$
          & $-4$
          & $\la\Sigma^0|\bar d_R d_L|\Lambda\ra$
          & $\sqrt3$
          \\
$\la\Sigma^0|\bar d_R d_L|\Sigma^0\ra$
          & $-2$
          & $\la \Lambda|\bar d_R d_L|\Sigma^0\ra$
          & $\sqrt3$
          \\
$\la\Lambda|\bar d_R d_L|\Lambda\ra$
          & $0$
          & $\la n| \bar d_R s_L|\Lambda\ra$
          & $3\sqrt{\frac{3}{2}}$
          \\
$\la\Sigma^0|\bar s_R u_L|p\ra$
          & $-\frac{1}{\sqrt2}$
          & $\la\Lambda|\bar s_R u_L|p\ra$
          & $3\sqrt{\frac{3}{2}}$
          \\
\end{tabular}
\end{ruledtabular}
\end{table}

From Eq. (\ref{eq: MM}), we see that the $\bar B_{q''}\to\bfBB'$ factorization amplitudes are related to the scalar and pseudo-scalar density matrix elements $\la \bfBB'|(\bar q q')_{S,P}|0\ra$. 
For example, we have
\be
\la\bfBB'| (\bar q q')_{V\mp A}|0\ra
\la 0| (\bar q'' b)_{V-A}|\bar B_{q''}\ra
&=&-i f_{B_{q''}}[(m_q-m_{\bar q'}) \la \bfBB'| (\bar q q')_{S}|0\ra
\non\\
&&\quad\hspace{18pt}
\mp (m_q+m_{\bar q'}) \la \bfBB'| (\bar q q')_{P}|0\ra].
\en
It is evident that each term in the above matrix element is proportional to light quark masses.
By neglecting higher order contributions from $m_{q}$ and $m_{q'}$, the quark mass dependence in $\la \bfBB'| (\bar q q')_{S,P}|0\ra$ can be ignored.

Following Ref.~\cite{Brodsky:1980sx}, we have
\begin{eqnarray}
\langle {\mathbf B}(p)| {\cal O}|{\mathbf B}^\prime(p^\prime)\rangle
&=&\bar u(p)\left[\frac{1+\gamma_5}{2}\,F^{+}(t)
                 +\frac{1-\gamma_5}{2}\,F^{-}(t)\right] u(p^\prime),
\nonumber\\
F^{\pm}(t)&=&e^{(\pm)}_{\parallel}
               ({\cal O}:{\mathbf B}^\prime\to{\mathbf B})
               \,F_{\parallel}(t),
\end{eqnarray}
in the large $t$ limit. 
For simplicity, we illustrate with the space-like case.
Coefficients of $F_{\parallel}$ for the
${\mathcal O}=\bar q_R q^\prime_L,\,\bar q_L q^\prime_R$ cases
are given by
\begin{eqnarray}
e^{-}_{\parallel}(\bar q_R q^\prime_L:{\mathbf B}^\prime\to{\mathbf B})
 &=&3(\langle {\mathbf B};\,\uparrow\uparrow\downarrow|
            O[q^\prime(1,\downarrow)\to q(1,\uparrow)]
            |{\mathbf B}^\prime\,;\downarrow\uparrow\downarrow\rangle
\nonumber\\
&&+\langle {\mathbf B};\,\downarrow\uparrow\uparrow|
            O[q^\prime(3,\downarrow)\to q(3,\uparrow)]
             |{\mathbf B}^\prime\,;\downarrow\uparrow\downarrow\rangle),
\nonumber\\
e^{+}_{\parallel}(\bar q_R q^\prime_L:{\mathbf B}^\prime\to{\mathbf B})
&=&0,
\nonumber\\
e^{\pm}_{\parallel}(\bar q_L q^\prime_R:{\mathbf B}^\prime\to{\mathbf B})
&=& 
       e^{\mp}_{\parallel}(\bar q_R q^\prime_L:{\mathbf B}^\prime\to{\mathbf B}),
\label{eq: e}
\end{eqnarray}
where the factor $3$ in the first line are introduced without lost of generality. 
Note that 
applying $O[q^\prime(1,\downarrow)\to q(1,\uparrow)]$ to $|{\mathbf B}';\downarrow\uparrow\downarrow\ra$ changes the 
parallel spin $q^\prime(1)|\downarrow\rangle$ part of 
$|{\mathbf B}^\prime;\downarrow\uparrow\downarrow\rangle$
to a $q(1)|\uparrow\rangle$ part and likewise for the operation of $O[q^\prime(3,\downarrow)\to q(1,\uparrow)]$ on $|{\mathbf B}';\downarrow\uparrow\downarrow\ra$. 
It is easy to see that flipping the anti-parallel spin
$|\uparrow\rangle$ part of 
$|{\mathbf B}^\prime;\downarrow\uparrow\downarrow\rangle$
to $|\downarrow\rangle$ will give a helicity $\lambda=-3/2$ state, 
where the transition amplitude is suppressed.
Hence we only need to consider the parallel spin case.

\begin{table}[t!]
\caption{\label{tab: ePA} The coefficients $e^{-}_{\parallel}$ for various $\la {\mathbf B}|(\bar u_R u_L,\bar d_R d_L,\bar s_R s_L)|{\mathbf B}'\ra$ in $PA^{(\prime)}$ matrix elements.
}
\begin{ruledtabular}
\begin{tabular}{cccccccc}
$\la {\mathbf B}|(\bar u_R u_L,\bar d_R d_L,\bar s_R s_L)|{\mathbf B}'\ra$
          & $e^{-}_{\parallel}$ in $PA^{(\prime)}$
          & $\la {\mathbf B}|(\bar u_R u_L,\bar d_R d_L,\bar s_R s_L)|{\mathbf B}'\ra$
          & $e^{-}_{\parallel}$ in $PA^{(\prime)}$
          \\
\hline $\la\Delta^{++}|(\bar u_R u_L,\bar d_R d_L,\bar s_R s_L)|\Delta^{++}\ra$
          & $(6,0,0)$
          & $\la\Delta^+|(\bar u_R u_L,\bar d_R d_L,\bar s_R s_L)|\Delta^+\ra$
          & $(4,2,0)$
          \\
$\la\Delta^0|(\bar u_R u_L,\bar d_R d_L,\bar s_R s_L)|\Delta^0\ra$
          & $(2,4,0)$
          & $\la\Delta^-|(\bar u_R u_L,\bar d_R d_L,\bar s_R s_L)|\Delta^-\ra$
          & $(0,6,0)$
          \\
$\la\Sigma^{*+}|(\bar u_R u_L,\bar d_R d_L,\bar s_R s_L)|\Sigma^{*+}\ra$
          & $(4,0,2)$
          & $\la\Sigma^{*0}|(\bar u_R u_L,\bar d_R d_L,\bar s_R s_L)|\Sigma^{*0}\ra$
          & $(2,2,2)$
          \\
$\la\Sigma^{*-}|(\bar u_R u_L,\bar d_R d_L,\bar s_R s_L)|\Sigma^{*-}\ra$
          & $(0,4,2)$
          & $\la\Xi^{*0}|(\bar u_R u_L,\bar d_R d_L,\bar s_R s_L)|\Xi^{*0}\ra$
          & $(2,0,4)$
          \\
$\la\Xi^{*-}|(\bar u_R u_L,\bar d_R d_L,\bar s_R s_L)|\Xi^{*-}\ra$
          & $(0,2,4)$
          & $\la\Omega^{-}|(\bar u_R u_L,\bar d_R d_L,\bar s_R s_L)|\Omega^{-}\ra$
          & $(0,0,6)$
          \\
\hline $\la p|(\bar u_R u_L,\bar d_R d_L,\bar s_R s_L)|p\ra$
          & $(4,-1,0)$
          & $\la n|(\bar u_R u_L,\bar d_R d_L,\bar s_R s_L)|n\ra$
          & $(-1,4,0)$
          \\
$\la \Sigma^+|(\bar u_R u_L,\bar d_R d_L,\bar s_R s_L)|\Sigma^+\ra$
          & $(4,0,-1)$
          & $\la\Sigma^0|(\bar u_R u_L,\bar d_R d_L,\bar s_R s_L)|\Sigma^0\ra$
          & $(2,2,-1)$
          \\
$\la\Sigma^-|(\bar u_R u_L,\bar d_R d_L,\bar s_R s_L)|\Sigma^-\ra$
          & $(0,4,-1)$
          & $\la \Lambda|(\bar u_R u_L,\bar d_R d_L,\bar s_R s_L)|\Lambda\ra$
          & $(0,0,1)$
          \\
$\la\Xi^0|(\bar u_R u_L,\bar d_R d_L,\bar s_R s_L)|\Xi^0\ra$
          & $(-1,0,4)$
          & $\la\Xi^-|(\bar u_R u_L,\bar d_R d_L,\bar s_R s_L)|\Xi^-\ra$
          & $(0,-1,4)$
          \\
\end{tabular}
\end{ruledtabular}
\end{table}

We can make use of Eq. (\ref{eq: e}) to obtain $e^\pm_\parallel$ for $\langle {\mathbf B}(p)| (\bar q q^\prime)_{S,P}|{\mathbf B}^\prime(p^\prime)\rangle$, 
\be
e^{\pm}_{\parallel}((\bar q q^\prime)_S:{\mathbf B}^\prime\to{\mathbf B})
&=&e^{\pm}_{\parallel}(\bar q_L q^\prime_R:{\mathbf B}^\prime\to{\mathbf B})
+e^{\pm}_{\parallel}(\bar q_R q^\prime_L:{\mathbf B}^\prime\to{\mathbf B})
\non\\
&=&e^{-}_{\parallel}(\bar q_R q^\prime_L:{\mathbf B}^\prime\to{\mathbf B}),
\nonumber\\
e^{\pm}_{\parallel}((\bar q q^\prime)_P:{\mathbf B}^\prime\to{\mathbf B})
&=&e^{\pm}_{\parallel}(\bar q_L q^\prime_R:{\mathbf B}^\prime\to{\mathbf B})
-e^{\pm}_{\parallel}(\bar q_R q^\prime_L:{\mathbf B}^\prime\to{\mathbf B})
\non\\
&=&\pm e^{-}_{\parallel}(\bar q_R q^\prime_L:{\mathbf B}^\prime\to{\mathbf B}),
\en
i.e. we have
\begin{eqnarray}
\langle {\mathbf B}(p)| (\bar q q^\prime)_S|{\mathbf B}^\prime(p^\prime)\rangle
&=&e^{-}_{\parallel}(\bar q_R q^\prime_L:{\mathbf B}^\prime\to{\mathbf B}) F_{\parallel}(t) \bar u(p) u(p^\prime),
\non\\
\langle {\mathbf B}(p)| (\bar q q^\prime)_P|{\mathbf B}^\prime(p^\prime)\rangle
&=&e^{-}_{\parallel}(\bar q_R q^\prime_L:{\mathbf B}^\prime\to{\mathbf B}) F_{\parallel}(t) \bar u(p)\gamma_5 u(p^\prime).
\end{eqnarray}
Therefore, the matrix elements are related with coefficients $e^{-}_{\parallel}(\bar q_R q^\prime_L:{\mathbf B}^\prime\to{\mathbf B})$.

The matrix elements $\langle {\mathbf B}(p)| (\bar q q^\prime)_{S,P}|{\mathbf B}^\prime(p^\prime)\rangle$ occur in topological amplitudes 
$A^{(\prime)}$, 
$E^{(\prime)}$,  
$PE^{(\prime)}$ and
$PA^{(\prime)}$
as shown in Eqs. (\ref{eq: TA=fac Delts S=0}) and (\ref{eq: TA=fac Delts S=-1}), with the help of equations of motion, Eq.~(\ref{eq: MM}).
By using Eq. (\ref{eq: e}), it is straightforward to obtain the coefficients $e^{-}_{\parallel}$ for various matrix elements $\la {\mathbf B}|\bar q_R q'_L|{\mathbf B}'\ra$ with results on coefficients $e^{-}_{\parallel}$ shown in Tables.~\ref{tab: eAeP}, \ref{tab: ePE} and \ref{tab: ePA}.
Note that the Clebsch-Gordan coefficients in Eqs.~(\ref{eq: TA=fac Delts S=0}) and (\ref{eq: TA=fac Delts S=-1}) canceled out with coefficients $e^{-}_{\parallel}$ and the asymptotic relations shown in Eqs. (\ref{eq: TA=fac asym Delta S=0}) and (\ref{eq: TA=fac asym Delta S=-1}) are established.



\begin{thebibliography}{99}

\bibitem{LHCb:2016nbc}
R.~Aaij \textit{et al.} [LHCb],
``Evidence for the two-body charmless baryonic decay $ {B}^{+}\to p\overline{\varLambda} $,''
JHEP \textbf{04}, 162 (2017)
doi:10.1007/JHEP04(2017)162
[arXiv:1611.07805 [hep-ex]].

\bibitem{Belle:2007gob}
Y.~T.~Tsai \textit{et al.} [Belle],
``Search for $B^0 \to p \bar p, \Lambda \bar\Lambda$ and $B^+\to p \bar\Lambda$ at Belle,''
Phys. Rev. D \textbf{75}, 111101 (2007)
doi:10.1103/PhysRevD.75.111101
[arXiv:hep-ex/0703048 [hep-ex]].

\bibitem{Belle:2007lbz}
M.~Z.~Wang \textit{et al.} [Belle],
``Study of $B^+ \to p \bar \Lambda \gamma, p \Lambda \pi^0$ and $B^0 \to p \bar \Lambda \pi^-$,''
Phys. Rev. D \textbf{76}, 052004 (2007)
doi:10.1103/PhysRevD.76.052004
[arXiv:0704.2672 [hep-ex]].

\bibitem{Belle:2007oni}
J.~T.~Wei \textit{et al.} [Belle],
``Study of $B^+ \to p \bar p K^+$ and $B^+ \to  p \bar p \pi^+$,''
Phys. Lett. B \textbf{659}, 80-86 (2008)
doi:10.1016/j.physletb.2007.11.063
[arXiv:0706.4167 [hep-ex]].

\bibitem{LHCb:2017swz}
R.~Aaij \textit{et al.} [LHCb],
``First Observation of the Rare Purely Baryonic Decay $B^0\to p\bar p$,''
Phys. Rev. Lett. \textbf{119}, no.23, 232001 (2017)
doi:10.1103/PhysRevLett.119.232001
[arXiv:1709.01156 [hep-ex]].

\bibitem{LHCb:2022lff}
 [LHCb],
``Search for the rare hadronic decay $B_s^0\to p \bar{p}$,''
[arXiv:2206.06673 [hep-ex]].


\bibitem{Belle:2019abe}
B.~Pal \textit{et al.} [Belle],
``Evidence for the decay $B^0\to p\bar{p}\pi^0$,''
Phys. Rev. D \textbf{99}, no.9, 091104 (2019)
doi:10.1103/PhysRevD.99.091104
[arXiv:1904.05713 [hep-ex]].

\bibitem{CLEO:1989xsn}
D.~Bortoletto \textit{et al.} [CLEO],
``A Search for $b \to u$ Transitions in Exclusive Hadronic $B$ Meson Decays,''
Phys. Rev. Lett. \textbf{62}, 2436 (1989)
doi:10.1103/PhysRevLett.62.2436


\bibitem{PDG} 
P.~A.~Zyla \textit{et al.} [Particle Data Group],
``Review of Particle Physics,''
PTEP \textbf{2020}, no.8, 083C01 (2020)
doi:10.1093/ptep/ptaa104


\bibitem{Cheng:2001tr} 
  H.~Y.~Cheng and K.~C.~Yang,
  ``Charmless exclusive baryonic $B$ decays,''
  Phys.\ Rev.\ D {\bf 66}, 014020 (2002)
  doi:10.1103/PhysRevD.66.014020
  [hep-ph/0112245].
  

  


\bibitem{Deshpande:1987nc}
N.~G.~Deshpande, J.~Trampetic and A.~Soni,
``Remarks On $B$ Decays Into Baryonic Modes And Possible Implications For V(Ub),''
Mod.\ Phys.\ Lett.\  {\bf 3A}, 749 (1988).


\bibitem{Jarfi:1990ej}
M.~Jarfi, O.~Lazrak, A.~Le Yaouanc, L.~Oliver, O.~Pene and
J.~C.~Raynal,
``Decays Of $B$ Mesons Into Baryon - Anti-Baryon,''
Phys.\ Rev.\ D {\bf 43}, 1599 (1991).


\bibitem{Cheng:2001ub}
H.~Y.~Cheng and K.~C.~Yang,
``Charmful baryonic $B$ decays $\bar B^0\to \Lambda_c \bar p$ and $\bar B \to  \Lambda_c \bar p \pi (\rho)$,''
Phys.\ Rev.\ D {\bf 65}, 054028 (2002) [Erratum-ibid.\ D {\bf 65},
099901 (2002)] [arXiv:hep-ph/0110263].


\bibitem{Chernyak:ag}
V.~L.~Chernyak and I.~R.~Zhitnitsky,
``$B$ Meson Exclusive Decays Into Baryons,''
Nucl.\ Phys.\ B {\bf 345}, 137 (1990).


\bibitem{Ball:1990fw}
P.~Ball and H.~G.~Dosch,
``Branching Ratios Of Exclusive Decays Of Bottom Mesons Into Baryon - Anti-Baryon Pairs,''
Z.\ Phys.\ C {\bf 51}, 445 (1991).


\bibitem{Chang:2001jt}
C.~H.~Chang and W.~S.~Hou,
``$B$ meson decays to baryons in the diquark model,''
Eur.\ Phys.\ J.\ C {\bf 23}, 691 (2002) [arXiv:hep-ph/0112219].


\bibitem{Gronau:1987xq}
M.~Gronau and J.~L.~Rosner,
``Charmless $B$ Decays Involving Baryons,''
Phys.\ Rev.\ D {\bf 37}, 688 (1988).


\bibitem{He:re}
X.~G.~He, B.~H.~McKellar and D.~d.~Wu,
``SU(6) Prediction Of $\Lambda_c$ Branching Ratio in $B$ Meson Decays,''
Phys.\ Rev.\ D {\bf 41}, 2141 (1990).


\bibitem{Sheikholeslami:fa}
S.~M.~Sheikholeslami and M.~P.~Khanna,
``$B$ Meson Weak Decays Into Baryon Anti-Baryon Pairs in SU(3),''
Phys.\ Rev.\ D {\bf 44}, 770 (1991).


\bibitem{Luo:2003pv}
Z.~Luo and J.~L.~Rosner,
  ``Final state phases in $B\to$ baryon anti-baryon decays,''
  Phys.\ Rev.\ D {\bf 67}, 094017 (2003)
  [hep-ph/0302110].

\bibitem{Chua:2003it} 
  C.~-K.~Chua,
  ``Charmless two body baryonic $B$ decays,''
  Phys.\ Rev.\ D {\bf 68}, 074001 (2003)
  [hep-ph/0306092].

\bibitem{Chua:2013zga}
C.~K.~Chua,
``Charmless Two-body Baryonic $B_{u,d,s}$ Decays Revisited,''
Phys. Rev. D \textbf{89}, no.5, 056003 (2014)
doi:10.1103/PhysRevD.89.056003
[arXiv:1312.2335 [hep-ph]].

\bibitem{Chua:2016aqy}
C.~K.~Chua,
``Rates and $CP$ asymmetries of Charmless Two-body Baryonic $B_{u,d,s}$ Decays,''
Phys. Rev. D \textbf{95}, no.9, 096004 (2017)
doi:10.1103/PhysRevD.95.096004
[arXiv:1612.04249 [hep-ph]].

\bibitem{He:2006vz}
X.~G.~He, T.~Li, X.~Q.~Li and Y.~M.~Wang,
``Calculation of $BR(\overline B^0 \to \Lambda_c^+ + \bar p$) in the PQCD approach,''
Phys. Rev. D \textbf{75}, 034011 (2007)
doi:10.1103/PhysRevD.75.034011
[arXiv:hep-ph/0607178 [hep-ph]].

\bibitem{Hsiao:2014zza} 
  Y.~K.~Hsiao and C.~Q.~Geng,
  ``Violation of partial conservation of the axial-vector current and two-body baryonic $B$ and D$_s$ decays,''
  Phys.\ Rev.\ D {\bf 91}, no. 7, 077501 (2015)
  doi:10.1103/PhysRevD.91.077501
  [arXiv:1407.7639 [hep-ph]].
  
\bibitem{Hsiao:2019wyd}
Y.~K.~Hsiao, S.~Y.~Tsai, C.~C.~Lih and E.~Rodrigues,
``Testing the $W$-exchange mechanism with two-body baryonic $B$ decays,''
JHEP \textbf{04}, 035 (2020)
doi:10.1007/JHEP04(2020)035
[arXiv:1906.01805 [hep-ph]].
  
\bibitem{Jin:2021onb}
X.~N.~Jin, C.~W.~Liu and C.~Q.~Geng,
``Study of charmless two-body baryonic $B$ decays,''
Phys. Rev. D \textbf{105}, no.5, 053005 (2022)
doi:10.1103/PhysRevD.105.053005
[arXiv:2112.13377 [hep-ph]].

\bibitem{Cheng:2014qxa} 
  H.~Y.~Cheng and C.~K.~Chua,
  ``On the smallness of Tree-dominated Charmless Two-body Baryonic $B$ Decay Rates,''
  Phys.\ Rev.\ D {\bf 91}, no. 3, 036003 (2015)
  [arXiv:1412.8272 [hep-ph]].



\bibitem{review}
H.~-Y.~Cheng and J.~G.~Smith,
  ``Charmless Hadronic B-Meson Decays,''
  Ann.\ Rev.\ Nucl.\ Part.\ Sci.\  {\bf 59}, 215 (2009)
  [arXiv:0901.4396 [hep-ph]].




\bibitem{Huang:2021qld}
X.~Huang, Y.~K.~Hsiao, J.~Wang and L.~Sun,
``Baryonic $B$ Meson Decays,''
Adv. High Energy Phys. \textbf{2022}, 4343824 (2022)
doi:10.1155/2022/4343824
[arXiv:2109.02897 [hep-ph]].


  
\bibitem{Zeppenfeld:1980ex}
D.~Zeppenfeld,
``SU(3) Relations For $B$ Meson Decays,''
Z.\ Phys.\ C {\bf 8}, 77 (1981).


\bibitem{Chau:tk}
L.~L.~Chau and H.~Y.~Cheng,
``Analysis Of Exclusive Two-Body Decays Of Charm Mesons Using The Quark Diagram Scheme,''
Phys.\ Rev.\ D {\bf 36}, 137 (1987).

\bibitem{Savage:ub}
M.~J.~Savage and M.~B.~Wise,
``SU(3) Predictions For Nonleptonic $B$ Meson Decays,''
Phys.\ Rev.\ D {\bf 39}, 3346 (1989) [Erratum-ibid.\ D {\bf 40},
3127 (1989)].

\bibitem{Chau:1990ay}
L.~L.~Chau, H.~Y.~Cheng, W.~K.~Sze, H.~Yao and B.~Tseng,
``Charmless Nonleptonic Rare Decays Of $B$ Mesons,''
Phys.\ Rev.\ D {\bf 43}, 2176 (1991) [Erratum-ibid.\ D {\bf 58},
019902 (1998)].


\bibitem{Gronau:1994rj}
M.~Gronau, O.~F.~Hernandez, D.~London and J.~L.~Rosner,
``Decays of $B$ mesons to two light pseudoscalars,''
Phys.\ Rev.\ D {\bf 50}, 4529 (1994) [arXiv:hep-ph/9404283].


\bibitem{Gronau:1995hn}
M.~Gronau, O.~F.~Hernandez, D.~London and J.~L.~Rosner,
``Electroweak penguins and two-body $B$ decays,''
Phys.\ Rev.\ D {\bf 52}, 6374 (1995) [arXiv:hep-ph/9504327].

\bibitem{Chiang:2004nm}
C.~W.~Chiang, M.~Gronau, J.~L.~Rosner and D.~A.~Suprun,
``Charmless $B\to P P$ decays using flavor SU(3) symmetry,''
Phys. Rev. D \textbf{70}, 034020 (2004)
doi:10.1103/PhysRevD.70.034020
[arXiv:hep-ph/0404073 [hep-ph]].

\bibitem{Cheng:2014rfa} 
  H.~Y.~Cheng, C.~W.~Chiang and A.~L.~Kuo,
  ``Updating $B\to PP,VP$ decays in the framework of flavor symmetry,''
  Phys.\ Rev.\ D {\bf 91}, no. 1, 014011 (2015)
  doi:10.1103/PhysRevD.91.014011
  [arXiv:1409.5026 [hep-ph]].



\bibitem{Brodsky:1980sx}
S.J.~Brodsky, G.P.~Lepage and S.A.~Zaidi,
``Weak And Electromagnetic Form-Factors Of Baryons At Large Momentum
Transfer,''
Phys.\ Rev.\ D {\bf 23}, 1152 (1981).

\bibitem{Aaij:2013fta} 
  R. Aaij {\it et al.}  [LHCb Collaboration],
  ``First evidence for the two-body charmless baryonic decay $B^0 \to p \bar{p}$,''
  JHEP {\bf 1310}, 005 (2013)
  [arXiv:1308.0961 [hep-ex]].


\bibitem{Li:2012cfa}
H.~n.~Li, C.~D.~Lu and F.~S.~Yu,
Phys. Rev. D \textbf{86}, 036012 (2012)
doi:10.1103/PhysRevD.86.036012
[arXiv:1203.3120 [hep-ph]].

\bibitem{Qin:2013tje}
Q.~Qin, H.~n.~Li, C.~D.~L\"u and F.~S.~Yu,
Phys. Rev. D \textbf{89}, no.5, 054006 (2014)
doi:10.1103/PhysRevD.89.054006
[arXiv:1305.7021 [hep-ph]].





\bibitem{Altmannshofer:2021qrr}
W.~Altmannshofer and P.~Stangl,
``New physics in rare $B$ decays after Moriond 2021,''
Eur. Phys. J. C \textbf{81}, no.10, 952 (2021)
doi:10.1140/epjc/s10052-021-09725-1
[arXiv:2103.13370 [hep-ph]].

\bibitem{Cornella:2021sby}
C.~Cornella, D.~A.~Faroughy, J.~Fuentes-Martin, G.~Isidori and M.~Neubert,
``Reading the footprints of the $B$-meson flavor anomalies,''
JHEP \textbf{08}, 050 (2021)
doi:10.1007/JHEP08(2021)050
[arXiv:2103.16558 [hep-ph]].

\bibitem{Buras} 
  A.~J.~Buras,
  ``Weak Hamiltonian, CP violation and rare decays,''
  hep-ph/9806471.


\bibitem{Beneke:2001ev} 
  M.~Beneke, G.~Buchalla, M.~Neubert and C.~T.~Sachrajda,
``QCD factorization in B $\to \pi K, \pi \pi$ decays and extraction of Wolfenstein parameters,''
  Nucl.\ Phys.\ B {\bf 606}, 245 (2001)
  [hep-ph/0104110].
 
 \bibitem{text}
T.~D.~Lee,
``Particle Physics And Introduction To Field Theory,''
Contemp.\ Concepts Phys.\  {\bf 1}, 1 (1981);
H.~Georgi, {\it Weak Interactions And Modern Particle Theory},
Benjamin/Cummings, 1984.
 

\bibitem{Chua:2002yd}
C.~K.~Chua and W.~S.~Hou,
``Three body baryonic $\bar B \to \Lambda \bar p \pi$ decays and such,''
Eur. Phys. J. C \textbf{29}, 27-35 (2003)
doi:10.1140/epjc/s2003-01203-8
[arXiv:hep-ph/0211240 [hep-ph]].

\bibitem{Chua:2001vh}
C.~K.~Chua, W.~S.~Hou and S.~Y.~Tsai,
``Understanding $B \to D^* N \bar N$ and its implications,''
Phys. Rev. D \textbf{65}, 034003 (2002)
doi:10.1103/PhysRevD.65.034003
[arXiv:hep-ph/0107110 [hep-ph]].

\bibitem{EM}
M.~Ambrogiani \textit{et al.} [E835],
``Measurements of the magnetic form-factor of the proton in the timelike region at large momentum transfer,''
Phys. Rev. D \textbf{60}, 032002 (1999)
doi:10.1103/PhysRevD.60.032002;
T.~A.~Armstrong \textit{et al.} [E760],
``Measurement of the proton electromagnetic form-factors in the timelike region at $8.9$ GeV$^{2} - 13$ GeV$^{2}$,''
Phys. Rev. Lett. \textbf{70}, 1212-1215 (1993)
doi:10.1103/PhysRevLett.70.1212;
M.~Castellano, G.~Di Giugno, J.~W.~Humphrey, E.~Sassi Palmieri, G.~Troise, U.~Troya and S.~Vitale,
``The reaction $e^+ e^- \to p \bar p$ at a total energy of 2.1 GeV,''
Nuovo Cim. A \textbf{14}, 1-20 (1973)
doi:10.1007/BF02734600;
G.~Bassompierre \textit{et al.} [Mulhouse-Strasbourg-Turin],
``First Determination of the Proton Electromagnetic Form-Factors at the Threshold of the Timelike Region,''
Phys. Lett. B \textbf{68}, 477-479 (1977)
doi:10.1016/0370-2693(77)90475-0;
G.~Bassompierre, M.~a.~Schneegans, G.~Binder, G.~Gissinger, S.~Jacquey, P.~Dalpiaz, P.~f.~Dalpiaz, C.~Peroni and L.~Tecchio,
``Electron positron pair production in anti-p p annihilation at rest and related determination of the electromagnetic form-factor of the proton in the timelike region,''
Nuovo Cim. A \textbf{73}, 347-363 (1983)
doi:10.1007/BF02724235;
B.~Delcourt, I.~Derado, J.~L.~Bertrand, D.~Bisello, J.~C.~Bizot, J.~Buon, A.~Cordier, P.~Eschstruth, L.~Fayard and J.~Jeanjean, \textit{et al.}
``Study of the Reaction $e^+ e^- \to p \bar{p}$ in the Total Energy Range 1925-{MeV} - 2180-{MeV},''
Phys. Lett. B \textbf{86}, 395-398 (1979)
doi:10.1016/0370-2693(79)90864-5;
D.~Bisello, S.~Limentani, M.~Nigro, L.~Pescara, M.~Posocco, P.~Sartori, J.~E.~Augustin, G.~Busetto, G.~Cosme and F.~Couchot, \textit{et al.}
``\ensuremath{>}A Measurement of $e^+ e^- \to \bar{p} p$ for 1975-{MeV} $\le \sqrt{s} \le$ 2250-{MeV},''
Nucl. Phys. B \textbf{224}, 379 (1983)
doi:10.1016/0550-3213(83)90381-4;
D.~Bisello \textit{et al.} [DM2],
``Baryon pair production in $e^+ e^-$ annihilation at $S^{1/2} = 2.4$ GeV,''
Z. Phys. C \textbf{48}, 23-28 (1990)
doi:10.1007/BF01565602;
G.~Bardin, G.~Burgun, R.~Calabrese, G.~Capon, R.~Carlin, P.~Dalpiaz, P.~F.~Dalpiaz, J.~P.~de Brion, J.~Derre and U.~Dosselli, \textit{et al.}
``Measurement of the proton electromagnetic form-factor near threshold in the timelike region,''
Phys. Lett. B \textbf{255}, 149-154 (1991)
doi:10.1016/0370-2693(91)91157-Q;
G.~Bardin, G.~Burgun, R.~Calabrese, G.~Capon, R.~Carlin, P.~Dalpiaz, P.~F.~Dalpiaz, J.~Derre, U.~Dosselli and J.~Duclos, \textit{et al.}
``Precise determination of the electromagnetic form-factor of the proton in the timelike region up to $s = 4.2$ GeV$^2$,''
Phys. Lett. B \textbf{257}, 514-518 (1991)
doi:10.1016/0370-2693(91)91929-P;
G.~Bardin, G.~Burgun, R.~Calabrese, G.~Capon, R.~Carlin, P.~Dalpiaz, P.~F.~Dalpiaz, J.~Derr\'e, U.~Dosselli and J.~Duclos, \textit{et al.}
``Determination of the electric and magnetic form-factors of the proton in the timelike region,''
Nucl. Phys. B \textbf{411}, 3-32 (1994)
doi:10.1016/0550-3213(94)90052-3;
A.~Antonelli, R.~Baldini, M.~Bertani, M.~E.~Biagini, V.~Bidoli, C.~Bini, T.~Bressani, R.~Calabrese, R.~Cardarelli and R.~Carlin, \textit{et al.}
``Measurement of the electromagnetic form-factor of the proton in the timelike region,''
Phys. Lett. B \textbf{334}, 431-434 (1994)
doi:10.1016/0370-2693(94)90710-2.

\bibitem{ParticleDataGroup:2018ovx}
M.~Tanabashi \textit{et al.} [Particle Data Group],
``Review of Particle Physics,''
Phys. Rev. D \textbf{98}, no.3, 030001 (2018)
doi:10.1103/PhysRevD.98.030001


\bibitem{Narison:2014ska}
S.~Narison,
``Improved $f_{D*_{(s)}}, f_{{B*}_{(s)}}$ and $f_{B_{c}}$ from QCD Laplace sum rules,''
Int. J. Mod. Phys. A \textbf{30}, no.20, 1550116 (2015)
doi:10.1142/S0217751X1550116X
[arXiv:1404.6642 [hep-ph]].

\bibitem{CKMfitter}
J.~Charles \textit{et al.} [CKMfitter Group],
``CP violation and the CKM matrix: Assessing the impact of the asymmetric $B$ factories,''
Eur. Phys. J. C \textbf{41}, no.1, 1-131 (2005)
doi:10.1140/epjc/s2005-02169-1
[arXiv:hep-ph/0406184 [hep-ph]];
updated results at: http://ckmfitter.in2p3.fr


\bibitem{FSI}
H.~Y.~Cheng, C.~K.~Chua and A.~Soni,
  ``Final state interactions in hadronic $B$ decays,''
  Phys.\ Rev.\ D {\bf 71}, 014030 (2005)
  doi:10.1103/PhysRevD.71.014030
  [hep-ph/0409317];
C.~K.~Chua,
  ``Rescattering effects in charmless $\bar B_{u,d,s}\to P P$ decays,''
  Phys.\ Rev.\ D {\bf 78}, 076002 (2008)
  doi:10.1103/PhysRevD.78.076002
  [arXiv:0712.4187 [hep-ph]];
  
\bibitem{Chua:2018ikx}  
C.~K.~Chua,
``Revisiting final state interaction in charmless $B_q\to PP$ decays,''
Phys. Rev. D \textbf{97}, no.9, 093004 (2018)
doi:10.1103/PhysRevD.97.093004
[arXiv:1802.00155 [hep-ph]].


\bibitem{Ciuchini:1997rj}
M.~Ciuchini, R.~Contino, E.~Franco, G.~Martinelli and L.~Silvestrini,
``Charming penguin enhanced B decays,''
Nucl. Phys. B \textbf{512}, 3-18 (1998)
[erratum: Nucl. Phys. B \textbf{531}, 656-660 (1998)]
doi:10.1016/S0550-3213(97)00768-2
[arXiv:hep-ph/9708222 [hep-ph]].




  
  
\bibitem{Moroi:1995fs} 
  T.~Moroi,
  ``Effects of the gravitino on the inflationary universe,''
  hep-ph/9503210.

\end{thebibliography}
\end{document}